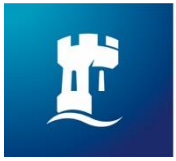
University of Nottingham
UK | CHINA | MALAYSIA

EEE Department of Electrical and Electronic Engineering

# EEEE4008

# Final Year Individual Project Thesis

## New Reservoir Computing Kernel Based on Chaotic Chua Circuit and Investigating Application to Post-Quantum Cryptography

**AUTHOR:**      **Mr Matthew John Cossins**

**ID NUMBER:**      **20133950**

**SUPERVISOR:**      **Dr Sendy Phang**

**MODERATOR:**      **Dr Christian Klumpner**

**DATE:**      **09/05/2024**

Fourth year project report is submitted in part fulfilment of the requirements of the degree of Master of Engineering.

# Executive Summary

The aim of this project was to develop a new Reservoir Computer implementation, based on a chaotic Chua circuit. In addition to suitable classification and regression benchmarks (Section 3.2), the Reservoir Computer was applied to Post-Quantum Cryptography, with its suitability for this application investigated and assessed, Section 3.4. The cryptographic algorithm utilised was the Learning with Errors problem, for both encryption and decryption. To achieve this, the Chua circuit was characterised, in simulation, Section 2.1.1c), and by physical circuit testing, Section 2.2.1c). The Reservoir Computer was designed and implemented using the results of the characterisation (Sections 2.1.2 and 2.2.2). As part of this development, noise was considered and mitigated (Sections 2.2.1c)iv and 3.8).

The benchmarks demonstrate that the Reservoir Computer can achieve current literature benchmarks with low error. However, the results with Learning with Errors suggest that a Chua-based Reservoir Computer is not sufficiently complex to tackle the high non-linearity in Post-Quantum Cryptography. Future work would involve researching the use of different combinations of multiple Chua Reservoir Computers in larger neural network architectures. Such architectures may produce the required high-dimensional behaviour to achieve the Learning with Errors problem.

This project is believed to be only the second instance of a Chua-based Reservoir Computer in academia, and it is the first to be applied to challenging real-world tasks such as Post-Quantum Cryptography. It is also original by its investigation of hitherto unexplored parameters, and their impact on performance. It demonstrates a proof-of-concept for a mass-producible, inexpensive, low-power consumption hardware neural network. It also enables the next stages in research to occur, paving the road for using Chua-based Reservoir Computers across various applications.

# Contents

# Subject-Specific Terms

- Artificial Neural Network (**ANN**): *computing system mimicking biological neurons to complete tasks.*
- Bifurcation: *in chaotic systems, adjusting a parameter may cause doubling from one solution to two.*
- Chaos: *sensitive system – a small perturbation in conditions disproportionately impacts the output.*
- Chaotic Attractor: *different states of a chaotic system: illustrated in Chua by parametric of outputs.*
- Chua Circuit: *simplest electronic circuit producing chaotic behaviour.*
- Classification: *discrimination of inputs into distinct output classes – a typical task for ANN.*
- Kernel: *method used to increase dimensions of ANN system – in this case a Chua circuit reservoir.*
- Learning With Errors (**LWE**) Cryptography: *basis for new public key encryption methods, including multiple algorithms forming the new encryption standard. Quantum-secure.*
- Multiplexing/Demultiplexing: *integrating multiple signals into one, and then segregating the output back into the original number of signals.*
- Normalised Mean Square Error (**NMSE**): *measure of scoring performance of ANN.*
- Perturbation: C*hange in parameters and/or conditions of a system.*
- Post-Quantum Cryptography (**PQC**): *cryptographic techniques resistant to quantum computers.*
- Public Key Encryption (**PKE**): *modern cryptographic technique for encrypting communications.*
- Regression: *determining the mathematical relationship between input variables and output variables.*
- Reservoir Computing (**RC**): *relatively new form of ANN based on black box reservoir with I/O.*

# Tables

# Figures

# 1) Introduction

This section will briefly introduce the project, explaining the aims, deliverables, and specification. It will also cover the methodology proposed in the project planning report.

## 1.1  Project Background

This project aimed to develop an original Reservoir Computer (**RC**), based on a chaotic Chua circuit, and to investigate its applicability to Post-Quantum Cryptography (**PQC**).

Traditional computers rely on digital semiconductor systems. At the lowest level data is restricted to binary. Processing is centralised in a Central Processing Unit (**CPU**) and kept separate from memory; this is von Neumann architecture [1]. This system undeniably works but is in opposition to how the human brain and nervous system operates. Computing in humans is distributed and analogue; memory and processing occurs together and there is no binary restriction on data [2]. It is also highly efficient, performing 100 TFlops per second with only 20 W of power [1]. It is arguable that instead of implementing artificial neural networks (**ANN**) in traditional digital computing, there are significant speed, memory, and power advantages to using analogue, distributed systems [1]. Artificial Intelligence (**AI**) is currently a highly prominent and socially relevant field, but most research is wedded to the digital computing paradigm [3]. This project aimed to address this lack of research into analogue computing through using an analogue Chua circuit as a RC kernel.

There is also a clear engineering need for the application to PQC; the on-going development of PQC algorithms is vital to protecting private data from the Quantum Computing (**QC**) threat [4]. The National Institute of Standards and Technology (**NIST**) have called for further research into the field [4]. With adoption of PQC algorithms expected from 2026 [5], now is the time to consider implementation. The PQC market is expected to grow at a staggering 95.2% per year between 2024-2030 [6], demonstrating that the threat is considered extremely serious. This project planned to develop an original implementation using a chaotic RC and to assess the broader viability of analogue reservoir computing to PQC. The Chua circuit is the simplest form of chaotic circuit – capable of generating deterministic but pseudo-random outputs that are significantly impacted by small perturbations [7, 8]. Whilst the circuit itself has been heavily researched, its usage in applications has been limited. Its use as a basis kernel for a RC has been demonstrated only once [9] and the application of such a RC to PQC is original. In the first stage of the project, the Chua circuit was characterised through numerical simulation and mathematical analysis, with a focus towards developing a Chua-based RC (**Chua-RC**). Potential issues (e.g., noise) were assessed and mitigated. The second stage was experimental realisation and testing of the circuit implemented on printed circuit board (**PCB**). Finally, the circuit was integrated with MATLAB in the third stage. It was designed to receive input from a MATLAB-controlled arbitrary function generator (**AFG**). Its outputs could then be processed and trained, via MATLAB, utilising Algorithm 1 specified in [10]. The RC was then applied to performing emerging front-runners of PQC, Lattice-Based (**LB**) and Learning with Errors (**LWE**) algorithms [4, 5].

## 1.2  Originality and Reservoir Computer State-of-Art

A RC is a relatively new concept and is a type of artificial neural network (**ANN**) that consists of an Input/Output (**I/O**), and a neuron kernel [11]. This can be seen in Figure 1(a). Examples include optical RCs using silicon photonic neuron kernels [10]. Such a kernel can also be based on a Chua circuit, as has previously been developed only once [9]. The paper developing this kernel has used a modified Murali-Chua circuit, whilst this project intends to start with the Chua circuit specified by Kennedy [7]. Moreover, the paper calls for the need for assessing more challenging applications, including time-varying inputs, and for further work investigating other RC parameters [9]. This project, with a differing Chua circuit implementation, attempted and achieved both: the first through the application to PQC, and the second through assessing the effect of resistance and input voltage range on performance. It also performed original tapping of physical signals at multiple points in the circuit. These were time-multiplexed to generate additional read-outs and then processed by an adaptation of the training algorithm in [10]. A high-level overview of the system can be seen in Figure 1(b).

This project believes that the Chua circuit is an ideal basis kernel for further exploration, due to it being based on low-cost and accessible components, and it being minimalistic in design [7]. In addition to being an original application of a chaotic RC to PQC, with an original, bespoke development approach, the Chua-RC is suitable for widescale use; a necessity as PQC algorithms become ubiquitous in modern society. It also reduces the limitations inherent in the project not being a commercial venture. The physical testing can be conducted in a university environment. Use of more affordable components may prevent the improvement of parameters such as noise compared to other solutions, however if a baseline acceptable range is achieved, this work will have provided an original, low-cost, and mass-producible template for applying a chaotic RC to PQC.

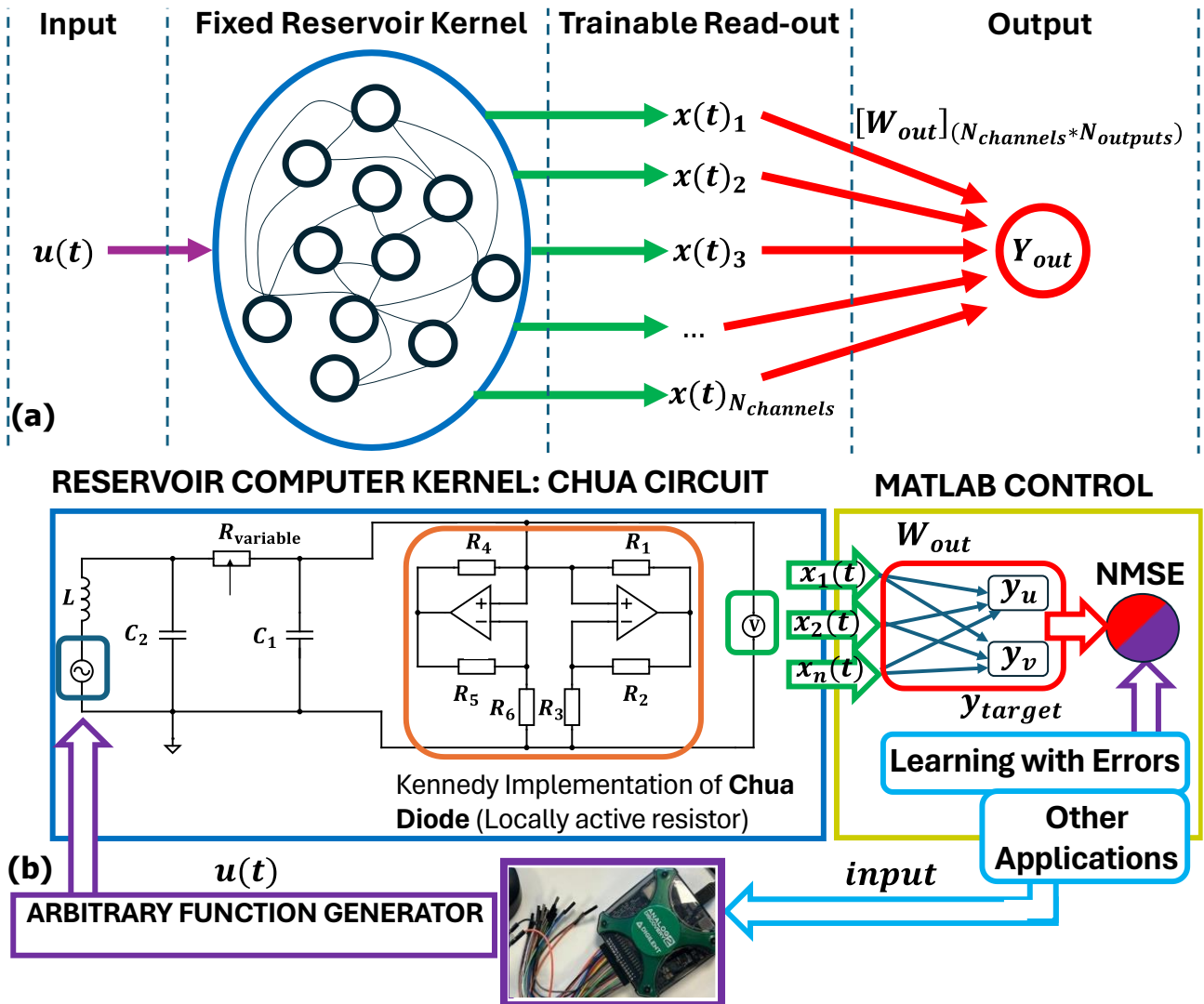

**Figure 1- (a)** Reservoir computing architecture. Processing of input $u(t)$ into neuron activation states $x(t)$. Use of weight $W_{out}$ to then obtain output $Y_{out}$, **(b)** System level block diagram: Chua circuit reservoir kernel with MATLAB I/O control. $Y_{out}$ adjusted to fit required output for LWE encryption.

## 1.3 Aim and Specification

The final deliverable goal was an extensively tested Chua-RC, benchmarked in performance using relevant tasks, and applied to PQC. Testing was to consist of both simulation and physical circuit testing, and design choices for the Chua-RC were to be supported by prior MATLAB/LTSpice simulations and experimental testing of a Chua circuit. The experiment aimed to achieve at least a Signal-Noise-Ratio (**SNR**) of 10 dB [8], (ideally as high as possible) and a kHz range bandwidth. Characterising the noise is an important step in the process as it allows assessment of any risk posed by noise to successful and reproducible Chua-RC results.

### 1.3.1 Deliverables

| Core Deliverable | Requirements/Testing |
|---|---|
| **D1.** Characterise Chua circuit (**Simulation**) - bifurcation and parametric voltage plots demonstrating chaos and displaying key regions and parameters for tuning and optimising future RC. | Bandwidth matching or exceeding kHz range[1]. SNR[2] exceeding 10 dB. This can be tested using scope/Fast Fourier Transform (**FFT**) of LTSpice/MATLAB. The usual $-3$ dB measurement of bandwidth is not appropriate for chaotic circuits so instead testing will be achieved by measuring "the frequency of the simplest unstable periodic orbit" [12]. Deliverable demonstrated in Section 2.1.1c). |
| **D2.** Chua circuit experiment with comparison to simulation (**Experiment**) | Match results from simulation as close as possible (bandwidth, SNR, bifurcation diagrams and parametric plots). Test using oscilloscope or OpenScopeMZ (**OSMZ**)/Analog Discovery 2 (**AD2**) [3]. Test both breadboard and Veroboard/PCB. Determine final choice of investigation regions for tuning of RC. Deliverable demonstrated in 2.2.1c). |
| **D3.** Chua Reservoir Computer with MATLAB (**Application**) | Successful perturbing initial conditions of RC using MATLAB-controlled input, followed by successful read-out/display of data into MATLAB and processing/training using read-out algorithm. Can be tested on suitable RC benchmarks such as the regression and classification tasks in [9]. A further objective would be to adapt into a real-time solution. Deliverable demonstrated in Section 3.2. |
| **D4.** Reservoir Computer performing PQ encryption. Investigate whether this is within the capability of Chua-RC (**Application**) | MATLAB implementation of PQC encryption [4]. Use of generated values as a teacher for Chua-RC. Training of RC and investigation of different tuneable parameters to optimise system to "learn" the PQC method. Further would be to adapt into a multi-bit message system. Deliverable demonstrated in Section 3.4. |
| **D5.** Ensure noise kept minimal through methods as required (e.g. PCB, filter, alternative Chua Diode) (**Simulation/Experiment**) | Deliverable if SNR is outside of the acceptable range ($< 10$ dB), or other concerns/additional time available. This is tested as above, e.g. with AD2 or OSMZ. Alternative Chua Diode implementations may provide side-benefit of greater bandwidth. The system can also be assessed through repeated Chua Diode V-I characteristic plotting to identify changes due to noise and assess reproducibility – the system must be sensitive to produce chaos but not be overly sensitive to noise. Deliverable demonstrated in Sections 2.2.1c)iv and 3.8. |

**Table 1-** Core Specification Deliverables.

[1] Chua bandwidth is typically limited by Operational-Amplifiers (**Op-Amp**), providing a kHz range [12]. Improving bandwidth is not within the project scope, the aim is application to Chua-RC for PQC, but it is important to provide bandwidth at least within this range.

[2] A Kennedy-based Chua circuit was simulated with an SNR of just over 10 dB [8]. This provides a baseline, but through choices such as superior Op-Amps (AD712), the $> 20$ dB range can be aimed at.

[3] AD2 has built-in SNR measurement [13]. OSMZ data can undergo Fast Fourier Transform to obtain SNR.

[4] The project is limited by sparse MATLAB implementations of PQC compared to conventional encryption, so this is developed as part of this project.

**D1** and **D2** aim for a realistic, attainable specification that will enable the original application to PQC. This specification can be tested by the chosen tools. The application was to be benchmarked against examples from literature [9] in **D3**, before ambitiously being assessed in **D4** with a PQC encryption implementation. This is a completely original application of a Chua-RC to PQC. Additional stretch goals were also devised and can be seen in Table 2 below. Originally, the goal of applying a full NIST algorithm, such as CRYSTALS-KYBER to the Chua-RC was intended. It became clear that this was over-ambitious and beyond the project scope. The decision to combine this objective with **D7** is discussed in Section 2.1.3c).

**Table 2-** Stretch Specification Deliverables.

| Stretch Deliverable | Requirements/Testing and Expectation |
|---|---|
| **D6.** Develop Reservoir Computer to perform decryption of PQC method. (**Application**) | This can be tested in the same way as encryption in **D4**. It is dependent on the results from **D4** as the Chua-RC performance in decryption will likely be closely related to its performance on encryption. This is a stretch deliverable as it may exceed the available time in the project. |
| **D7.** Apply final system to additional PQC | This can be tested as in **D4**. The application to additional algorithms will likely not be possible due to time constraints. This would require time to understand each algorithm, create MATLAB implementations, and translate the encryption |
| encryption algorithms (**Application**) | process to Chua-RC. This deliverable would be contingent on early results showing Chua-RC is well-suited to PQC problems. |
| **D8.** Publishing results to an academic journal paper | This would be a significant, but rewarding, challenge. This will depend on time constraints and the success of hitting deliverables/the usefulness of the investigation's results. This will be considered only after the completion of the project. |

### 1.3.2 Methodology

For project management, the plan was to use a weekly sprint structure, managed by the Jira tool and bookended by each supervisor meeting. Each meeting would set the next sprint's objectives. Sprints may be expanded to cover multiple weeks, depending on the tasks involved. Within Jira, tasks can be designated a stage. Prior to each meeting, slides summarising the previous sprint's work should be emailed to the Supervisor. A Microsoft Team for additional file-sharing and communication was set-up to be shared by both supervisor and student, prior to the project proposal. This adapts from the software development Scrum methodology. It also implements aspects from the Agile methodology: with documentation periods provided for reflection and making changes. The overall planned methodology: an agile, scrum-style approach split into three stages (**Simulation**, **Experimentation**, and **Application**), demonstrated a bespoke approach customised to the project.

#### a) Numerical Simulation

The Simulation stage was planned to cover the completion of **D1**. The tools required were LTSpice and MATLAB. Initial simulations were conducted in LTSpice, but MATLAB proved advantageous for producing bifurcation diagrams due to its handling of steady-state and its flexibility in programmability. The programs are limited in their simulation of real-world conditions, but can incorporate some realistic practical parameters, including non-idealities inherent in components and noise. They were ideal due to student licenses available through the University. A refresher on using MATLAB for simulating circuits was required for this stage. The challenge posed was medium due to prior experience with MATLAB/LTSpice, however the mathematics involved in dynamical systems required additional time to understand. In the case that noise was an issue, or the project was progressing ahead of schedule, this stage planned additional work such as designing a noise filter in LTSpice/MATLAB or a PCB in KiCAD (**D5**). This challenge was medium due to experience in PCB/Filter Design, but strict time constraints. A KiCAD refresher (approx. ½-day) was planned for the PCB Design.

#### b) Experimental Approach

The experiment stage was identified as critically dependent on the simulation. **D2** requires simulation for determining the circuit design. The circuit construction plan involved designing a Veroboard/PCB, ordering appropriate components, and soldering. Initial testing was to use oscilloscopes and AFGs. Further testing will use the Digilent OSMZ – a freely available tool, but limited due to lacking support from Digilent [14]. If this proved problematic, the purchase of a high-quality AD2 was identified as an alternative. If noise was an issue, additional mitigation designs, completed for **D5**, were to be constructed. The challenge was medium because chaotic features could easily be prevented by imperfect construction. There is a limitation on quality with "by-hand" compared to commercial machine soldering, but there was intent to source multiples of components, allowing for a draft and final construction, in addition to breadboard drafts.

#### c) Application

The final planned stage was the application to Chua-RC and investigating its ability to perform PQC. This stage was limited by dependency on the simulation and experiment and was expected to be the most challenging. This was mitigated by ensuring half of the stage would operate independent of earlier stages. The initial work would involve researching and understanding LB/LWE algorithms and attempting MATLAB implementation. It would also include developing a MATLAB system for handling the RC I/O. The I/O control would be limited by whether the OSMZ could be MATLAB-controlled, so this work would establish whether an AD2 was needed. The I/O control was identified to require familiarisation with the Data Acquisition, and Instrument Control Toolboxes, challenging due to no prior experience with these toolboxes. However, MATLAB's versatility and wealth of options made it the logical choice. The second half of the stage, integration with the circuit to form an RC (**D3**) was dependent on the completion of the experiment stage. **D4** (applying the RC to PQC) was dependent on **D3**. Extensive research time was provided early to reduce difficulty and ensure obstacles arose before the conclusion of prior stages. An extensive final testing stage was also planned. Many different tuneable parameters could be both simulated and experimentally tested to attempt finding an optimal region for Chua-RC for the PQC problem. The stretch goals **D6**, **D7**, and **D8** would be considered afterwards.

# 2) Design and Implementation

## 2.1 System Modelling

### 2.1.1 Chua Circuits for Reservoir Computing

#### a) Reservoir Computing

A Reservoir Computer is an implementation of a recurrent neural network (**RNN**), a type of ANN [2, 15]. A reservoir, in this context, refers to a non-linear dynamical system, see Section 1.2 and Figure 1(a). The reservoir is underlined; a black box with inputs and outputs [2]. A trainable read-out processes the reservoir output, compares with a teacher, and determines an ideal weight. This weight can be used on validation input data to generate an estimate for the correct output. This estimate can be compared with the desired target output to determine performance. The black box nature of RC is a key advantage over alternative neural networks, as the kernel can remain fixed, with training only performed at the read-out [16]. There is no training performed on the input or in the kernel. This means a dataset can be processed in the kernel once to generate an output. This output can be repeatedly retrained with different teachers to perform different tasks [1]. The non-linearity of the kernel transforms inputs into high-dimensional outputs (this can be increased through time-multiplexing), allowing a training read-out to discriminate [1]. It contains only a single hidden layer, the kernel, reducing the complexity of the system whilst able to compute difficult non-linearly separable tasks [16].

Many different reservoir kernels have been implemented. Whilst a non-linear system can be virtualised with traditional computers, non-linear systems can also be found within nature, or within physical mechanical or electrical analogue systems [16]. Using a physical analogue kernel reduces the processing performed in software and assigns the non-linear behaviour to analogue hardware, exploiting the speed advantages of such hardware [16] and utilising its intrinsic computation [9]. One family of reservoirs currently used widely in academia are those implemented via optical circuits [10]. Another set of reservoirs are chaotic and memristor circuits – including the Chua circuit [9]. A key requirement of a reservoir kernel is being sensitive to perturbation and thereby having a good separation property. This means that there will be significant variance between distinct inputs, due to the transformation to higher dimensions [10]. One aspect of this paper will be the investigation of the Chua circuit dynamics as different inputs are applied, and for different fixed conditions. This will assess the ideal conditions for optimal performance.

#### b) Kernel based on Chua Circuit

Chaos Theory is the study of chaotic systems, which can arise from a variety of sources. These systems are sensitive to changes in initial conditions; very small perturbations in the system can have a disproportionate impact on the resultant behaviour. Lorenz discovered that small changes to initial conditions when modelling weather led to extremely different predictions [17]. This is an example of bifurcation and has been popularised as the Butterfly Effect: a butterfly flapping wings causing an extreme weather event in another region on the globe [18]. Chaos, in this context, is a scientific term referring to output data that appears random and therefore indeterministic. It appears this way because of the sensitivity of a chaotic system to perturbation [18]. In the case that it was possible to exactly know the initial conditions, the result would be completely deterministic. The study of Chaos theorises that were it possible to know all initial conditions, everything could be predicted. Chaotic behaviour occurs in a range of natural and artificial systems, and this thesis will focus upon the Chua circuit, discovered by Leon Chua [17].

The Chua circuit has been demonstrated as the basis kernel of a RC only once in literature [9]. Jensen implemented a form of Chua circuit as a RC for applications with a non-time-varying input: non-linear polynomial regression and concentric circles classification. The paper calls for further investigation into Chua circuits and their application to Reservoir Computing, such as with time-varying inputs, and with more complex applications. This thesis aims to reproduce Jensen's work, but with a different Chua circuit (Kennedy implementation) and to extend this to a real-world application with time-varying inputs – PQC. It will also perform extensive investigation into areas not considered in the Jensen paper, such as other fixed conditions of the Chua circuit.

The Chua circuit (also called Chua's circuit) is the simplest electronic circuit capable of exhibiting chaotic phenomena. It enables an inexpensive and practical implementation of a chaotic system [19]. With fixed conditions but variable driven input, it can be used as a chaotic black box [20]. As stated by Kennedy [19], to produce chaos, a circuit must consist of:

  **Condition 1.**          At least one nonlinear component
  **Condition 2.**          At least one locally active resistor

**Condition 3.**          At least three energy storage elements

The Chua circuit has two capacitors and an inductor – achieving Condition 3. It also has a Chua diode (which is both a non-linear element and a locally active resistor [19]), achieving Condition 1 and Condition 2. Chua's circuit evolved from a linear parallel RLC circuit [19], demonstrated in Figure 2(b). Figure 2(a) shows a Kennedy implementation of the Chua circuit.

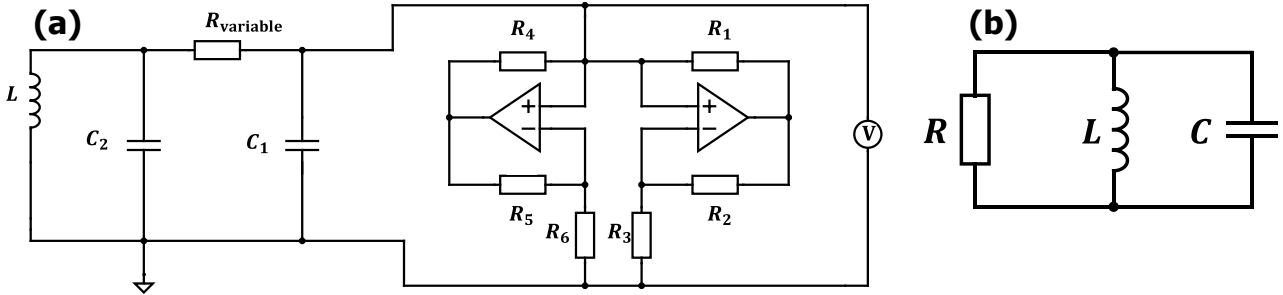

**Figure 2- (a)** Chua circuit (Kennedy Chua diode), **(b)** Parallel RLC circuit.

The linear parallel RLC can be characterised by the following ordinary differential equations,

$$\frac{dI_l}{dt} = -\frac{1}{L}V_c$$

$$\frac{dV_c}{dt} = \frac{1}{C}I_l - \frac{1}{C}I_r = \frac{1}{C}I_l - \frac{1}{CR}V_c$$

(1)

The Chua circuit can be described with the following state equations [21],

$$\frac{dI_l}{dt} = -\frac{1}{L}V_{c2}$$

$$\frac{dV_{c2}}{dt} = \frac{1}{C_2}I_l - \frac{1}{RC_2}(V_{c2} - V_{c1})$$

(2)

$$\frac{dV_{c1}}{dt} = \frac{1}{RC_1}(V_{c2} - V_{c1}) - \frac{1}{C_1}f(V_{c1})$$

where $f(V_{c1})$ represents is the non-linear function of the Chua diode.

There are various implementations of a Chua circuit, including those designed to have high frequency responses [12], those implemented in Field Programmable Gate Arrays (**FPGA**) [22], and many different variants of how best to construct the Chua diode component – not an off-the-shelf component [21]. Whilst this thesis will focus primarily on the Kennedy implementation, the work conducted can theoretically be applied to other implementations, albeit with adjustment to tuneable parameters to ensure optimal conditions. As will be discussed (Section 4.1.3), high frequency Chua implementations offer the potential for input voltage signals to be applied at much higher frequencies, allowing the system to scale up its speed. FPGA implementations also offer a potential route for Chua circuits to be used in a mass-produced device and can reduce noise.

It should be noted that a common use of Chua circuits is to utilise the generated chaos directly to perform communications and image encryption [8, 23]. This is not the goal of this thesis. This thesis will instead attempt to use a Chua circuit as the kernel of a RC. This RC will then be applied to performing PQC encryption by learning the mathematical operations that translate an input plaintext message into an encrypted output.

To use the Chua circuit for an application, its properties must be understood. There are several tuneable circuit parameters. This thesis will examine varying the resistance and the capacitance of the circuit. It will also look at varying the amplitude range of input voltage. To obtain the characteristics of the circuit for varying these parameters, bifurcation diagrams can be produced. These show the circuit voltage response as a parameter is varied and show points of bifurcation – where the circuit "jumps" into a more chaotic region. They are a central task to characterising a dynamical system [24]. It is commonly believed that RC systems perform best on the edge of chaos. A small perturbation, such as that caused by input voltage, pushes the RC into chaos. Carroll demonstrates that this is not true in general, however it is a good starting point when assessing optimal performance regions [25]. Ideal parameters can be selected to achieve this based on the work performed in this thesis.

The Chua circuit contains a variety of complex bifurcations [20]. The different outputs can be categorised loosely into six main states. The first, "DC equilibrium", is above the region of chaos. The circuit behaves linearly in this region. As the system is pushed into the chaotic region, it forms a "single period" state. As parameters are further adjusted, period-doubling occurs and the state transforms through various "multi-period" outputs and then into the distinctive "Rössler-type" single scroll. This followed by bifurcation into a "double scroll", before finally becoming a "large limit cycle" as the limits of the Chua diode are reached [21]. These states are demonstrated through simulation and experimental testing in Sections 2.1.1c) and 2.2.1c).

### c) Simulation

An important first step in the project was to understand the underlying circuit dynamics of the Chua circuit. Simulation enables more accurate modelling of the circuit than the ordinary differential equations outlined in the previous section. A first principles mathematical model is an acceptable choice, but in addition to being less representative of real circuitry, there are also limited options for modelling real-world noise. Moreover, developing the model from first principles with sufficient customisation options could take more time than producing a circuit simulation. The key objectives from the simulation of the circuit were:

**Sim Objective 1.**   Understand the circuit dynamics when varying component values.
**Sim Objective 2.**   Generate a typical expected output for defined initial conditions.
**Sim Objective 3.**   Investigate and understand the impact of an input voltage signal on the circuit output.
**Sim Objective 4.**   Generate bifurcation diagrams for resistance, capacitance, and voltage amplitude.
**Sim Objective 5.**   Assess expected bandwidth.
**Sim Objective 6.**   Assess the impact of noise and component non-idealities.

Two options were considered: MATLAB and LTSpice. The benefits of using each are summarised in Table 3.

| LTSpice Advantages | MATLAB Advantages |
| --- | --- |
| Accurate manufacturer component models | Programmability/automation, Parallel simulation |
| More intuitive to use and easier to assess different areas of the circuit | Transition between simulation and application stage (interchangeability of simulation and circuit) |
| FFT frequency spectrum mode | Wide array of options for customising noise |

**Table 3-** LTSpice vs MATLAB tools for simulating Chua circuit.

Initially, the simulation was developed in LTSpice. This was a natural choice due to prior experience with the software. Crucially, the LTSpice software comes with many pre-built models of industry components. In the case a component is not preset in the software, there are freely available models that can be downloaded. This means the simulation is realistic to the real-world – generic component models can be ignored in favour of exact, manufacturer specification, models that capture any intricacies of a particular component's characteristics. This meant that for Simulation Objectives 1, 2, 3, and 5, the LTSpice software was ideal and best-placed to provide accurate data. LTSpice was therefore critical in the early characterisation of the Chua circuit.

However, LTSpice is limited in its customisation and in its programmability. Basic instruction options do exist; for example, LTSpice can be configured to repeat a simulation multiple times whilst varying a parameter, with a plot against this varying parameter generated. This is very rudimentary and does not allow subsequent post-processing of the obtained data within LTSpice. When generating bifurcation diagrams, it is important to remove initial transient behaviour and instead sample the steady state output. Since the initial transient behaviour will vary in length as circuit parameters are adjusted, this is difficult to automate within LTSpice. It can be achieved more effectively with MATLAB Simulink; a superior choice for Simulation Objective 4. There are also limitations on the types of voltage signals that can be generated within LTSpice. It is far easier to generate arbitrary waveforms within MATLAB than LTSpice. In MATLAB it is computationally efficient to pass an array of arbitrary data into Simulink as a voltage signal. The Chua-RC application will require arbitrary waveforms, so whilst LTSpice is well-placed to assess conventional waveforms such as sine, square, DC, or sawtooth, investigating more complex voltage function for Simulation Objective 3 requires MATLAB.

The use of MATLAB Simulink has been considered carefully. Its accuracy is likely to be less than LTSpice since it uses only generic component models that require parameterisation from datasheets. Compared with manufacturer models in LTSpice, this is inferior. However, the programmability makes it extremely useful. The Simulink model can be customised and launched from within a MATLAB script. This means that scripts can be written to automate initialising different inputs, launching the simulation a desired number of times, performing appropriate post-model-processing on outputs, and generating graphs. Small tweaks and changes can easily be made. In the case of generating bifurcation diagrams, for example, the script can obtain the data output from Simulink and then sample only the steady-state behaviour.

Another reason for using MATLAB in the simulation stage is that it made the simulation to application transition far more seamless. MATLAB is used as the control code for the reservoir computer in the application. This code performs initial processing on the input data and interfaces with an external function generator to establish a voltage signal. This voltage signal is input to the Chua circuit, and an external oscilloscope captures circuit outputs and feeds them back to MATLAB. The post-processing and reservoir computer training follow before performance metrics are established. Additionally using MATLAB for the simulation stage enables the simulated Reservoir and the physical circuit Reservoir to be interchanged for the same I/O control code.

Another significant benefit of MATLAB is the ability to simulate non-idealities and noise in components. Since the Chua circuit is a system very sensitive to initial conditions, it is important to assess the effects of uncertainty and thermal noise. MATLAB allows each component to be assigned a tolerance, reflecting that passive components will not have exact desired values, and this tolerance can be assigned a random distribution such as Gaussian. The components can also be assigned Johnson noise [26]. Two identical circuits can be simulated – however one with noise and the other without, and their results compared to understand the impact of non-ideal components. Additionally, input noise can be simulated by parameterising a voltage signal to the circuit with the noise data of the chosen function generator, with an estimate of Signal to Noise Ratio (**SNR**) calculated. Finally, by characterising the Chua diode with a voltage-current plot under ideal and noisy conditions, a benchmark can be achieved that can easily be compared against experimental results. As such, MATLAB will be utilised for Simulation Objective 6.

### i. LTSpice Simulation and Results

The Kennedy implementation of the Chua circuit is simulated in LTSpice in Figure 3. The circuit is simulated using the AD712 Op-Amp model. This is one of several potential Op-Amps that can be used to produce the correct V-I characteristics for the Chua diode. Compared with alternatives such as the TL072, the AD712 is a higher quality (albeit more expensive) component, so this was chosen as the best Op-Amp to utilise. A voltage supply of $\pm 9$ V is provided to the Op-Amps and the inductor is simulated with a series resistance of $17 \, \Omega$ [27].

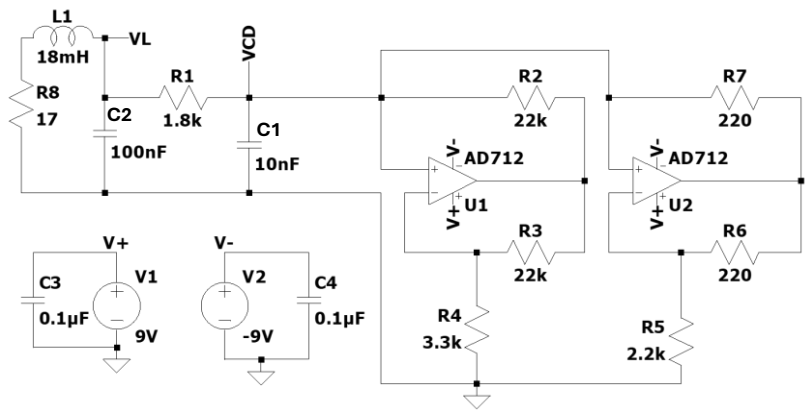

**Figure 3-** LTSpice simulation of Chua circuit based on Kennedy implementation.

Two Chua circuit outputs are considered in this thesis, referred to henceforth as the Inductor Voltage, and the Chua Diode Voltage. They can be seen in Figure 3 as $V_L$ and $V_{CD}$ respectively. Adjusting $R_{\text{variable}}$ ($R_1$ in Figure 3) allows the circuit to generate a wide range of distinct time domain voltage outputs. The capacitance $C_1$ is fixed at 10 nF, as in Figure 3. Figure 4(d-i) shows each time domain voltage output as resistance is increased. Small changes of only around 100 to 150 $\Omega$ cause significant changes to the waveforms. In particular, the jump from 1.6 k$\Omega$ to 1.75 k$\Omega$, Figure 4(e) to Figure 4(f) and Figure 4(h) to Figure 4(i), changes both voltage outputs from appearing highly chaotic to closely resembling sine waves. The parametric plots of Inductor Voltage against Chua Diode Voltage are also provided in Figure 4(a-c). Parametric plots are commonly used to represent the chaotic states in 2D [21]. The different chaotic states are explained in Section 2.1.1b). Figure 4(a), 1.5 k$\Omega$, shows a double scroll. This is on the lower end of the chaotic region, close to approaching the limits of the Chua diode. As will be discussed in Section 2.1.1c)ii, decreasing the resistance further than 1.5 k$\Omega$ will eventually result in the large limit cycle state being reached.

As the resistance is increased to 1.6 k$\Omega$, the double scroll remains but has increased in size (both voltages ranges are larger), illustrated in Figure 4(b). The chaotic behaviour demonstrated by the time domain plots of Inductor Voltage and Chua Diode Voltage in Figure 4(h) and Figure 4(e) respectively is very typical of a Chua circuit. Figure 4(e) helps to show how the double scroll is formed. The Chua Diode Voltage oscillates between two clear stable points, one positive and one negative. In the 2D representation, they form each scroll. As the resistance is then further increased to 1.75 kΩ, the double scroll halves into a single chaotic attractor, known as a Rössler-type [21]. This halving of the double scroll occurs almost instantaneously with a resistance change as small as 1 Ω. This is demonstrable from the bifurcation diagram in Figure 13(a), Section 2.1.1c)ii. This highlights the high non-linearity of the circuit; a seemingly insignificant change to the circuit's initial conditions can cause a significant output difference. If the resistance is increased further beyond 1.75 kΩ, period-halving (the inverse of bifurcation) will occur. The outputs will enter "multi-period" states and keep halving until a single period is reached. If the resistance is increased more, eventually DC equilibrium will be reached.

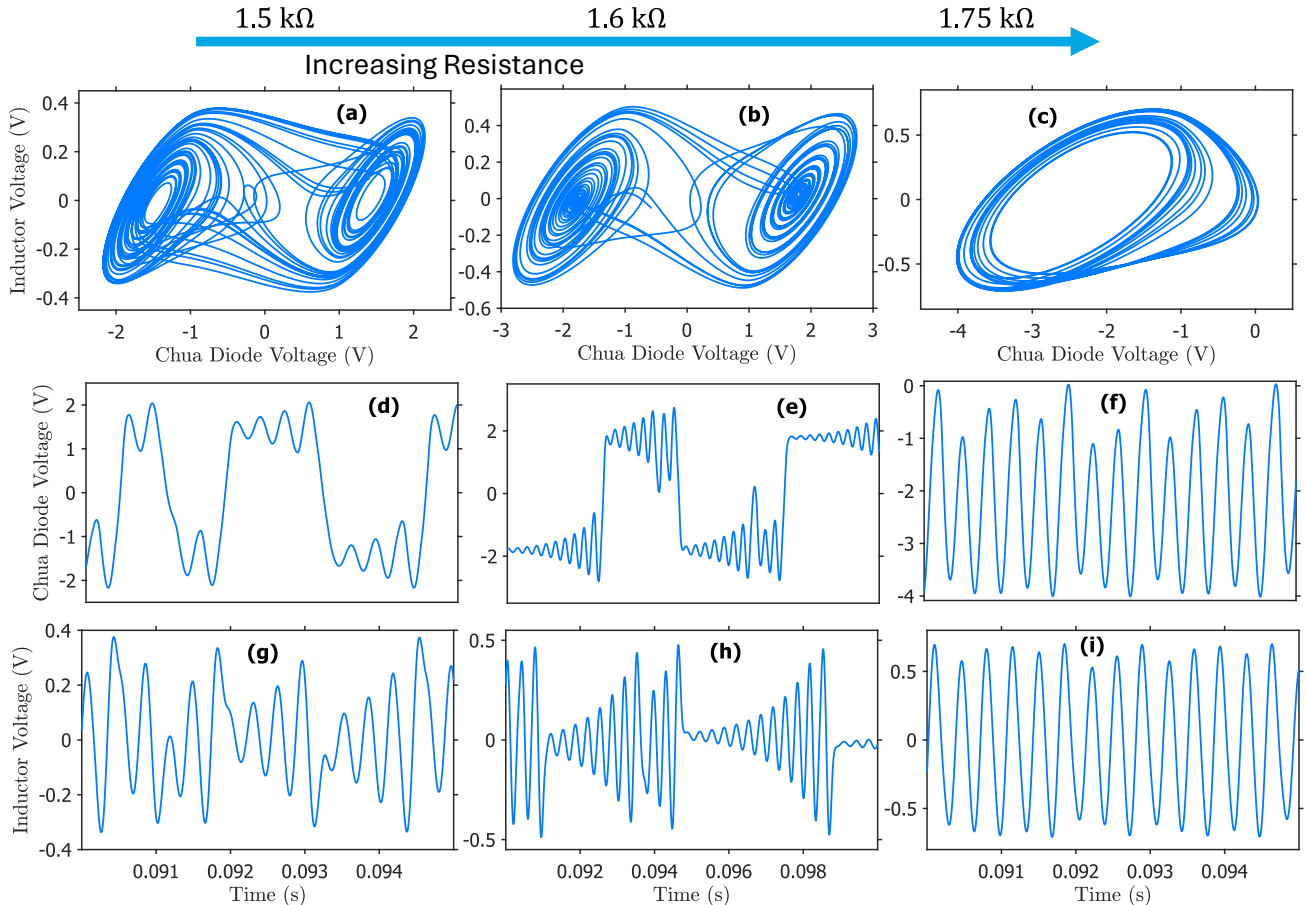

**Figure 4-** Parametric and time domain plots from varying resistance $R_{variable}$ of Kennedy Chua circuit modelled in LTSpice – fixed capacitance of 10 nF. **(a)** Chaotic Double Scroll Attractor 1.5 kΩ (close to Chua Diode limits), **(b)** Double Scroll Attractor 1.6 kΩ, **(c)** Rössler-type Attractor 1.75 kΩ, **(d)** Chua Diode Voltage 1.5 kΩ, **(e)** Chua Diode Voltage 1.6 kΩ, **(f)** Chua Diode Voltage 1.75 kΩ, **(g)** Inductor Voltage 1.5 kΩ, **(h)** Inductor Voltage 1.6 kΩ, **(i)** Inductor Voltage 1.75 kΩ.

A similar result can be produced through varying other parameters of the circuit. For example, varying the capacitance $C_1$ across the Chua diode produces the results seen in Figure **5**. This was performed with resistance fixed at 1.7 kΩ, as this is broadly within the middle of the chaotic region for a fixed 10 nF capacitance. The range either side of 10 nF is assessed. This produces a similar left-to-right process of period-halving, though the values for capacitance shown do not align exactly with the results from the resistance plots. The 8 nF plots of Figure 5(a), Figure 5(d), and Figure 5(g), and the 9 nF plots of Figure 5(b), Figure 5(e), and Figure 5(h), align well with the results from 1.5 kΩ, and 1.6 kΩ respectively. However, the 11 nF plots of Figure 5(c), Figure 5(f), and Figure 5(i) show only a single period parametric, rather than a Rössler-type. This is much further along in the period-halving process and is at the upper extreme of the Chua circuit dynamics. If the circuit is pushed any further, the outputs will be non-varying DC voltages. Changes in capacitance of only 1 or 2 nF have a significant impact on the circuit dynamics. This will make experimental testing difficult, since it will be hard to obtain many different capacitors with sufficiently small value increments. This difficulty, and the steps taken to address it, are discussed in Section 2.2.1c)ii.

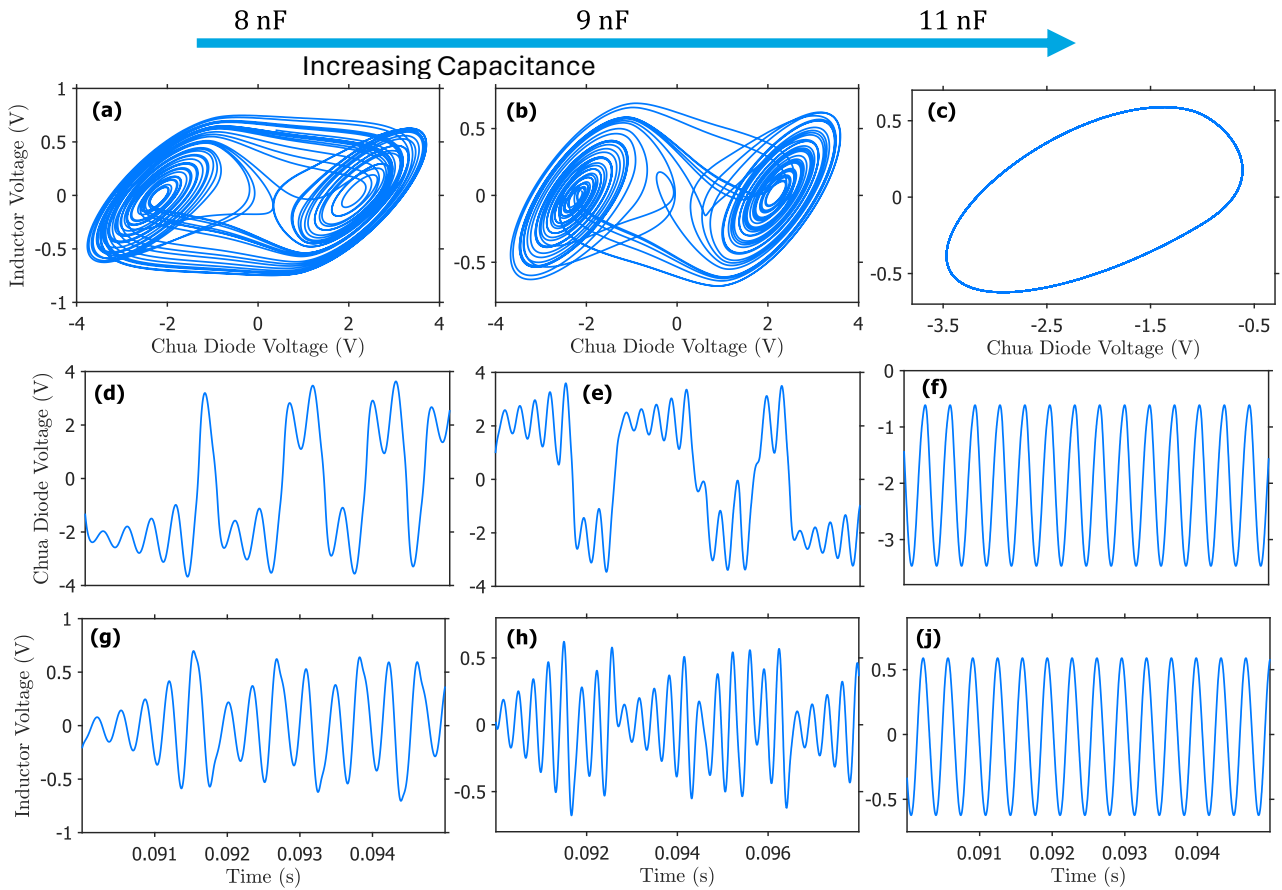

**Figure 5-** Parametric and time domain plots from varying capacitance $C_1$ across Chua Diode of Kennedy Chua circuit modelled in LTSpice – fixed resistance of 1.7 kΩ. **(a)** Double Scroll (very close to limit of Chua Diode) 8 nF, **(b)** Double Scroll 9 nF, **(c)** Single Period 11 nF, **(d)** Chua Diode Voltage 8 nF, **(e)** Chua Diode Voltage 9 nF, **(f)** Chua Diode Voltage 11 nF, **(g)** Inductor Voltage 8 nF, **(h)** Inductor Voltage 9 nF, **(i)** Inductor Voltage 11 nF.

The final parameter assessed was a driving input voltage amplitude. The circuit can be modified to include a voltage source in series with the inductor, $V_3$ in Figure 6. A sine wave of 100 Hz can be generated and the effect on the circuit dynamics viewed. The circuit is initially fixed to 10 nF, and 1.92 kΩ. This ensures that the circuit starts in DC equilibrium, and the input voltage can perturb it into chaotic behaviour. 100 Hz was chosen as the frequency as this is within the Chua circuit bandwidth, discussed at the end of this section.

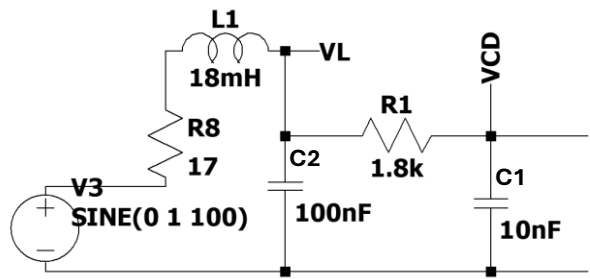

**Figure 6-** Modified LTSpice circuit for sinusoidal AC voltage input.

Figure **7** shows the circuit voltage outputs for input voltage amplitudes of 0.1 V, 0.3 V, and 0.7 V. In this case, the sequence of chaotic states is reversed. As the amplitude is increased, the bifurcation period-doubling occurs. This means that Figure **7**(a) broadly approximates a Rössler-type. The difference in appearance is due to the impact of the 100 Hz sine wave, visible in the time domain plots Figure **7**(d) and Figure **7**(g). This trend continues with Figure **7**(b-c), which both show double scrolls warped due to the 100 Hz sine wave, with the second clearly approaching the limits of the Chua diode. By adjusting input voltage amplitude across approximately a 1 V range, the Chua circuit can be perturbed from DC equilibrium into various chaotic states.

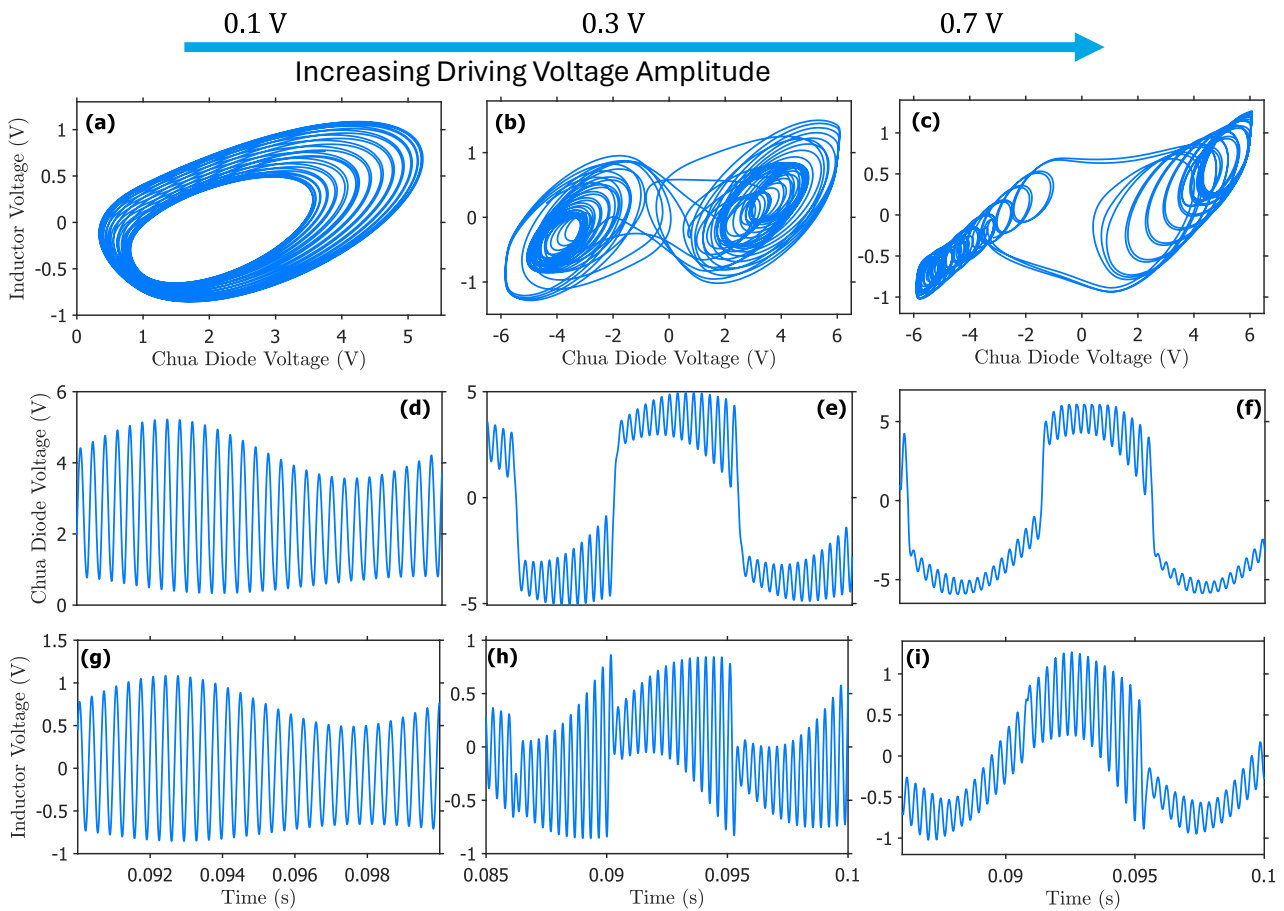

**Figure 7-** Parametric and time domain plots from varying input voltage amplitude to Kennedy Chua circuit modelled in LTSpice – fixed capacitance $C_1$ of 10 nF, fixed resistance of 1.92 kΩ. **(a)** Approximately Rössler-type Attractor 0.1 V, **(b)** Approximately Double Scroll 0.3 V, **(c)** Approximately Double Scroll approaching Chua Diode limits 0.7 V, **(d)** Chua Diode Voltage 0.1 V, **(e)** Chua Diode Voltage 0.3 V, **(f)** Chua Diode Voltage 0.7 V, **(g)** Inductor Voltage 0.1 V, **(h)** Inductor Voltage 0.3 V, **(i)** Inductor Voltage 0.7 V.

This data allowed for plotting the bifurcation diagram, explained in Section 2.1.1b). However, as mentioned in Section 2.1.1c), it is challenging to remove initial transients from the output data in LTSpice, and difficult to programmatically execute many different simulations. This makes producing bifurcation diagrams with LTSpice difficult. However, the data can be extracted and processed within MATLAB more seamlessly. Thus, the modelling is done via Simulink, with automated processing and plot generation occurring in MATLAB. The initial LTSpice attempt for the bifurcation with changing resistance is shown in Figure 8, and demonstrates the presence of different chaotic states in the Chua circuit dynamics. The highlighted red circle on Figure 8(a), for example, shows adjusting resistance only slightly can cause a significant change in output voltage. The highlighted green circle on Figure 8(b), however, is not the expected result, and includes initial transient behaviour. As will be discussed in Section 2.1.1c)ii, this is fixed using the MATLAB model.

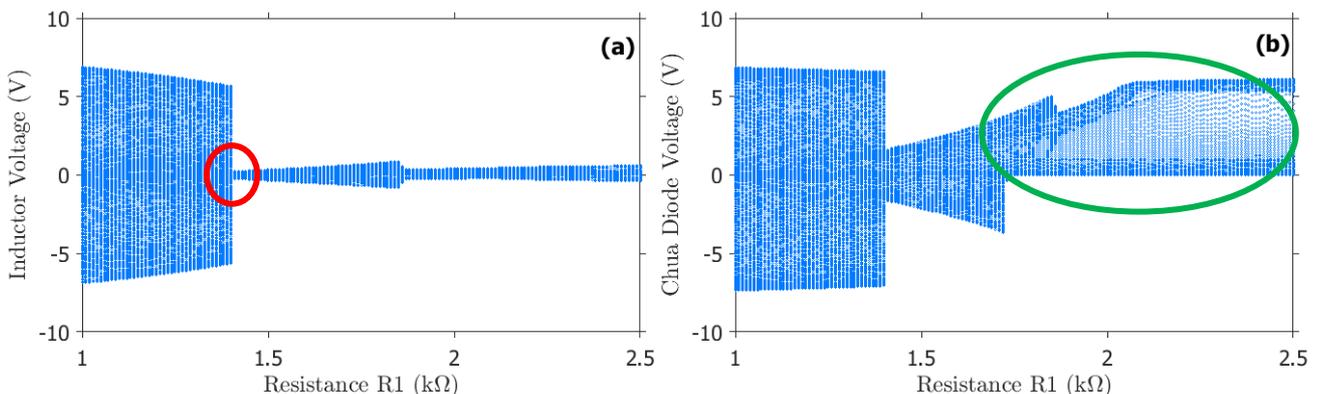

**Figure 8-** LTSpice bifurcation for varying resistance $R_{variable}$. **(a)** Inductor, **(b)** Chua diode.

The final use of the LTSpice simulation is to determine the expected bandwidth of the Chua circuit. This can be done by generating a Fast Fourier Transform (**FFT**) of the time-domain outputs. This will produce frequency-domain plots from which bandwidth can be assessed. As recommended in literature [12], the frequency response of a Chua circuit is best obtained using the FFT of the simplest unstable period of the Chua circuit. This FFT can be seen in Figure 9(b). There is a prominent spike just above 1 kHz, and the amplitude is broadly greater than −3 dB up to close to 10 kHz, after which it degrades. It is challenging to obtain accurate exact measurements for the bandwidth of chaotic oscillations [12], but this clearly provides a kHz range bandwidth of up to 10 kHz and allows for comparison against other Chua circuit implementations. The simulated V-I characteristic of the Chua Diode can be seen in Figure 9(c). This corresponds with the expected characteristic demonstrated by Kennedy [21].

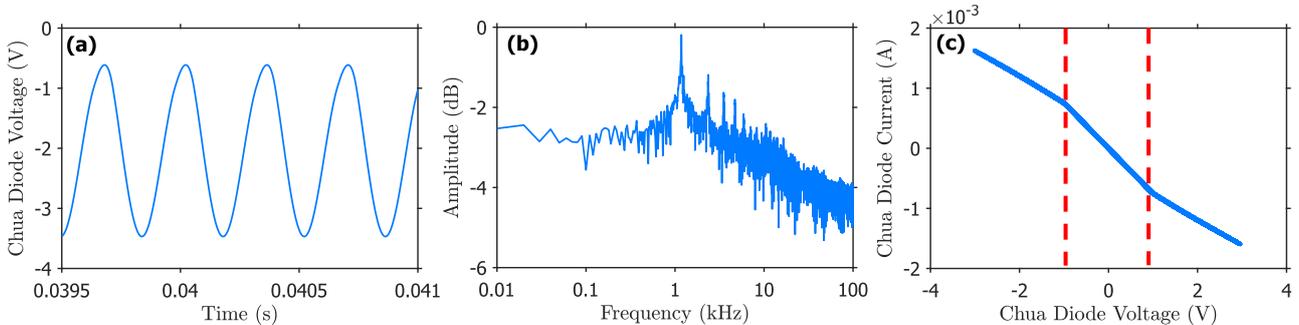

**Figure 9- (a)** Time domain Chua Diode Voltage at simplest unstable period of Chua circuit, **(b)** FFT of Chua Diode Voltage, **(c)** Chua diode V-I characterisation (1.63 kΩ) with each gradient separated.

### ii.   MATLAB Simulink Simulation and Results

The MATLAB Simulink implementation of the Chua circuit can be seen in Figure 10, utilising the built-in Simscape electrical models. Whilst direct Simulink/Simscape models for components such as the AD712 do not exist, the circuit was initialised with parameters reflecting the characteristics of this components.

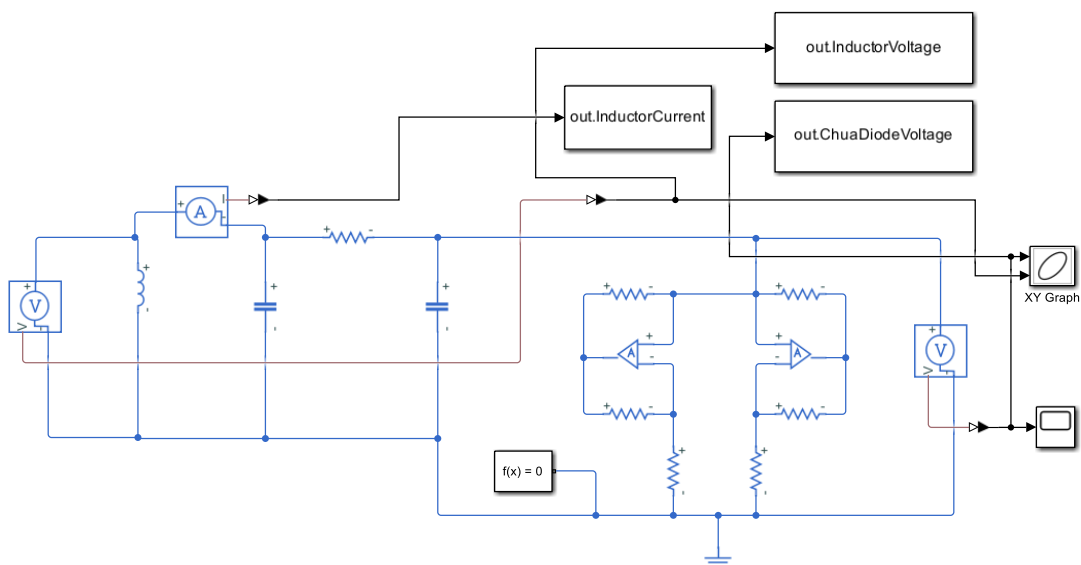

**Figure 10-** MATLAB Simulink schematic of Chua circuit (based on Kennedy).

This circuit can be compared with the LTSpice simulation by generating time-domain and parametric plots. Figure 11 provides six parametric plots with varying resistance. Figure 11(b) clearly demonstrates a double scroll attractor at 1.6 kΩ, achieving the same result as the LTSpice simulation from Figure 4(b). Figure 11(c) and Figure 11(d) produce Rössler-type/multi-period attractors at 1.77 kΩ and 1.8 kΩ. The LTSpice simulation produced a similar Rössler-type attractor at 1.75 kΩ in Figure 4(c). This is a slightly lower resistance, and in the Simulink circuit this would produce a double scroll. Figure 11(e) and Figure 11(f) show a double-period and single-period result, respectively. Further increasing the resistance will result in DC equilibrium. Figure 11(a) demonstrates the other end of the range, where the limits of the Chua diode are reached. These plots demonstrate that the Simulink model can accurately simulate the Chua circuit, but that there are small discrepancies in what parametric is produced by a given resistance – particularly with a higher resistance required in Simulink to move from a double to single scroll.

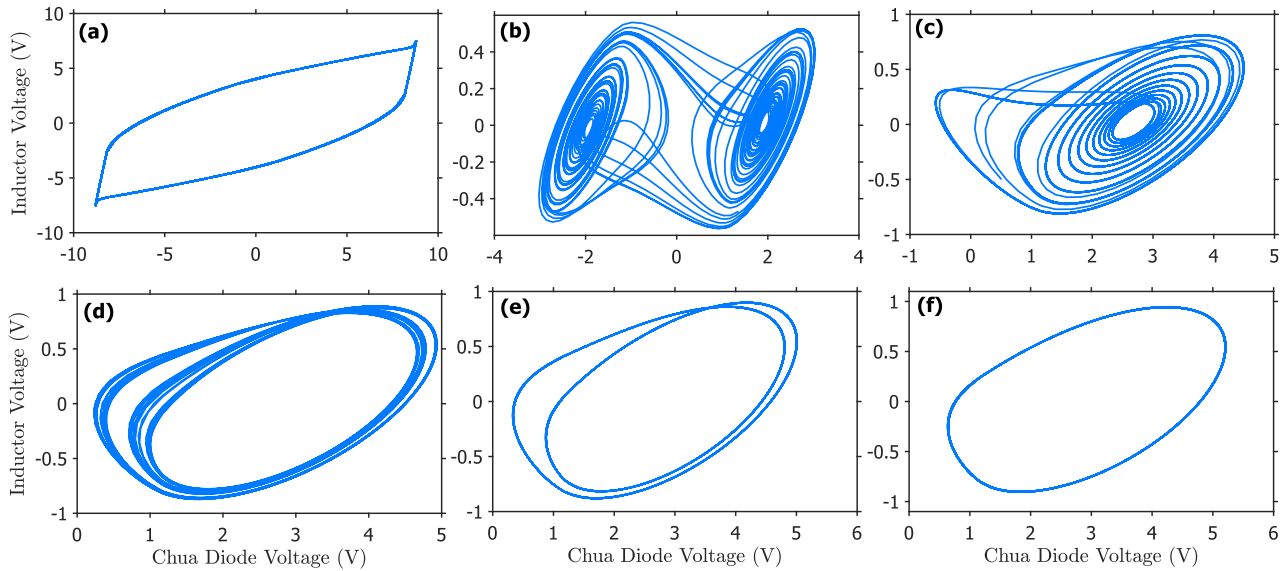

**Figure 11-** Simulink parametric plots for varying resistance, **(a)** Limit of Chua Diode 1.4 kΩ, **(b)** Double Scroll Attractor 1.6 kΩ, **(c)** Rössler-type Chaotic Attractor 1.77 kΩ, **(d)** Multi-period 1.8 kΩ, **(e)** Double-period 1.81 kΩ, **(f)** Single-period 1.836 kΩ.

The discrepancies are caused by two main factors. Firstly, the parameterisation of Op-Amp components in the Simulink model will not be as accurate as the manufacturer models provided in LTSpice. Secondly, the underlying circuit solvers will differ in algorithm and will likely differ in how rounding errors are handled. Due to the sensitive, chaotic nature of the circuit, small changes such as these will have larger impacts on the output, enough to shift slightly the resistance required to achieve a particular state. The effect of rounding errors and different circuit solver techniques can be seen in Figure 12. The outputs are undeniably similar but are shifted in time and the Simulink output is also smoother.

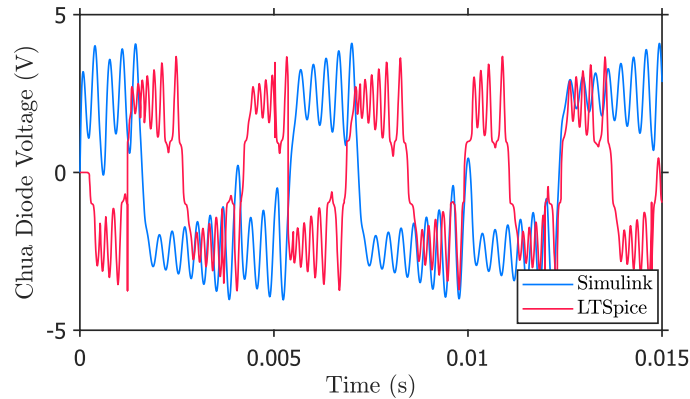

**Figure 12-** LTSpice vs MATLAB Chua Diode Voltage 1.73 kΩ.

The previous plots broadly show the whole range of the Chua circuit dynamics, but it is useful to be able to see more clearly the transition between different states. MATLAB can be utilised to generate bifurcation diagrams for varying resistance, capacitance, and input voltage amplitude, displayed in Figure 13. The upper row, Figure 13(a-c), demonstrates the change in Chua Diode Voltage as each of the three parameters are varied. The lower row, Figure 13(d-f), repeats the same, but for the Inductor Voltage. Figure 13(a) can be split into 5 regions. The first, rightmost, region is the single line, representing DC equilibrium. As the resistance is lowered to approximately 1.92 kΩ, a single-period parametric is achieved. Within this region, period-doubling occurs as resistance is lowered, until the third region is reached – the single scroll/Rössler-type (approximately 1.77 kΩ to 1.9 kΩ). At around 1.75 kΩ to 1.77 kΩ, the single scroll bifurcates into a double scroll. This double scroll is achieved until approximately 1.55 kΩ, after which the limits of the Chua diode are reached. This differs from the LTSpice model, where the limits of the Chua diode were not reached until approximately 1.4 kΩ, see Figure 8. This is explained again by rounding errors, different circuit solvers, and less accurate models in Simulink.

The same regions can also be seen on Figure 13(b), where capacitance is varied from 5 to 15 nF. The other difference is that, whilst in the same region, increasing the resistance noticeably increases voltage. Interestingly, both Figure 13(d) and Figure 13(e), the inductor voltages, share four out of five of the regions, but do not show a noticeable change from single to double scroll – this change in parametric occurs due to the Chua Diode Voltage and is an additionally major bifurcation that the Inductor Voltage lacks. This is also observable in Figure 8. Aside from slight changes in at what exact resistance major bifurcations occur, the key difference between the MATLAB and LTSpice bifurcation diagrams is the rightmost region of Figure 8(b) and Figure 13(a). This difference is entirely down to additional transient behaviour in the LTSpice data and using a programmable Simulink model allowed this to be easily removed.

The final assessed parameter for bifurcation was input voltage amplitude, with the diagrams shown in Figure 13(c) and Figure 13(f). This is a good parameter for deliberately perturbing the signal as it can be easily varied with automated processes. A sine wave AC voltage was used and the amplitude varied from 0 V to 1 V, with the resistance fixed to 1.95 kΩ. By fixing the resistance to a point in DC equilibrium, the voltage amplitude's impact across the whole range of the Chua circuit dynamics can be viewed.

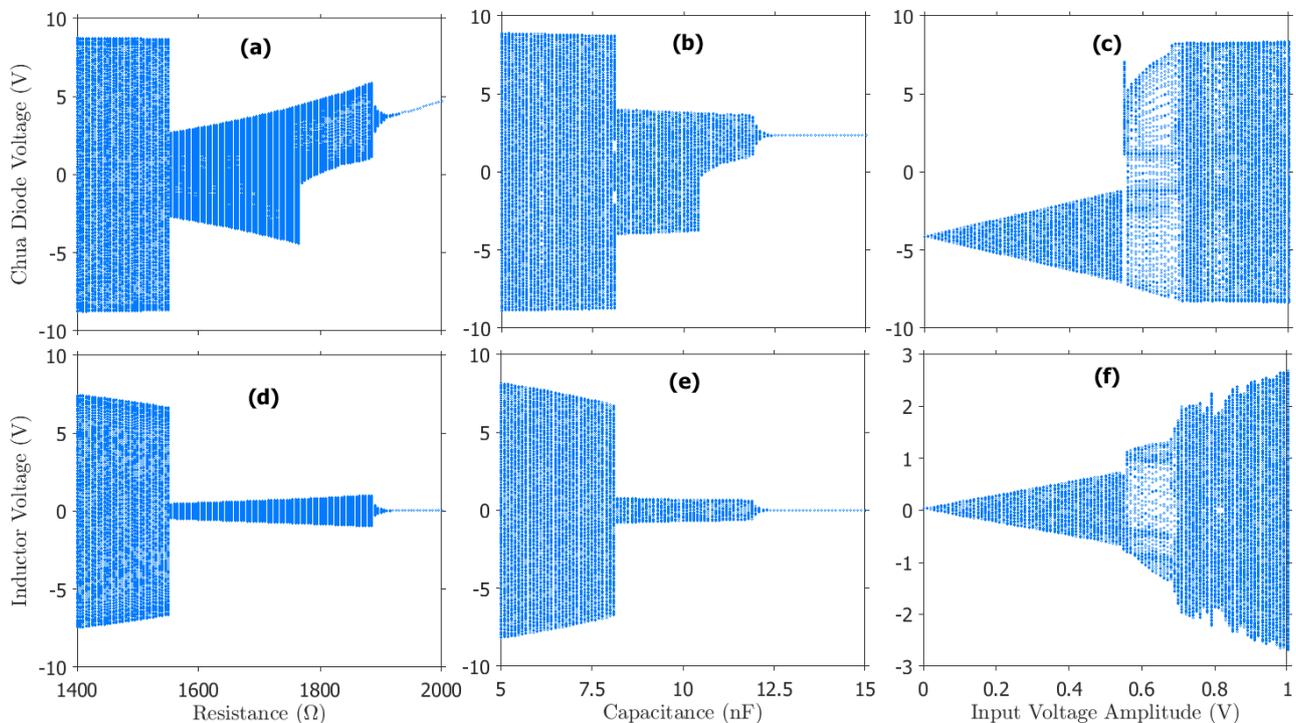

**Figure 13- (a)** Resistance bifurcation Chua Diode Voltage – circuit fixed at 10 nF, **(b)** Capacitance bifurcation Chua Diode Voltage – circuit fixed at 1.95 kΩ, **(c)** Amplitude bifurcation Chua Diode Voltage – circuit fixed at 1.95 kΩ and 10 nF, **(d)** Resistance bifurcation Inductor Voltage – circuit fixed at 10 nF, **(e)** Capacitance bifurcation Inductor Voltage – circuit fixed at 1.95 kΩ, **(f)** Amplitude bifurcation Inductor Voltage – circuit fixed at 1.95 kΩ and 10 nF.

The only known example of Chua-RC in literature was produced by Jensen [9]. In order to confirm the process used for this paper, the same simulation techniques were applied to the optimal circuit designed by Jensen. A square wave input voltage was then applied, at the specified resonant frequency of 9.4 kHz, to produce the voltage bifurcation diagram. Figure 14(b) provides the diagram from Jensen and Figure 14(a) provides the diagram reproduced by this thesis – a clear validation of the simulation techniques used for the Kennedy circuit.

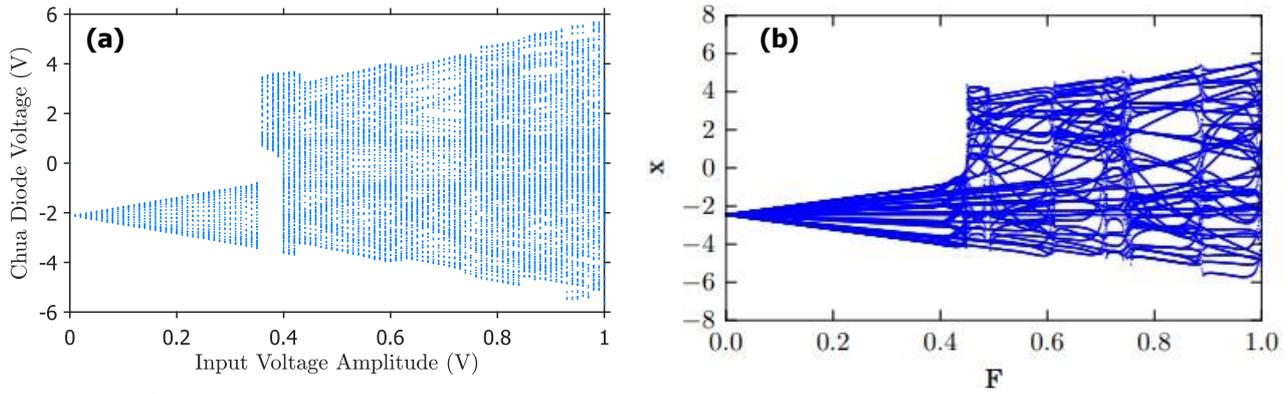

**Figure 14-** Comparison of driving voltage bifurcation diagram generated by this work, and the diagram generated by Jensen [9], **(a)** Driving voltage bifurcation diagram with Jensen circuit simulated in Simulink, **(b)** Driving voltage bifurcation diagram from original Jensen paper.

Another parameter is the frequency of an input voltage at a fixed amplitude. This paper does not assess this parameter but recognises that there is potential for further investigation. Although this parameter was not investigated directly, the voltage bifurcation diagrams for varying amplitude were considered at different frequencies. The results demonstrate that lower frequency causes the major bifurcations to require greater amplitude, and that they become steeper changes. Figure 15(a-b) shows the same bifurcation diagram as Figure 13(c), but with frequencies of 500 Hz and 1000 Hz respectively. At 1000 Hz, the major bifurcation occurs at approximately 0.1 V, compared with approximately 0.3 V for 500 Hz, and 0.55 V for 100 Hz.

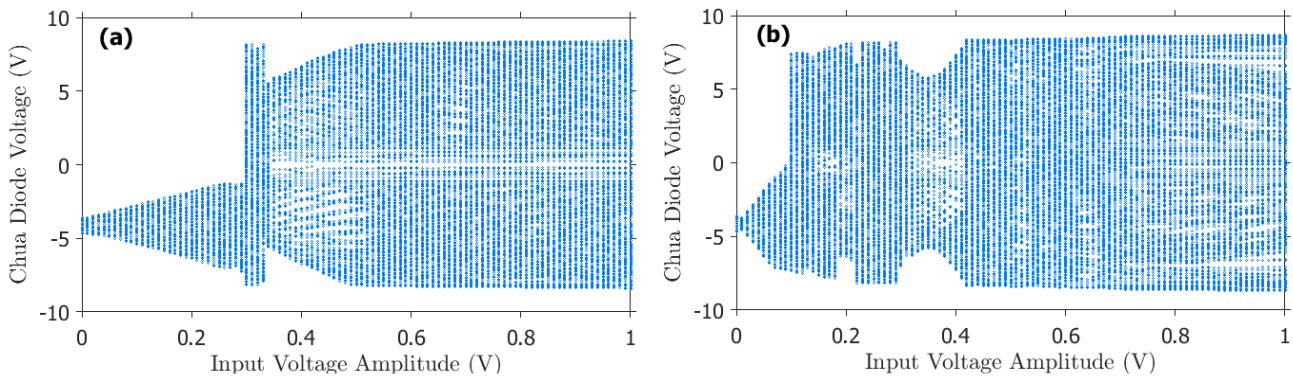

**Figure 15-** Driving voltage bifurcation diagrams (Chua Diode Voltage), **(a)** 500 Hz, **(b)** 1000 Hz.

MATLAB can be used to simulate non-idealities and noise in the Chua circuit. This can be used to identify if noise or small imperfections are significantly detrimental to circuit performance. Appropriate mitigation steps can be taken as required. Passive components can be replaced with versions including tolerance and noise options. For example, each resistor, capacitor, and inductor can be simulated with a manufacturer tolerance (e.g., 5% with a random gaussian distribution) included. Resistors can also be simulated with thermal Johnson noise. The Op-Amps can be parameterised with noise information from their datasheets. The capacitors were assumed to have negligible series resistance (**ESR**). Parallel conductances of capacitors and the inductor were set to the Simulink defaults. The Op-Amp models were configured to have a noise voltage density of 22 nV/$\sqrt{\text{Hz}}$ and a noise current density of 0.01 pA/$\sqrt{\text{Hz}}$, the same at both inputs, as per the datasheet [27]. This noise was sampled up to 10 kHz – as per the previously determined system bandwidth. The thermal Johnson noise on each resistor was also sampled up to 10kHz and an ambient temperature of 20 ℃ was assumed. Figure 16(a) shows a double scroll produced at 1.73 k$\Omega$ for a noisy circuit whereas Figure 16(b) shows the same result for the noiseless circuit. Whilst they are not identical, the chaotic state remains the same and they are close to identical. This suggests that thermal noise and minor component deviations, commonly found in practical off-the-shelf components, are not sufficient to cause problematic changes to circuit dynamics.

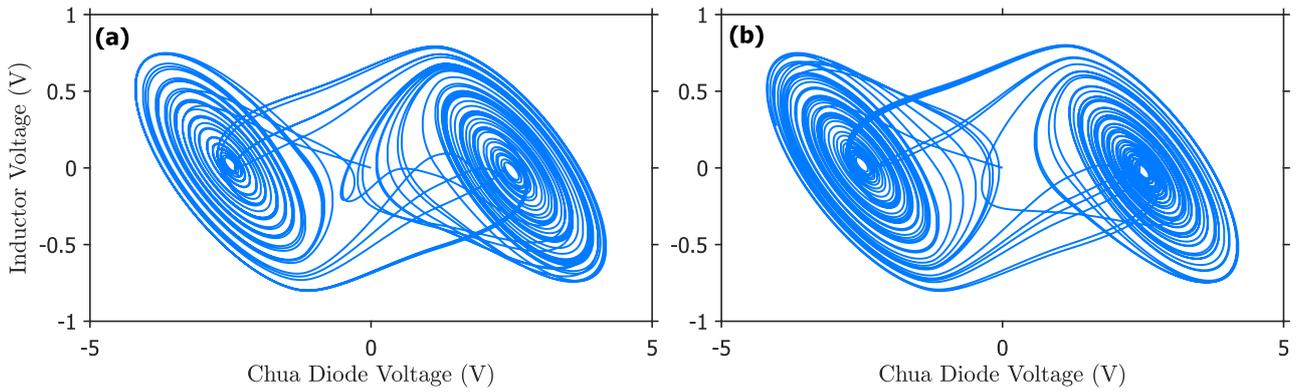

**Figure 16-** **(a)** Double scroll at 1.73 kΩ (Noisy), **(b)** Double scroll at 1.73 kΩ (Noiseless)

The noise on a voltage input to the circuit can also be modelled by taking the parameters from the function generator/oscilloscope component – the Analog Discovery 2 (**AD2**). The choice of this device is discussed in Section 2.1.2a)v. The AD2 has a noise voltage density of $6.6\,\text{nV}/\sqrt{\text{Hz}}$ and a noise current density of $0.6\,\text{fA}/\sqrt{\text{Hz}}$ [13]. Figure 17 shows the difference in output voltage between a noiseless circuit and a circuit with noise on the input voltage. Again, the difference is very small and does not actually correspond to a significant change in overall amplitude or frequency. Rather it just shifts the signal.

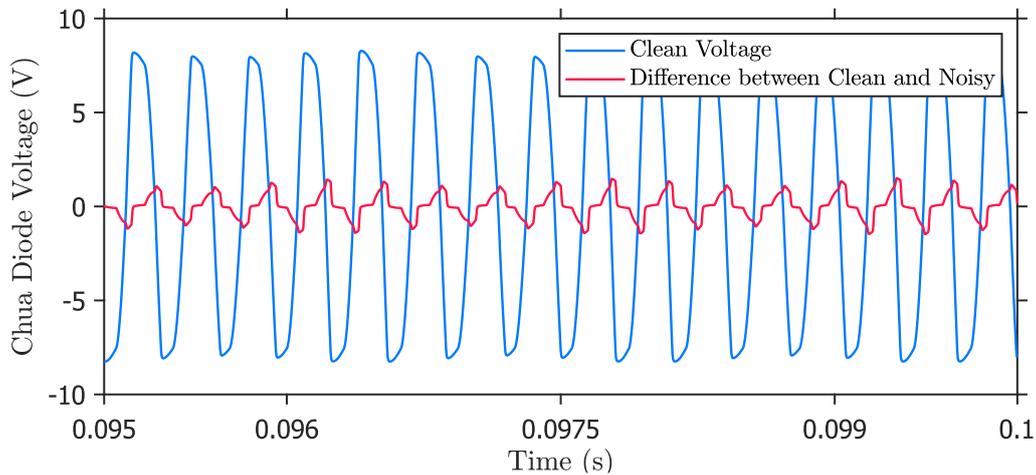

**Figure 17-** Difference between noisy and noiseless circuits overlaid on top of noiseless circuit voltage.

The MATLAB SNR function determines a value of 25.6 dB. This exceeds the minimum desired specification, and supports the suggestion that noise is not a significant problem in the Chua-RC. By using appropriate equipment and good practice when constructing the circuit, there should be no issues.

### 2.1.2  Reservoir Computer Architecture

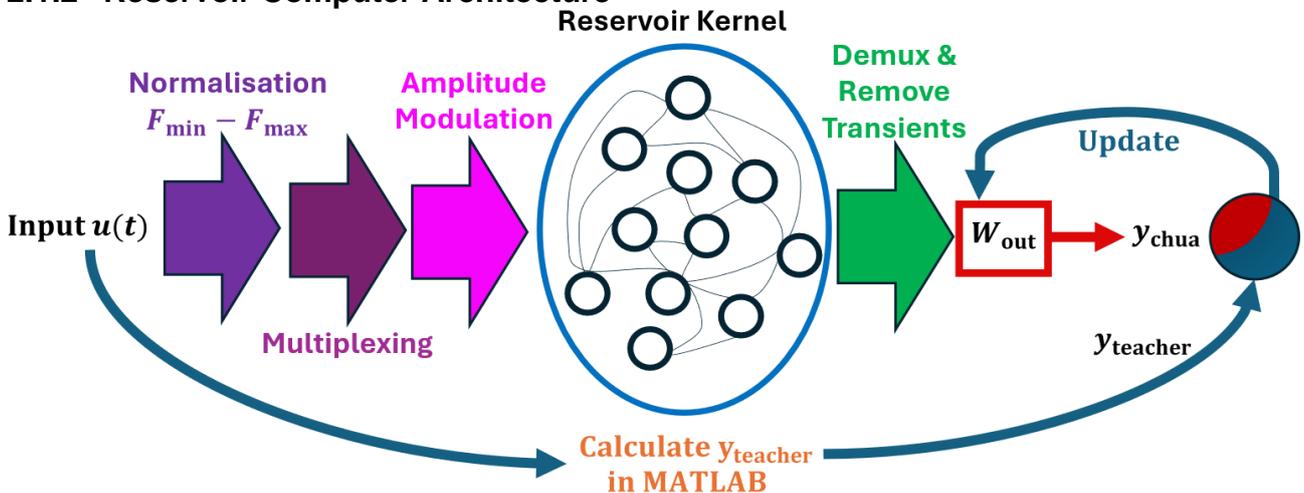

**Figure 18-** Block diagram of full reservoir computer process.

The three key parts of a RC are the input, the output, and reservoir [10]. The reservoir in this case is a Chua circuit, and full simulation has been completed, detailed in Section 2.1.1c). Whilst the input and output stages will vary for different applications, requiring changes in pre/post-processing and different training, the reservoir can remain fixed [1]. Figure 18 shows the full RC process design for this project. An input, $u(t)$, is pre-processed through first normalising to a chosen voltage range, then multiplexing, before modulating a carrier wave at a specific frequency. The resultant voltage waveform is input to the Chua circuit. At the output of the Chua circuit, the voltage is measured. Initial transients are removed, and the signal is demultiplexed into multiple distinct outputs. These outputs are used in a training algorithm, using a $y_{\text{teacher}}$ calculated in MATLAB. This design is explained across the following sections.

### a) Reservoir Computer Input
### i. Input Voltage Amplitude Range

In the case of the input, a voltage signal must be provided to the Chua circuit by an arbitrary function generator (**AFG**) with a Digital to Analogue Converter (**DAC**). The key characteristic of this signal that must be considered is the amplitude range, $F_{\min}$ to $F_{\max}$ in Figure 18. As demonstrated in Figure 13(c) and Figure 13(f), changing the amplitude of a voltage input to a Chua circuit causes bifurcation and changes the output state of the circuit. For a given fixed Chua circuit, it follows there could be an optimal voltage range that results in the maximum number of distinct output states. Figure 13(c) and Figure 13(f) suggests this range to be around the 0 to 1 V order of magnitude, though Figure 15 demonstrates that the voltage bifurcation diagram is impacted by signal frequency. The diagram would also vary with other circuit parameters (such as capacitance and resistance), as they define the starting state. Additionally, the amplitude must be large enough to perturb the circuit into changing states, particularly when moving from DC equilibrium into chaotic states. If the amplitude range includes values too small to cause bifurcation, the outputs corresponding to these values will not be useful as there will be no non-linear behaviour. This suggests a lower limit on the usable voltage amplitude.

Another consideration is that the amplitude must be small-enough to not force the circuit into a stable state (outside of the chaotic region). Figure 19 shows the time domain and parametric plot of the Chua circuit at the limits of the Chua diode. This can be obtained in different ways, for example by sufficiently lowering the resistance or capacitance. It can also theoretically be obtained by using too great a voltage amplitude. Once in this state, the circuit will not exit it without being "reset". In the Jensen paper, this was acceptable – this was because single values were sent one at a time and between runs a forcing amplitude could be applied to "reset" the circuit [9]. In the case of a time-varying signal, a "reset" signal could be applied between runs but not between different values in the same run. If one value is too large and causes the circuit to enter a fixed stable state, the circuit output will be consistent for the rest of the run and will not properly vary alongside the time-varying input. This suggests an upper limit on the usable voltage amplitude.

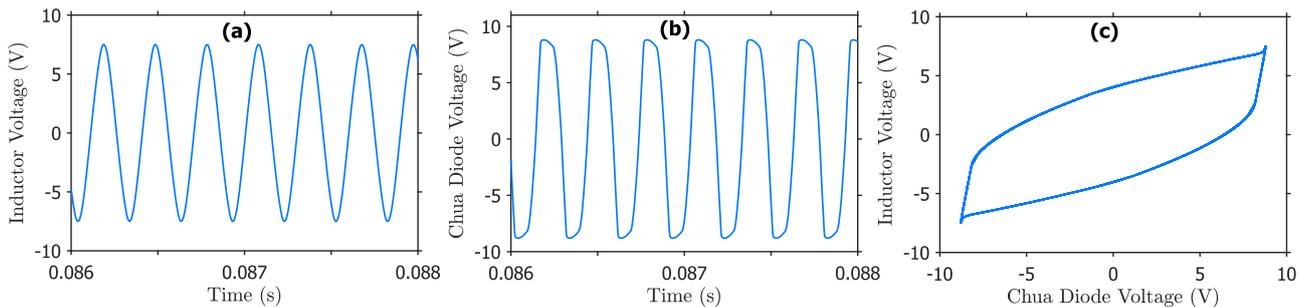

**Figure 19**- Time domain and parametric plots at the stable state at the limits of the Chua diode $R_1 = 1.4\ \text{k}\Omega$, **(a)** Inductor Voltage, **(b)** Chua Diode Voltage, **(c)** Parametric

Finally, the voltage range must be able to represent the input data. If the application is PQC, the input data will likely be a message, and encryption parameters (this is discussed in Section 2.1.3b)). The voltage range should allow sufficient resolution to distinctly encode all possible input values. It must also be a voltage range that can be accurately reproduced by a DAC, at a resolution and sampling frequency within the specification of the AFG. This is discussed in Section 2.1.2a)v in the context of selecting an appropriate AFG. Jensen fixed the input voltage amplitude range to 0.5 to 1 V, but made a call for work investigating other ranges [9]. This paper answers this call, assessing voltage range as one of two primary tuneable parameters for optimisation. This is discussed in Section 3.2.2.

### ii.  Multiplexing with Random Mask

A key benefit of a RC is the ability to transform low-dimensional inputs to higher-dimensional outputs [11]. This allows the read-out training to better separate the inputs according to patterns [16]. Generally, the greater the number of output dimensions, the more challenging the problems that can be solved, and the more superior the RC [1]. To increase the output dimensions, the number of outputs can be increased [16].

The Chua circuit kernel can be physically tapped at multiple outputs, but this is limited by the equipment used to perform signal acquisition as each output will require an oscilloscope channel. As will be discussed in Section 2.1.2b)i, the limit for this project is two physical taps: chosen to be the Inductor Voltage, and the Chua Diode Voltage. The number of physical taps is referred to as $N_{\text{taps}}$. In the case that spatial realisation of additional states is limited, time multiplexing can be utilised [1, 2]. In the case that the input signal is multiplexed with a randomised mask signal of size $N_{\text{mask}}$, the output can be demultiplexed to generate $N_{\text{mask}}$ variants. This leads to $N_x$ virtual neurons where $N_x = N_{\text{mask}} \cdot N_{\text{taps}}$. The time multiplexing process has artificially increased the number of neurons at the read-out of the Reservoir Computer, without requiring additional physical taps [2]. The input message, originally of length $N_{\text{message}}$, will now be of length $N_{\text{mux}} = N_{\text{message}} \cdot N_{\text{mask}}$. This can be observed in Figure 20. Figure 20(a) demonstrates a message, with $N_{\text{message}} = 11$. It is multiplexed with a random mask of size $N_{\text{mask}} = 4$. This results in four distinct signals, Figure 20(b), combined into a signal of size $N_{\text{mux}} = 44$ in Figure 20(c).

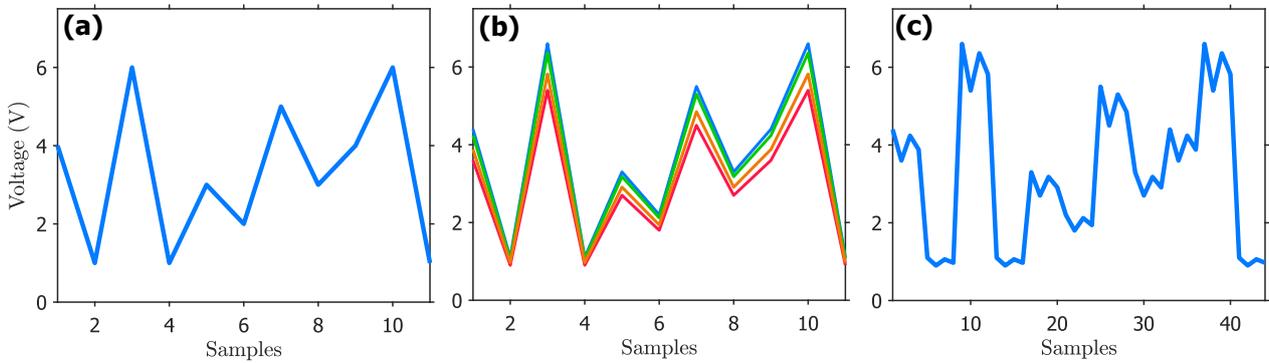

**Figure 20- (a)** Example message, **(b)** Example message variants (four masks), **(c)** Multiplexed message.

By increasing the number of virtual neurons, the requirements for additional physical taps are reduced [1]. This is extremely beneficial as it simplifies the signal acquisition process. In this project, only two output signals will be acquired, and this can be achieved with a single device. This reduces the required components, keeping costs and complexity lower. It may be that multiplexing is not required, and that just the two physical taps will be sufficient for training the RC – in this case $N_{\text{mask}} = 1$. This is discussed in Section 3.2.1.

### iii.  Sample Hold

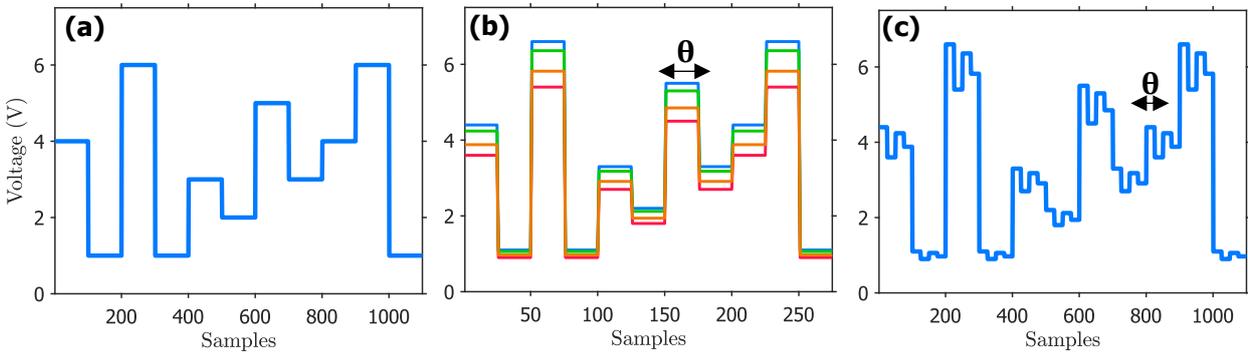

**Figure 21- (a)** Example message with sample hold, **(b)** Example message variants with sample hold, **(c)** Multiplexed message with sample hold.

After the input message is multiplexed, it can be thought of as a new message of greater size. Each piece of data within the message $u$ is a discrete sample: $u = [u_1, u_2, u_3, \ldots, u_{N_{\text{mux}}}]$. It should be noted that $u$ is the typical notation for a RC input. It should not be confused with the notation $[u, v]$ for an encrypted bit using LWE. To ensure that this signal can be accurately reproduced by a DAC, and that each piece of data is applied to the kernel input for sufficient time to perturb the system, the samples must be held for a total number of data-points $\theta$. The original message in Figure 20(a) would appear as in Figure 21(a). The multiplexed message seen in Figure 20(c) would become that shown in Figure 21(c). This was implemented into the design following feedback at the proposal defence, and is performed alongside the multiplexing process [11].

### iv.  Amplitude Modulation

Murali used a carrier sine wave of fixed frequency and variable amplitude to perturb a Chua circuit [20]. Jensen adapted this to use square waves for Chua-RC, due to the ease of generating square waves experimentally [9]. The input data, normalised to a time-varying DC voltage, acts as an envelope, varying the amplitude of a fixed frequency carrier waveform, encoding the data. This is Amplitude Modulation (**AM**). Future work could assess the use of Pulse-Width Modulation (**PWM**) to encode the input data; however, this is out of the scope of this thesis. Since prior work has been demonstrated using AM, its use is logical.

For a multiplexed input message such as the envelope featured in Figure 22(b), the resultant signal, with a 100 Hz sinusoidal carrier, see Figure 22(a), is shown in Figure 22(c). In this demonstration, the message is created to last 1 s. Since the carrier is 100 Hz, there are 100 periods in Figure 22(c). Figure 22 can be adjusted to use a square wave (oscillating between $-1$ and $+1$). It can also be adjusted to use a DC voltage of 1 V as the carrier. This means that final waveform will be just a variable amplitude DC signal– i.e. the green signal in Figure 22(b). This project will use sine waves when characterising the Chua circuit, as per Murali [20], but will primarily utilise square waves for the Chua-RC application, as per Jensen [9], as they are proven in application. Section 3.6 investigates the impact of changing the carrier for Chua-RC to sine waves or 1 $V_{DC}$.

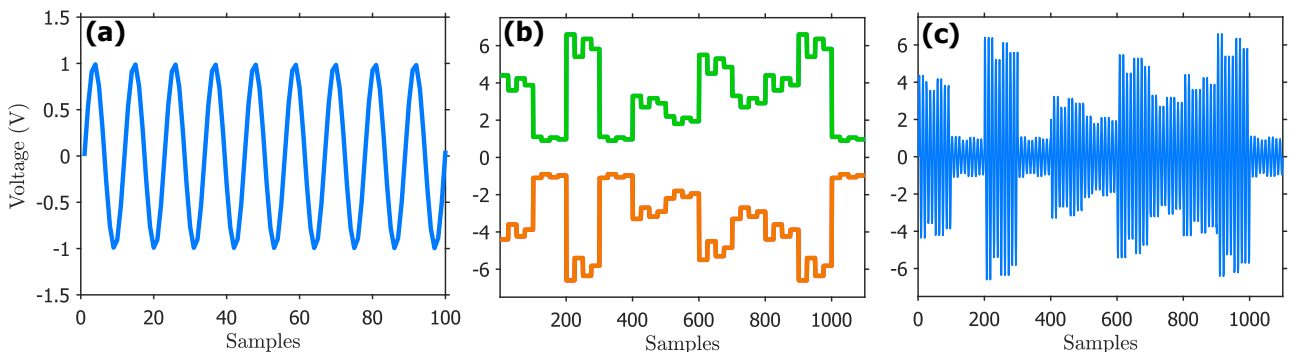

**Figure 22-** **(a)** Carrier sine wave, **(b)** Modulation envelope constructed from multiplexed message, **(c)** Sine wave carrier modulated by multiplexed message envelope.

### v.  Signal Generation Device

The choice of device for generating the input signal is important. The device must be capable of generating arbitrary waveforms defined by the input for an application of the Chua-RC. For example, in the case of encryption, the input will be a plaintext message, pre-processed with normalisation, multiplexing and AM. It must be possible to vary the amplitude and frequency of the waveform. Crucially, the signal must be produced within desired voltage limits and at an acceptable resolution. Additionally, noise and uncertainty must be minimised to ensure the generated signal is accurate to the desired signal. Finally, it should be possible to generate this signal through an automated process linked to the control system of the RC. In this case, this means the device must be controllable, either directly or indirectly, by MATLAB. Two main options were identified:

- **Digilent OpenScopeMZ**
- **Digilent Analog Discovery 2**

| | **Analog Discovery 2** [13] | **OpenScopeMZ** [14] |
|---|---|---|
| **Resolution** | 166 µV – accurate to $\pm 10$ mV $\pm 0.5\%$ | No data |
| **Sampling Rate** | 100 MSamples/s | 6.25 MSamples/s |
| **Amplitude** | $\pm 5$ V | $\pm 3$ V |
| **Bandwidth** | > 12 MHz | 1 MHz |
| **Noise Voltage Density** | 6.6 nV/$\sqrt{Hz}$ | No data |
| **Noise Current Density** | 0.6 fA/$\sqrt{Hz}$ | No data |

**Table 4-** Comparison between specifications of Analog Discovery 2 and OpenScopeMZ

Table 4 compares the specification of each device. The OpenScopeMZ (**OSMZ**), pictured in Figure 23(a), was considered since it is a low-cost, open-source oscilloscope and function generator already stocked by the university workshop. This supports the low-cost goal of the project. The Analog Discovery 2 (**AD2**) is a higher-specification device, considered in the case that the OSMZ proved inadequate.

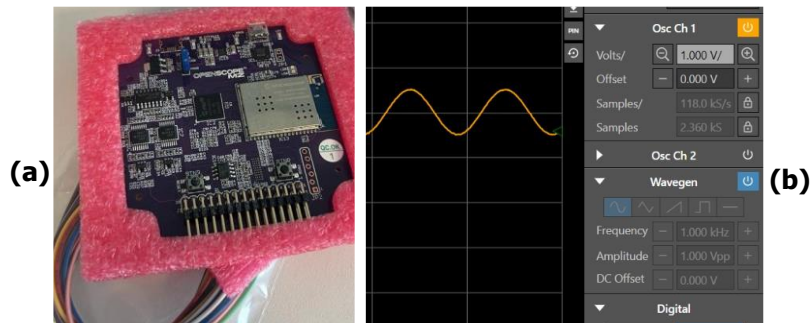

**Figure 23- (a)** Digilent OpenScopeMZ, **(b)** Sine wave generation and acquisition using OpenScopeMZ.

Whilst the OSMZ specification could theoretically achieve the aims of the project, it was concerning to see the OSMZ had a poor reputation. Digilent have discontinued the product line and halted all support. This means learning resources are very limited and it was unclear exactly what features the OSMZ can achieve: support for automated use (directly via MATLAB or indirectly via scripting), and arbitrary function generation were two key requirements that could not be confirmed. Data on resolution and noise performance was also missing.

Since the university workshop already stocked OSMZ devices, its capabilities could be tested. Figure 23(b) demonstrates the successful generation of a sine wave using the Digilent Waveforms Live software with an OSMZ. The generated signal is recorded using the in-built oscilloscope of the OSMZ. This was successful, and variables such as amplitude, DC offset, and frequency could be adjusted. However, there were no options for arbitrary signal generation, only typical waveforms. Moreover, there was no method of automating the process with scripts or via MATLAB directly. This meant that an AD2, pictured in Figure 24(c), was required.

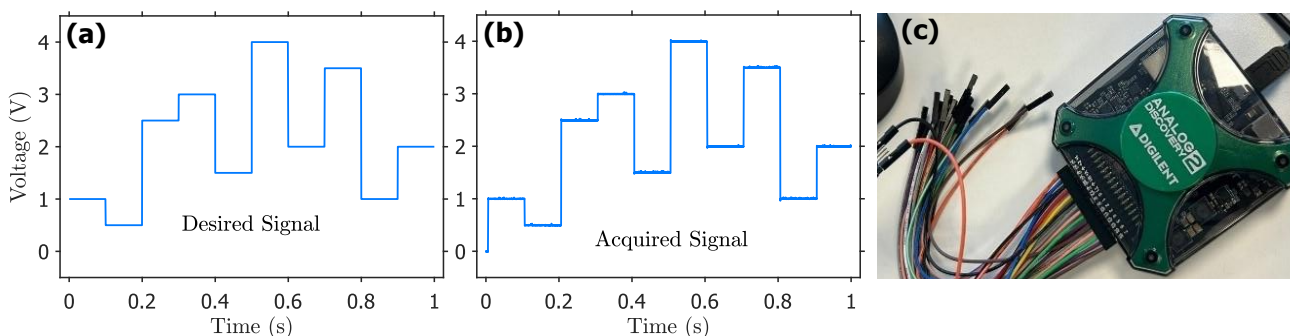

**Figure 24- (a)** Desired arbitrary function, **(b)** Generation and acquisition of arbitrary function using Analog Discovery 2 via MATLAB, **(c)** Analog Discovery 2 oscilloscope and function generator

The AD2 is one of Digilent's flagship products. It comes with extensive support and has MATLAB toolboxes allowing MATLAB automation and control – perfect for this project. The cost of this component was £239.34, which would have exceeded the project budget. However, since this is a piece of equipment that can easily be re-used for other projects (a non-consumable item), it was acceptable to buy it separately from the project budget. Figure 24(b) demonstrates using the AD2 to generate and then acquire a desired arbitrary function – pictured in Figure 24(a). This is performed using MATLAB and meets the requirements of the project. The AD2 has two AFG channels [13]. This provides the option for future work involving multiple circuit inputs, something not considered in this thesis. The input voltage can be applied in series with the inductor, similar to the circuit developed in [20]. Future work could assess different entry points to the kernel, but to restrict the scope of this project, the circuit configuration and Chua diode design were fixed.

### b)   Reservoir Computer Output
### i.   Signal Acquisition

As demonstrated in Section 2.1.2a)v, the AD2 and OSMZ can both be utilised as oscilloscopes. This is shown in Figure 24 and Figure 23 where the generated signals are acquired and displayed using the oscilloscopes of each device. Since the OSMZ lacks MATLAB support, and the AD2 is required for signal generation, it makes sense to utilise the AD2 again for signal acquisition. It has two oscilloscopes so can obtain the Inductor Voltage and the Chua Diode Voltage simultaneously. Furthermore, there were numerous issues encountered with getting the OSMZ to work consistently and reliably, corroborated by other's shared experiences online. Without further support available from Digilent, this would distract from the rest of the project. Therefore, the AD2 is used for both signal generation and signal acquisition. The MATLAB interfacing of the AD2 is particularly useful as it enables automated testing and is a practical method for obtaining and processing experimental data.

Using the AD2 limits the number of physical taps on the reservoir to two. However, this supports the two primary Chua outputs (Chua Diode Voltage and Inductor Voltage). Additional physical taps are not required through the use of multiplexing to artificially increase the number of output channels. A third output option is inductor current, however this was decided against since tapping current would extract power from the circuit.

### ii. Demultiplexing and Post-Processing

As detailed in Section 2.1.2a)ii, the multiplexing process results in $N_{mask}$ variants of the input to the Chua circuit kernel, processed as a single signal. The output can be demultiplexed into $N_{mask}$ number of distinct voltage signals per physical tap. In the case of $N_{mask} = 4$, each physical tap will have 4 variant outputs, broadly similar, but with slight differences induced by the multiplexing. The demultiplexing can be achieved by determining the proportion of the input signal accorded to each mask. The acquired output signal can be divided into distinct groups of samples according to each of the masks. For example, Figure 25(a) shows a multiplexed signal being demultiplexed into 4 distinct signals, shown in Figure 25(b). To avoid initial transients and to ensure that only the correct part of the signal is sampled, a central percentage is sampled. For example, the middle 80% can be sampled, as in Figure 25(a). Originally, the plan was to take the average or maximum of these samples, as in Figure 25(c). However, it was decided to use the waveforms, Figure 25(b), directly. Averaging or taking a maximum would reduce the data processed in the training stage, improving computational speed, and reducing memory requirements, but it would likely also remove non-linearity from the kernel.

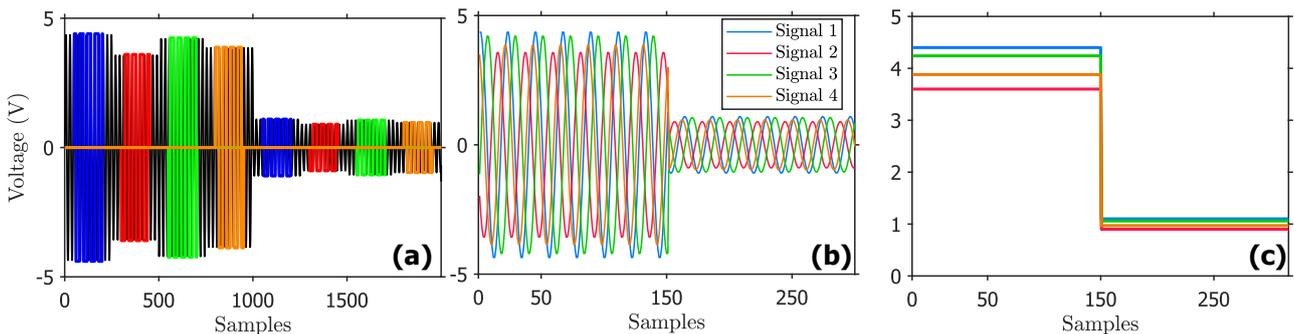

**Figure 25-** **(a)** Demultiplexing process, **(b)** Demultiplexed signals (taking a central sample), **(c)** Demultiplexed signals (taking the positive envelope).

It is important to ensure that both multiplexing and demultiplexing are implemented correctly and properly synchronised. Figure 26(a) shows an example of incorrect multiplexing. Masks are unevenly distributed across a message. The first and third message values are multiplexed with five masks each, whereas the second is only multiplexed with three.

During demultiplexing, it is important to ensure that samples are taken at correct points in the signal. The output from the kernel is demultiplexed into $N_{mask}$ signals. Each signal must correspond with only one mask. In a signal, the samples corresponding to each part of the original message must also align in time with the same mask. An example incorrect result is demonstrated in Figure 26(b). A single message value, which has been multiplexed with four masks and used to modulate the carrier, is oversampled by the demultiplexing process. The first mask (blue) is not properly synchronised and begins to sample the same message point for a second time, at the position of the fourth and final mask.

There will naturally be a small delay between generating an input signal and acquiring the output signal. This is demonstrated in Figure 26(c), using the AD2. This delay must be accounted for. In addition to causing synchronisation issues, it results in the loss of data at the end of the signal. This is because the initial delay occupies buffer space. To solve the synchronisation issue, the end of the delay can be identified. This can be achieved by checking for the presence of the signal compared with noise. This is practical in the example in Figure 26(c), as initial noise is approximately 0 V. In the Chua application, this is more challenging due to small noise having a non-linear impact on the output signal. However, it is still less significant than the signal.

After experimentation, a suitable threshold can be determined. This is discussed further in Section 2.2.2c)i. To avoid the data loss issue, a dummy message value can be sent at the end of the signal. After synchronisation has occurred, the post-processing code can sample the desired number of message values, and cut off whatever part of the dummy value remains. This ensures that none of the actual message has been lost.

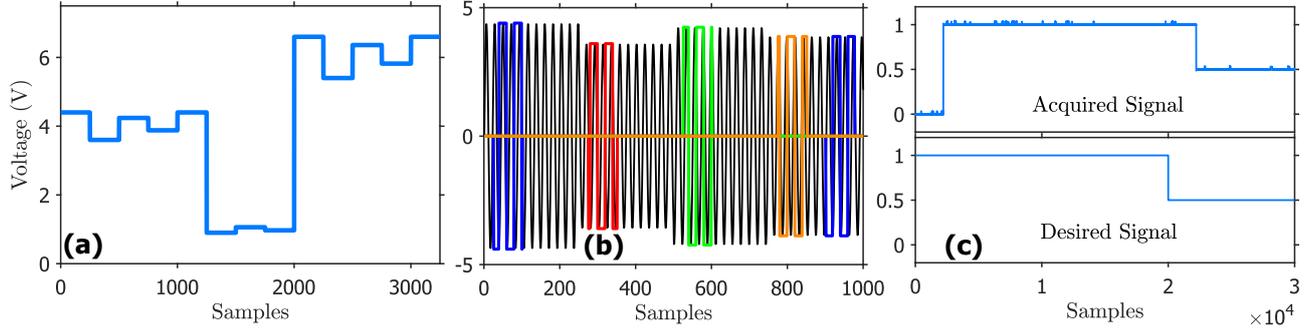

**Figure 26- (a)** Incorrectly multiplexed signal (uneven distribution of masks across the message), **(b)** Incorrect demultiplexing (the sampling for the next message value occurs too early and is therefore at the end of the prior message value), **(c)** Signal generation/acquisition delay (causing loss of data at the end of the message and a lack of synchronisation).

### iii.  Read-out Training

After the demultiplexing process, $N_x$ virtual neuron nodes are achieved [11]. Each of these has $N_{\text{message}} \cdot N_{\text{samplesPerMessageValue}}$ activation states. Each activation state is a sample associated with one of the original message values. A "teacher" result is provided to train the system read-out. This teacher can be generated by performing the desired application, such as PQC, in software, discussed in Section 2.2.2a)i. Once the Chua-RC has been trained for a specific application, a weight can be generated. The weight is multiplied by the matrix $\boldsymbol{X}$ of the output signals from the kernel. This generates the estimated result $\boldsymbol{Y}$, shown in Equation (3)

$$\boldsymbol{W} \times \boldsymbol{X} = \boldsymbol{Y} \tag{3}$$

Various training methods can be utilised to obtain the weight. This thesis has used an adapted Tikhonov Ridge Regression method (the typically utilised method for RC [1, 9, 25]), based on Algorithm 1 in [10]. This method requires matrix division – given $\boldsymbol{Y}$ and $\boldsymbol{X}$, determine $\boldsymbol{W}$. Since $\boldsymbol{W}$ will not be a square matrix, its inverse cannot be computed, complicating this process. Two numerical methods were considered: pseudo-inverse (implemented in MATLAB via $pinv$), and least-squares (implemented in MATLAB via $lsqminnorm$). The least-squares method was chosen since it is more efficient than pseudo-inverse [28]. The training read-out performance can be assessed by comparing the estimate and the desired result and calculating the Normalised Mean Square Error (**NMSE**). This is outlined in Equation (4):

$$\text{NMSE} = \frac{\sum_{i=1}^{n}\left(Y_{chua_i} - Y_{target_i}\right)^2}{n\sum_{i=1}^{n}\left(Y_{target_i}\right)^2} \tag{4}$$

$$\text{NRMSE} = \frac{RMSE}{\sigma} = \frac{\sqrt{\sum_{i=1}^{n}\left(Y_{chua_i} - Y_{target_i}\right)^2}}{\sigma\sqrt{n}} \tag{5}$$

This differs slightly from the metric utilised in the Jensen paper which is Normalised Root Mean Square Error (**NRMSE**), outlined in Equation (5). NMSE was chosen over NRMSE due to its common usage in other RC literature [2, 10, 11]. The training method utilises a bias in addition to the outputs from the Chua circuit; this is common practice. It also provides the option for an offset to ensure the output voltages don't average to 0 V. Pseudo-code for the training method, adapted from [10]) is as follows.

- Create empty matrices $XX$ and $YY$
- For each test case
  - Obtain $x_{chua}$, the demultiplexed matrix ($n_{timesteps} \cdot n_{channels}$)
  - For each row $p$ in $x_{chua}$
    - Add a bias of 1.
    - Add any offset required to the channel voltages to ensure non-zero average.
    - Calculate self-correlation matrix of row: $p_s = pp^T$
    - $XX = XX + p_s$ – this aggregates the neuron activation states for each test case
    - $YY = YY + y_{teacher} * p^T$ – this aggregates the desired output combined with the neuron activation states
- Calculate weight $= lsqminnorm(XX, (YY)^T)^T$

### 2.1.3 Post-Quantum Cryptography Application

This section will cover an introduction to Post-Quantum Cryptography (**PQC**), including explaining current state-of-the-art of cryptographic techniques, and why they are under threat from quantum computing. It will then illustrate an implementation of PQC that can be used as a test application for the Chua-RC.

### a)   Background of Public-Key Encryption and Post-Quantum Cryptography

PQC is the field of cryptographic techniques designed to be resistant to quantum computing decryption techniques [4]. Quantum computers are an emerging technology that utilise the ability of quantum bits (**qubits**) to simultaneously be both a 1 and a 0. This enables a vast increase in computation compared to classical computing [5]. It is believed that once quantum computers are designed with enough qubits, they will have the capability to crack the main public key encryption methods utilised across the world. Therefore, it is important to replace these methods with PQC [4]. It is important to note that PQC is distinct from Quantum Cryptography, which uses quantum computing to create more secure cryptosystems [29]. PQC algorithms are <u>not</u> quantum algorithms – they are classical mathematical problems that cannot be solved by quantum algorithms.

Two of the most important encryption methods utilised in the present day are the Advanced Encryption Standard (**AES**), and Rivest-Shamir-Adleman (**RSA**) [4]. Whilst very different, they each fulfil key criteria that mean, when utilised together, they form the backbone of modern encryption. AES is a symmetric private key algorithm based on the Rijndael Cipher [29]. The term symmetric in this context means that AES has a single key – used to both encrypt and decrypt a message [29]. Any party that wishes to access the plaintext (unencrypted) data from an encrypted AES ciphertext will require the same key. To ensure security, this key must be kept secret – a private key.

However, for a password to be used, it must often be shared. For example, in the case of accessing an online bank account, a user provides the password, and an authenticator checks the password. If correct, the decrypted data is provided. The bank's server is separate from your personal computer, and therefore the password must be transmitted over the internet. To do this securely, the password must be encrypted. However, if this password was to be encrypted using AES, this would require another private key. Once again, the same issue would occur – the issue being: how can a user tell a remote authenticator their password in order to gain access, without revealing their password to anyone who might intercept the communication.

In the cryptography community this problem is referred to as Alice (**A**) communicating one-way with Bob (**B**), whilst Eve (**E**), the eavesdropper, intercepts all communication [29]. The communication must be encrypted such that only Bob can decrypt it. This is achieved through asymmetric public key encryption (**PKE**) such as RSA. In this context, asymmetric means that there will be two distinct but related keys. One will be used for encryption – the public key, and the other for decryption – the private key. Since the public key is only used for encryption, anyone, including interceptors like Eve, can know it. It allows anyone to send an encrypted message to Bob. However, only Bob can decrypt the message as only Bob knows the private key [29]. Figure 27 illustrates the PKE process where a user (**A**) wishes to communicate login details with their Bank (**B**).

The bank's server generates a public key and communicates this to the user's computer. This is unencrypted, but that is acceptable since it is a public key. The computer encrypts the AES password using the RSA public key and sends it, encrypted, to the bank. Only the bank holds the RSA private key and therefore only the bank can decrypt the message, obtain the AES password, and authenticate whether the password is correct. For clarity, the reason symmetric (e.g., AES) is used for the encryption of the actual data and asymmetric (e.g., RSA) is used only for encrypting the keys, is because symmetric encryption/decryption is much faster [29].

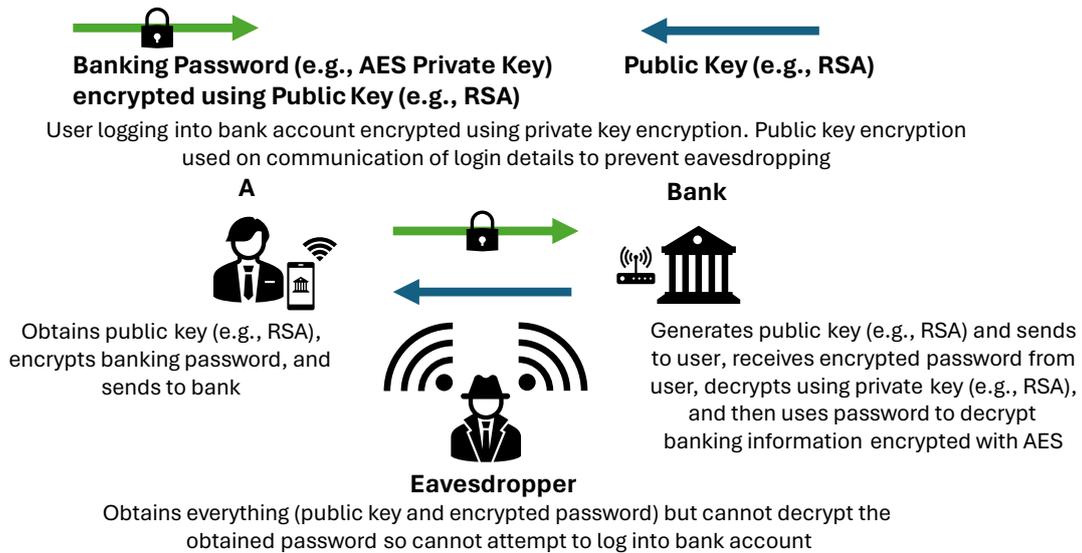

**Figure 27-** Public Key Encryption example - end user and bank with eavesdropper.

The RSA algorithm is based on the factorisation of numbers into secret primes. This is a very challenging mathematical problem, unless one factor is already known, in which case it becomes simple division. Equally, it is simple to factorise if the prime number is small. To be practical RSA must use extremely large numbers [29]. Knowing the secret factors allows only one party, Bob/Bank, to solve the problem. The precise RSA algorithm, and the way in which private and public keys are constructed is more complex than this, but this is sufficient for understanding this work. Whilst it is infeasible for even supercomputers to crack RSA, quantum computers using Shor's algorithm are predicted to have the capability within a decade [4]. Any current secure communications are also potentially at risk. "Harvest now, decrypt later" is being widely used, with the understanding that the data stored in current private communications, will likely still be relevant only a decade in the future. This includes financial data, military, political or corporate secrets, and private, personal data [5]. It is essential that the transition to PQC occurs promptly. Similar PKE methods such as Diffie-Hellman are also considered quantum-insecure [4, 5], but broad consensus is that symmetric encryption (AES) is quantum-resistant. Grover's algorithm is feared to pose a risk, prompting a call from NIST for larger AES key sizes, but at this time it is not considered a threat [4]. It is the threat of Shor's to PKE that must be addressed.

This has led NIST to identify potential replacements for RSA and other quantum-insecure PKE methods [5]. It requires similar mathematically "hard" one-way problems to RSA (easy to encrypt, near impossible to crack). Like RSA, they must have a trap-door method of decryption using a private key. They must, of course, not be susceptible to Shor's algorithm. Lattice-based (**LB**) algorithms have been identified as quantum-secure, and one sub-section of these algorithms is Learning with Errors (**LWE**) [4, 30]. NIST have selected several algorithms to form its new standards, with the main public key method being CRYSTALS-KYBER (based on LWE) [31]. They have called for academic research into implementations of these algorithms, and whilst this work will not focus on CRYSTALS-KYBER directly, it will investigate and attempt to provide a proof-of-concept for dedicated analogue reservoir computers performing LWE.

### b) Learning with Errors Problem

The Learning with Errors (**LWE**) problem is a one-way mathematical problem with a trap-door method. It proposes that, if presented with an array of $m$ tuples of data $(\boldsymbol{a_i}, b_i)$ where $i = 1, 2, \dots m$, each $\boldsymbol{a_i} \in \mathbb{Z}_q^n$, and each $b_i = \boldsymbol{a_i} \times \boldsymbol{s} + error \ (mod \ q) \in \mathbb{Z}_q$, it is mathematically hard to obtain $\boldsymbol{s}$ where $\boldsymbol{s} \in \mathbb{Z}_q^n$ [30, 32].

The notation translates as follows: $\boldsymbol{a_i} \in \mathbb{Z}_q^n$ is as a vector of size $n$ of any real integers $\mathbb{Z}$, reduced modulo $q$ (referred to as the modulus). This means that $\boldsymbol{a_i}$ is a vector of size $n$, consisting of integers from 0 to $q - 1$. The secret $\boldsymbol{s}$ is also a vector of size $n$ with integers in the range of 0 to $q - 1$. The error is a noise value that ensures the hardness of the problem by preventing the use of standard techniques for solving simultaneous equations. Note, in the original paper on LWE, the notation is $x \ (mod \ q)$ not $x \ \% \ q$ [30]. This thesis uses both the notation from Regev $x \ (mod \ q)$, and the MATLAB notation $mod(x, q)$.

In the case $n = 1$, each $\boldsymbol{a_i}$ can be treated as a scalar value. This is also true for the secret $\boldsymbol{s}$. If the modulus $q$ was chosen to be 5, each $\boldsymbol{a_i}$ can be any value from 0 to 4. If for one tuple, $\boldsymbol{a_i} = 4$, $\boldsymbol{s} = 3$ and $error = 2$, the value of $b_i = 4 \times 3 + 2 \ (mod \ q) = 14 \ (mod \ 5) = 4$ Simply presented with $a_i = 4$, $b_i = 4$, it is difficult to obtain the value of $s$ used to generate $b_i$ from $a_i$. Even with $m$ number of tuples, this cannot be simply determined mathematically – particularly as the $error$ value will vary between each tuple. Figure 28(a) demonstrates the calculation for $m$ vectors – this creates an $(m \times n)$ matrix.

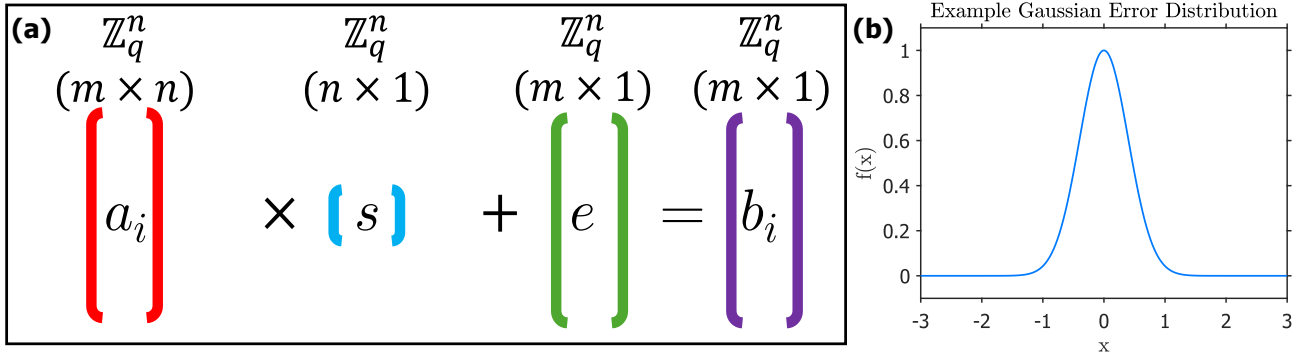

**Figure 28- (a)** Learning with Errors Problem, **(b)** LWE error distribution $\chi$ for $\alpha = 1$

The problem is defined with size parameter $n \geq 1$, modulus $q \geq 2$, and $error \in \mathbb{Z}_q$ chosen using a probability distribution $\chi$, defined by error parameter $\alpha$. An example error probability distribution can be seen in Figure 28(b), where $\alpha = 1$. The probability distribution $\chi$ is a non-standard gaussian distribution [30]. The gaussian distribution is defined as the following:

$$f(x) = \frac{1}{\sigma\sqrt{2\pi}} e^{-0.5\left(\frac{x-\mu}{\sigma}\right)^2}$$

(6)

This is where $\sigma$ is the standard deviation and $\mu$ is the mean. For a chosen error parameter $\alpha$, the standard deviation of the probability distribution $\chi$ is $\sigma = \frac{\alpha}{\sqrt{2\pi}}$. The mean is set as $\mu = 0$. Regev states that LWE is contingent on a real (non-complex) $\alpha > 0$ [30].

The LWE problem is translatable into a lattice-based problem. If the Gap Shortest Vector Problem (**GapSVP**) is hard in the worst-case, Regev states that LWE, due to its relationship with this pre-existing problem, is hard in the average case [30]. Since GapSVP has been assumed to be quantum-resistant, this in turn makes LWE quantum-resistant [32]. No quantum algorithms exist that can solve it. This allows LWE to be applied to public key cryptography, and this has occurred across numerous variants and examples. This thesis will focus upon an initial implementation of LWE, acknowledging that further work would be required to scale up and support more advanced algorithms within the LWE cryptosystem.

Regev proposes the following cryptosystem: the private key is the secret $\boldsymbol{s}$. As with all public key systems, the private key is retained by one party (e.g. Bob). Bob utilises the private key to generate the public key. The public key is an array of length $m$ consisting of the tuples: $(\boldsymbol{a_i}, b_i)$, where $i = 1,2,3,...,m$ [32]. Each $\boldsymbol{a_i}$ is generated randomly, and each $\boldsymbol{b_i}$ is generated according to the LWE equation, using $\boldsymbol{s}$. The public key is published freely for all to see and use. Once the public key has been generated, anyone can encrypt messages using it, and send them to Bob. Only Bob, with knowledge of the private key, has the capability to decrypt these messages. LWE cryptosystems are resistant against "chosen plaintext" and "chosen ciphertext" attack methods, so a malicious user with the public key cannot use the encryption process to attempt to reverse-engineer other encrypted messages [30].

A binary plaintext message (an array of 0s and 1s) can be encoded into an array $(\boldsymbol{u}, v)$ tuples by the following equations, where $k$ corresponds to randomly chosen indexes of the public key array, $\phi$ is the individual bit of the plaintext message being encoded, and $q$ is the modulus [30, 33]:

$$\boldsymbol{u} = \sum \boldsymbol{a_k} \ (mod \ q)$$

$$v = \sum b_k + \frac{q}{2} \times \phi \ (mod \ q)$$

(7)

If the public key array is size $m$, then its indexes (if indexing from 1 as per mathematical notation and MATLAB) are from 1 to $m$. Therefore, the values of $k$ can be any of the $2^m$ possible subsets of the values 1 to $m$. If $m = 20$ and it is chosen that there will be $N_{samples} = 5$ random samples, an example of the values of $k$ could be $[9, 18, 1, 7, 5]$. In the case where $n = 1$, both $s$ and $a_i$ can be considered scalars, resulting in:

$$u = \sum a_k \ (mod \ q)$$

$$v = \sum b_k + \frac{q}{2} \times \phi \ (mod \ q)$$

(8)

It is reasonable, given that Jensen has demonstrated the application of Chua-RC to non-linear polynomial regression [9], to investigate whether it can be applied to other non-linear operations such as modulo. LWE is a good real-world application for this test. Moreover, since the calculation of the encrypted ciphertext requires multiple pieces of data, $2N_{samples} + 1$ per message bit, this application requires a time-varying input signal, allowing the Chua-RC's capability with time-varying inputs to be assessed. The case $n = 1$ is chosen to restrict the scope of the project. Supporting vector private keys would be a later stage in development, but it is important to first test a proof-of-concept, and only scale-up after initial success.

For each message bit encrypted, the public and private keys remain the same. However, the samples used from the public key vary randomly, ensuring that two identical bits encrypted one after the other and using the same keys, will have differing ciphertext. This prevents a simple cryptanalysis attack: if every 1 was encrypted as the number $x$, and every 0 was encrypted as the number $y$, it would be embarrassingly simple to identify two possible messages $x = 0, y = 1$ and $x = 1, y = 0$, and then assess which, in context, is likely correct. Each message bit $\phi$ has its own $A_{samples}$ and $B_{samples}$ where $A_{samples} = [a_1, a_2, ..., a_{N_{samples}}]$ and $B_{samples} = [b_1, b_2, ..., b_{N_{samples}}]$. Each message bit is also encrypted to its own, distinct, $(u, v)$ tuple.

A 4-bit message (e.g., $[0, 0, 1, 0]$) would have an output of $[(u_1, v_1), (u_2, v_2), (u_3, v_3), (u_4, v_4)]$. Without knowledge of the private key $s$, it is mathematically hard to determine the original message. To decrypt this ciphertext, with knowledge of the private key, the following equation is used [32, 33]:

$$bitDecryption = v - u \cdot s \ (mod \ q)$$

(9)

Where $bitDecryption$ is closer to $q$, the message bit is a 1. Where $bitDecryption$ is closer to 0, the message bit is a 0. This can be written as:

$$\textbf{\textit{if}} \left( bitDecryption > \frac{q}{2} \right) \textbf{\textit{then}} \ bit \ \phi = 1$$

$$\textbf{\textit{else if}} \left( bitDecryption < \frac{q}{2} \right) \textbf{\textit{then}} \ bit \ \phi = 0$$

(10)

Two key criteria for a cryptosystem are correctness, and security [30, 32]. Correctness ensures that the encryption and decryption processes work. Put simply: do a set of chosen parameters, when used in the methods outlined above, ensure that the decrypted message will be equal to original plaintext. Security, concerns whether a chosen set of parameters will retain the hardness of the problem or open the problem to a weakness or flaw that can be exploited by an attacker.

There are many different conditions under which LWE can satisfy both correctness and security. Regev gives one example as: the modulus $q$ is a prime number between $n^2$ and $2n^2$, $m = c \cdot n \log(q)$ (where $c$ is an arbitrary constant), and $\alpha = \frac{1}{\log^2(n)\sqrt{n}}$ [30]. To investigate the ability of Chua-RC to perform LWE, some cryptosystem parameters (such as the modulus) will be fixed. To train the Chua-RC, a software implementation of the LWE cryptosystem must be created. This will provide the teacher data for the RC. This is discussed in Section 2.2.2a). This section will also cover how the input to the Chua-RC can be constructed using required encryption parameters from the LWE cryptosystem.

### c) Alternative Cryptosystem

Another lattice-based cryptosystem considered was NTRU (pronounced, "n-true", not an abbreviation). This differs from LWE and was primarily considered due to an already existing MATLAB implementation of both encryption and decryption [34]. The NTRU system is the basis of several high-performing PQC algorithms, including finalists in the NIST process. However, an NTRU algorithm was not selected as a future NIST standard [5, 31], whereas LWE-based algorithms such as CRYSTALS-KYBER were. It is believed to be less secure than LWE, despite having advantages in speed [31]. This means its prominence is less, and it is instead better to focus this thesis on LWE – the likely future of public key encryption. Another concern was that the MATLAB implementation was not from a properly published work, but rather a GitHub Repository of a student – so it would require thorough validation to ensure it was correct, costing time. Finally, NTRU, like many cryptosystems involved a 1-to-many encryption process. This is also true for LWE, where a single bit is encrypted into two values $u$ and $v$. However, in the case of the existing MATLAB implementation of NTRU, this process was more complicated – the output size would vary depending on the input and was far greater than the input size – for example 1 number being encrypted to 47 numbers. This is very unwieldy for the Chua-RC application, compared with the fixed 1-to-2 LWE cryptosystem.

Originally, the option of implementing a NIST process algorithm, such as CRYSTALS-KYBER, was considered. This was clearly an over-ambitious goal. It would require a much deeper knowledge of cryptographical mathematics and would be more challenging to translate to Chua-RC. It would also be a significant challenge to implement in MATLAB. Instead, the specified LWE cryptosystem is an ideal choice. This still represents a significant challenge but is an achievable goal. Since LWE is the basis of complex algorithms like CRYSTALS-KYBER, it will indicate whether Chua-RC implementations are feasible.

## 2.2  System Development (Realisation and Sub-System Testing)

This section covers the development of the design into a functional system. This primarily consists of three sub-systems: the RC input, output, and kernel. The development of the Chua circuit kernel is outlined first.

### 2.2.1  Chua Circuit Development

#### a)  Breadboard Design and Implementation

A breadboard-based Chua circuit was designed using TinkerCAD and then constructed using the components in Table 5. TinkerCAD was used due to its versatility and ease of use. It was also useful for developing a PCB case, as will be discussed. The design and construction can be seen in Figure 29(a-b). This allowed initial testing of the circuit prior to a more robust implementation on Veroboard or PCB. As in Figure 29, the circuit is designed with a fixed resistor rather than utilising a variable potentiometer. This allowed for an initial result to be achieved, and the resistor could later be swapped out for a 2 k$\Omega$ potentiometer. Figure 29(c) shows two power supply units (**PSU**) used to achieve $\pm 9$ V. The bottom PSU achieves $-9$ V through the connection of its positive terminal to the ground of the upper PSU. These power supplies can be substituted for two 9 V batteries with a two-pole switch, which was later completed to make a demonstration circuit, but having a stable power supply is superior. One missing feature on the breadboard design was the addition of bypass capacitors to help maintain a constant supply voltage. This is rectified in future Veroboard/PCB designs. An LCR meter was used to verify the inductance – seen in Figure 30(a-b). The inductance is specifically tested at 100 Hz and 1 kHz, to ensure that it is consistent across different possible frequencies.

**Table 5-** Required components and cost exc. VAT.

| Component | Cost (Exc. VAT) |
|---|---|
| 18 mH Radial Inductor (17 $\Omega$ series resistance, 5% tolerance, 21 mA max DC) [35] | £1.22 |
| 22 k$\Omega$, 2.2 k$\Omega$, 3.3 k$\Omega$, 220 $\Omega$, 100 nF, 10 nF Resistors/Capacitors (and others) | Free |
| Dual Op-Amp IC AD712KNZ (4 MHz Gain Bandwidth Product) [27] | £11.30 |
| Analog Discovery 2 (**AD2**) (12 MHz Bandwidth, 100 MSamples/s AFG, 30 MHz Oscilloscope) [13] | £238.94 |
| Nidec 2 k$\Omega$ Rotary Pot 10-Turns 1-Gang, M-22E10 (5% tolerance) [36] | £16.45 |

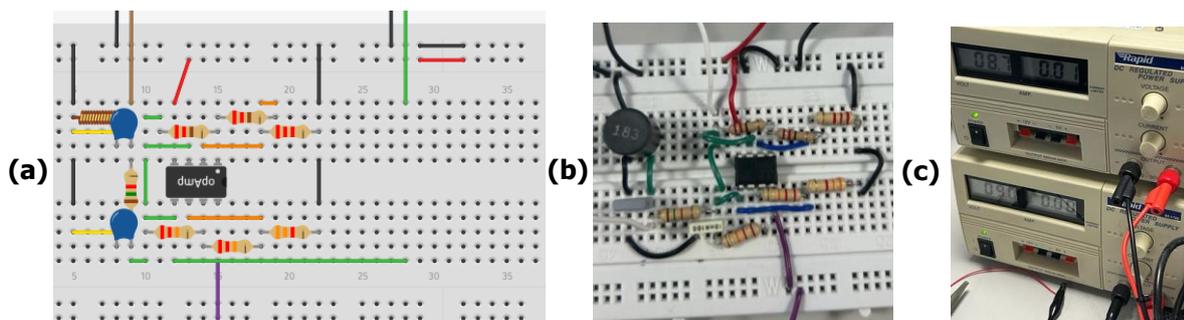

**Figure 29- (a)** TinkerCAD breadboard design of Chua circuit, **(b)** Breadboard construction in lab, **(c)** $\pm 9$ V achieved via two 9 V power supplies with bottom PSU + connected to top PSU GND.

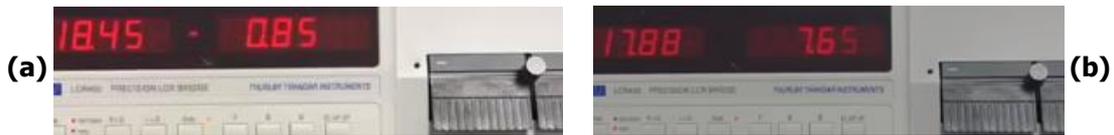

**Figure 30-** Inductance measured using LCR meter **(a)** 100 Hz, **(b)** 1 kHz.

A recurring issue when constructing the circuit was static in the laboratory. Small amounts of static were sufficient to destroy the expensive AD712 ICs. To compensate for this, all future laboratory work was conducted using an earthed anti-static wristband, seen in Figure 31(c). Another issue was the poor connectivity of the breadboard connections. This caused the output to be inconsistent, regularly preventing the circuit from exhibiting chaos. This meant the breadboard circuit was not practical for further testing or research beyond an initial validation that the circuit was correct. The initial breadboard results are shown in in Figure 31(a-b). These are two Chua double scroll chaotic attractors, achieved by a 2 kΩ potentiometer at approximately 1.55 kΩ and 1.6 kΩ respectively. They validate that the circuit exhibits chaotic behaviour. To reduce additional lead resistance in wires and to ensure strong, reliable electrical connections, the decision was taken to design a PCB, rectifying the problems. A Veroboard design was produced first, shown in Figure 31(d). This would, to an extent, ensure superior connections (due to soldering), but would require additional wires compared to a PCB design. A PCB design is more reliable, ideal for future mass-production, and better guards against parasitics and noise, in-line with the specification goals.

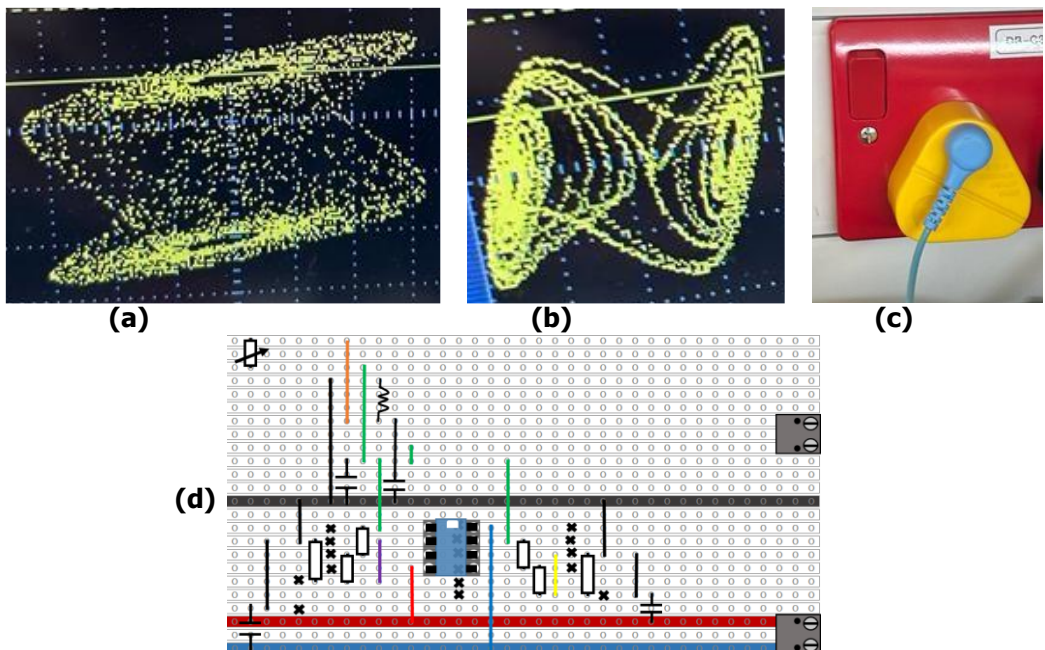

**Figure 31- (a)** Double scroll chaotic attractor (approx. 1.55 kΩ), **(b)** Double scroll chaotic attractor (approx. 1.6 kΩ), **(c)** Earthed anti-static wristband, **(d)** Veroboard design (Kennedy Chua circuit).

### b) PCB Design and Implementation

The PCB was designed using KiCAD. The electrical connections, the 3D rendering, and the final soldered PCB circuit can be seen in Figure 32(a-c). As shown, aside from wires to external connections (PSU, oscilloscope, etc.), the PCB eliminated additional sources of lead resistance. It also introduced two bypass capacitors to maintain the supply voltage. Finally, terminal blocks were added to allow the use of different potentiometers or fixed resistors, and to allow for read-out of both the Inductor Voltage and the Chua Diode Voltage. In the design, the track width was set to 0.5 mm, reducing parasitic inductance and noise. Minor issues were found with the PCB design. For example, the silkscreen for the terminal blocks was placed back-to-front. However, this caused no issue as the blocks could still be soldered in the correct direction. Similarly, the inductor footprint chosen was slightly larger than required; this too caused no issue. KiCAD was chosen due to previous experience with the software and enabled the use of manufacturer footprints for components. The order time for the PCB was only 2-3 weeks; this was ordered immediately after Christmas to arrive by the new term.

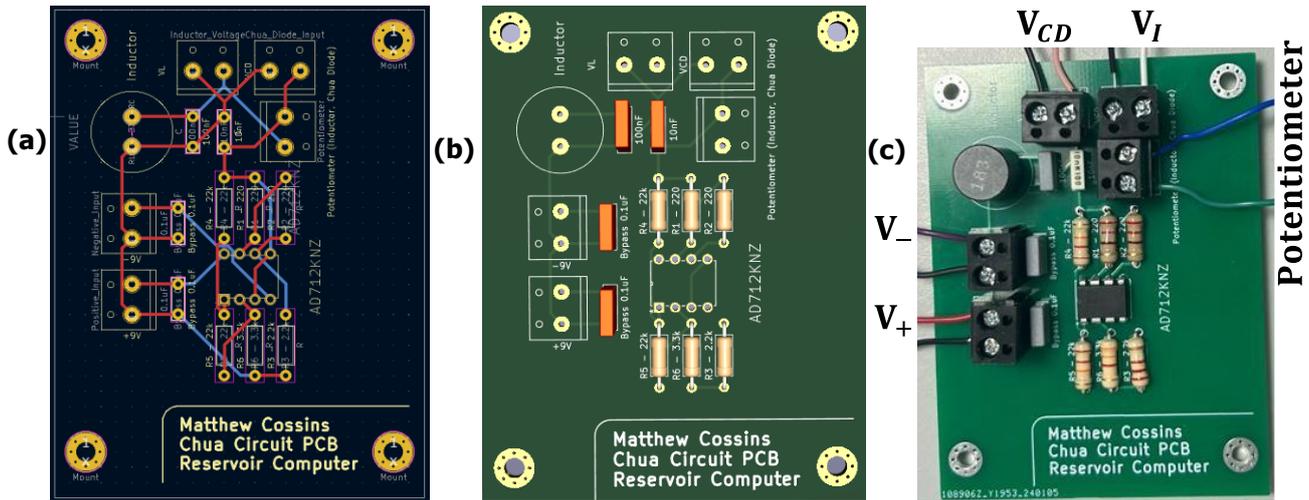

**Figure 32- (a)** PCB design (KiCAD), **(b)** 3D rendering of PCB, **(c)** Soldered circuit board.

It was logical to produce multiple versions of the circuit to enable different components to be varied when performing testing and obtaining bifurcation data. This was possible since the PCB order required a minimum of five. As such, additional components were ordered. To keep costs down, the AD712KNZ was substituted for the inexpensive TL072. To allow both the additional Op-Amps and additional inductors to be ordered from the preferred supplier of Farnell, the inductor was substituted for a COILCRAFT RFC1010B-186KE. These components, their specification, and their cost, can be seen in Table 6.

**Table 6-** Additional components and cost exc. VAT.

| Component | Cost (Exc. VAT) |
|---|---|
| 18 mH Radial Inductor, 23 Ω Series Resistance, 210 mA [37] | £1.64 |
| TL072 IC Dual-JFET 3 MHz Bandwidth [38] | £0.64 |

Another method for reducing static or other damage to the circuit is to provide protection for the circuit elements. A PCB case was designed, using TinkerCAD to edit an existing freely available design [39]. The design is pictured in Figure 33(a). This case was then 3D-printed and Figure 33(b) shows it with the circuit inside.

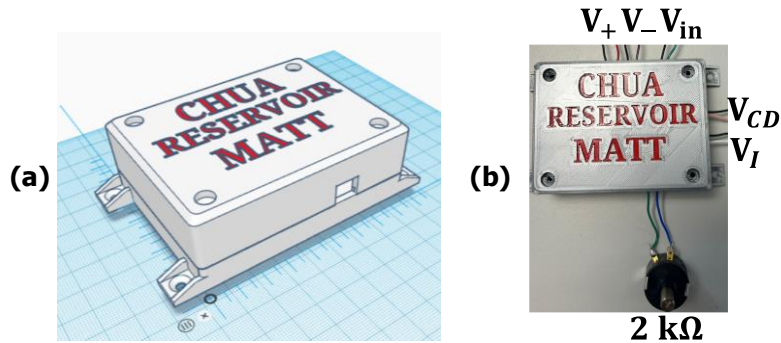

**Figure 33- (a)** 3D CAD model of PCB case, **(b)** 3D-printed case with PCB inside.

### c) Chua Circuit Sub-System Testing

The circuit was tested to compare against the simulations. This meant producing parametric plots of the Chua Diode Voltage against the Inductor Voltage, and time-domain plots. Resistance, capacitance, and input voltage amplitude were varied to generate bifurcation diagrams for each. Noise and bandwidth were also assessed through taking an FFT of results and using the AD2 SNR function. Finally, the V-I characteristic of the implemented Chua Diode was measured, using the technique demonstrated by Kennedy [21]. This could be repeated at different times to show the variance between readings and therefore the impact of noise – Section 3.8.

|  | **Analog Discovery 2** [13] | **GwInstek** [40] |
|---|---|---|
| **Sampling Rate** | 100 MSamples/s | 1 GSamples/s |
| **Range** | ±25 V | ±15 V |
| **Bandwidth** | > 30 MHz | 25 MHz |

**Table 7-** Comparison of specification between Analog Discovery 2 and GwInstek.

The GwInstek Digital Oscilloscope GDS-1022 was used for the initial validation testing of the breadboard circuit, but the AD2 is utilised for proper testing. This is the superior choice as it allows data to be easily stored within MATLAB. Moreover, the AD2 is analogue and has a sufficient sampling rate to ensure that output plots are clear and high resolution. Table 7 shows that the device specifications are broadly similar, despite the AD2 being cheaper and portable. It is also logical to use the AD2 since it will later be used for the RC.

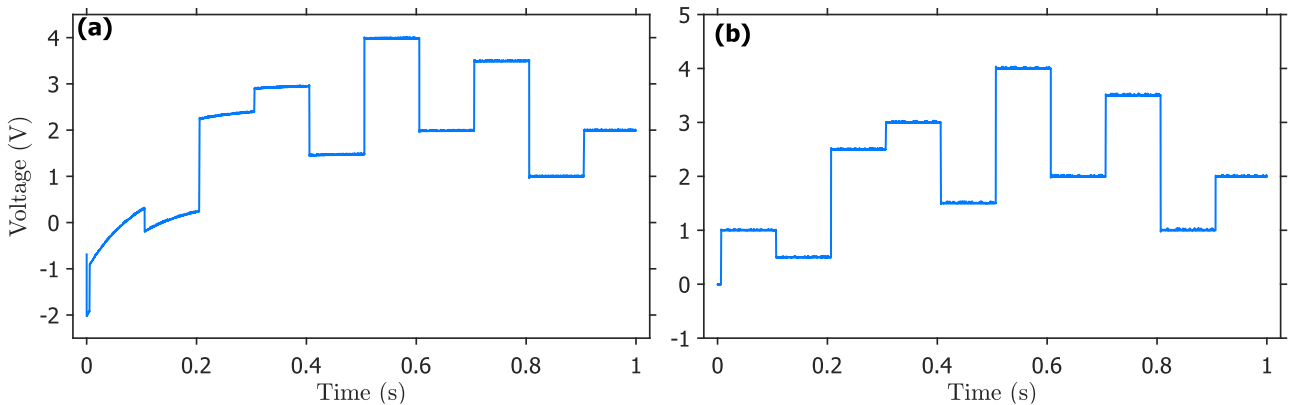

**Figure 34**- (a) Degradation of signal, (b) Correct generation/acquisition of signal with AD2.

An initial issue encountered when using the AD2 with MATLAB was a strange degradation in performance. This signal degradation can be seen in Figure 34(a) and was determined to be due to the simultaneous use, and conflict of, two different MATLAB toolboxes designed to work with Digilent devices. Removing some legacy code ($daqreset$) resolved the issue, seen in Figure 34(b). Importantly, this allowed for automated tests without having to manually reset the AD2 each time. Originally, it was anticipated that there would be a restriction on the amount of buffer data that could be used to create the arbitrary function on the AD2. This would restrict the option to adapt the work in the future to encrypt multi-bit messages, as opposed to the single-bit encryption focussed on by this project. The use of the newer toolbox removed any restriction on buffer size, addressing this concern. This is discussed in Section 2.2.2a)i.

### i.   Varying Resistance

Figure 35 demonstrates time-domain and parametric plots for the physical circuit when varying resistance. This demonstrates the same characteristics as with the previous LTSpice and Simulink models. Figure 35(a) shows a single-period parametric at 1.8 kΩ. As resistance is decreased, a Rössler-type attractor is achieved at 1.7 kΩ, seen in Figure 35(b). Decreasing the resistance further produces a double scroll at 1.6 kΩ, seen in Figure 35(c). This is shifted compared with Figure 4 in Section 2.1.1c)i, but shows the same process of different states of the Chua circuit as resistance is increased or decreased. The difference is that lower resistances are required in the physical circuit to achieve the same outputs. This is clearer in the resistance bifurcation diagram (Figure 36).

To plot the resistance bifurcation diagram, a PCB was populated with a fixed capacitance of 10 nF. No input voltage was applied. The resistance bifurcation diagram is limited in its resolution. The resistance had to be manually adjusted using a potentiometer, measured with a multimeter, and then the voltage reading taken. This made the process slow and inefficient – if this work was to be repeated, a programmable digital potentiometer would likely be utilised. However, the tests taken are sufficient to clearly show the bifurcation diagram. Figure 36(a) and Figure 36(b) both show good agreement with the diagrams achieved in Figure 13(a) and Figure 13(d). The key differences are firstly that the resistances at which major bifurcations occur are shifted, and secondly, the additional negative voltages in Figure 36(a).

The shift in resistances is likely due to non-idealities in the physical circuit. Additional series resistances in components such as the terminal blocks, and in the wires and PCB tracks could mean that despite the potentiometer being set to a value $R$, the actual resistance "seen" by the circuit is $R + r$ where $r$ is additional series resistance. This means that the discrepancy between the simulation and the physical circuit is likely smaller than appears. This discrepancy is not a problem. It means that chosen parameters for Chua-RC in the physical circuit must be shifted lower than those used in simulation to achieve the same voltage outputs.

The additional negative voltages are explainable by the fact that the position of the single scroll of the Chua circuit (positive or negative) is dependent upon the direction of change – a hysteresis effect. The tests that resulted in negative voltages will have arisen from adjusting the resistance in the other direction from the other end state (DC equilibrium or Chua diode limit), than the positive voltages. This has no impact, and the diagrams are essentially symmetrical about 0 V.

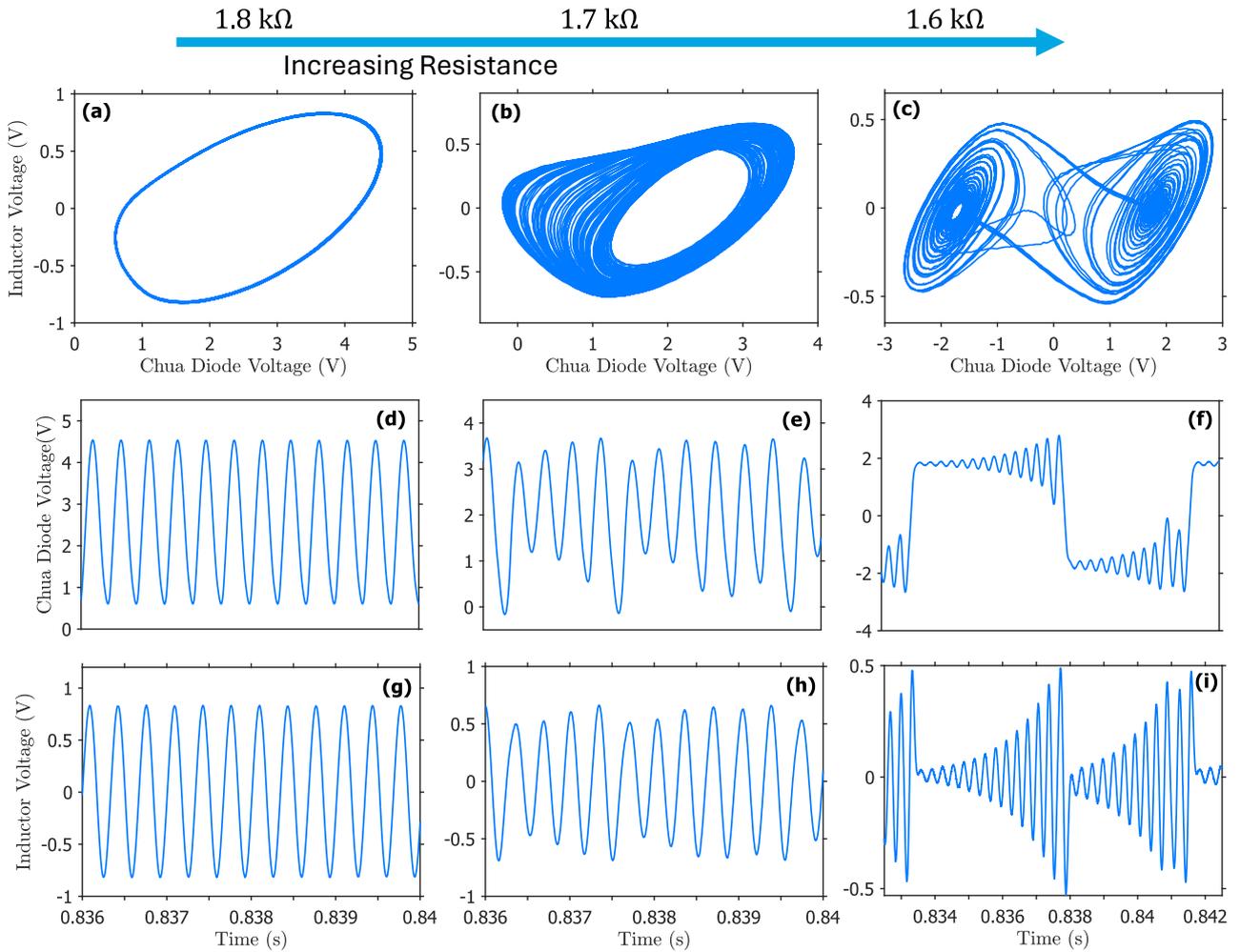

**Figure 35-** Time-domain and parametric voltage plots from physical circuit for varying resistance, **(a)** Single period 1.8 kΩ, **(b)** Rössler-type attractor 1.7 kΩ, **(c)** Double scroll attractor 1.6 kΩ, **(d)** Chua Diode Voltage 1.8 kΩ, **(e)** Chua Diode Voltage 1.7 kΩ, **(f)** Chua Diode Voltage 1.6 kΩ, **(g)** Inductor Voltage 1.8 kΩ, **(h)** Inductor Voltage 1.7 kΩ, **(i)** Inductor Voltage 1.6 kΩ.

As stated, the results achieved in Figure 36 are extremely close to those achieved in simulation in Figure 13.

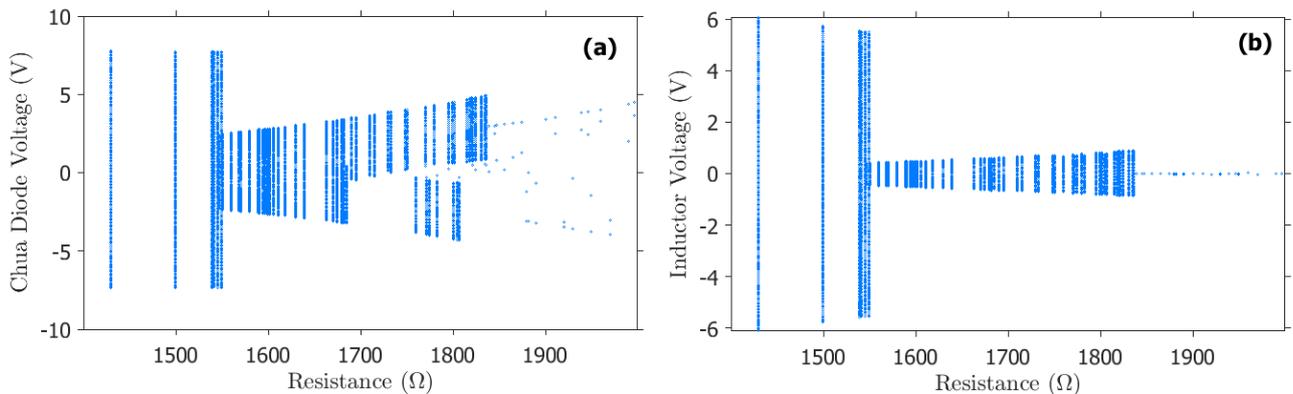

**Figure 36-** Bifurcation diagram of physical circuit for varying resistance, **(a)** Chua Diode Voltage bifurcation - fixed capacitance of 10 nF, **(b)** Inductor Voltage bifurcation – fixed capacitance of 10 nF.

## ii. Varying Capacitance

Varying capacitance was more difficult than the prior work with resistance. Resistance can be altered through a 2 kΩ potentiometer, but similar components for capacitance are less common. Since it was clear from the simulations that varying capacitance offers a very similar bifurcation diagram to resistance, it was deemed not necessary to invest in any specialist equipment. Instead, a breadboard testing set-up was utilised to construct different parallel configurations of available workshop capacitors. 14 tests were performed. Despite being a low number of tests, this was sufficient to reproduce, at low resolution, the bifurcation diagrams in Figure 13(b) and Figure 13(e). A dedicated circuit board was created for testing capacitance, with a female header used to replace the capacitor. Wires were attached from the header to the breadboard capacitors. The potentiometer was set to 1.7 kΩ to provide the middle of the resistance bifurcation in Figure 36. No input voltage was present. The testing set-up is shown in Figure 37(a). The breadboard capacitances were verified using the LCR meter, seen in Figure 37(b). This shows a 10 nF capacitor in parallel with a 3 nF capacitor, achieving 13 nF.

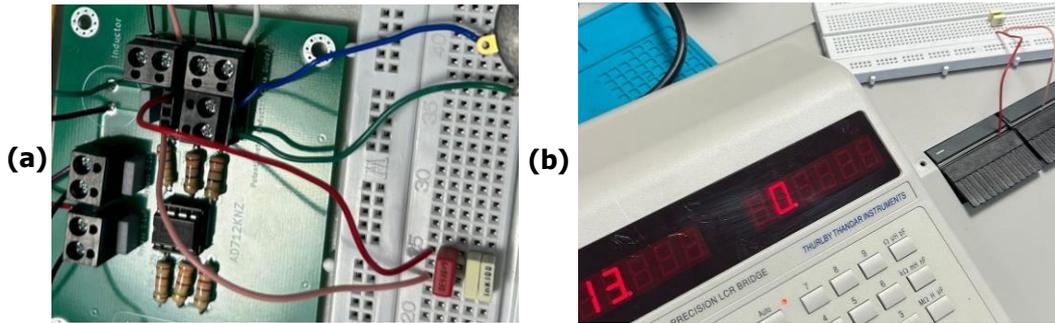

**Figure 37- (a)** Varying capacitance using breadboard extension to circuit, **(b)** LCR meter used to confirm that parallel capacitance on the breadboard achieves desired capacitance.

Figure 38(a-b) provides the bifurcation diagrams produced from these tests. These are low resolution, but they clearly show good agreement with the results from Figure 13. Despite the rudimentary testing set-up, the key changes in the diagram occur at approximately 8 nF and 12 to 13 nF for the physical circuit, the same as the simulation results in Figure 13.

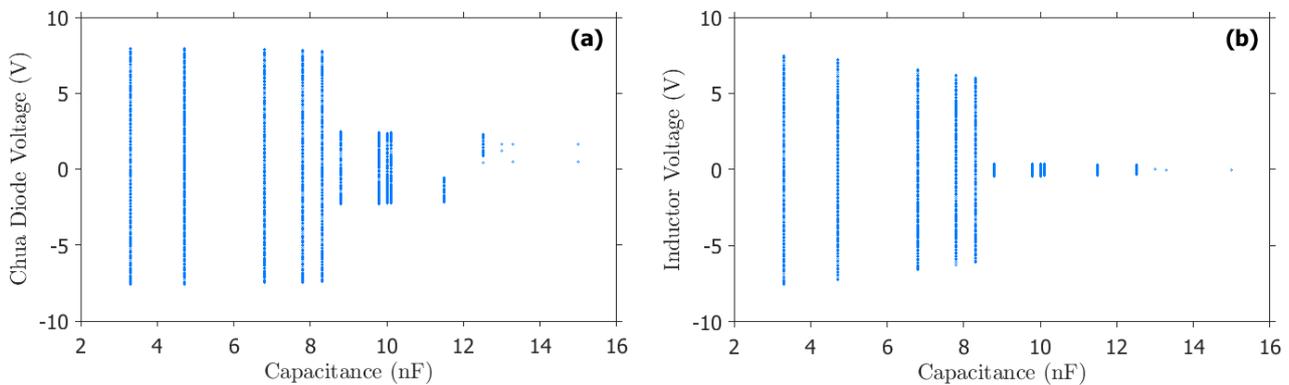

**Figure 38-** Bifurcation diagram of physical circuit for varying capacitance, **(a)** Chua Diode Voltage bifurcation - fixed resistance of 1.6 kΩ, **(b)** Inductor Voltage bifurcation – fixed resistance of 1.6 kΩ.

## iii. Varying Input Voltage

After the PCB design had already been completed and sent for manufacture, it was realised that an oversight meant there was not a port for providing the input voltage to the circuit. A fix was quickly implemented to rectify this mistake. By removing the inductor from the board and rewiring the inductor's through-holes to a secondary daughter board, a series voltage could be added between the inductor and ground. This voltage can be provided by the AD2, controlled by MATLAB. The daughter board and PCB combination can be seen in Figure 39(a). This board was used with fixed capacitance of 10 nF. For testing the bifurcation, the potentiometer was used to set the resistance to 1.85 kΩ – this is just into DC equilibrium (on the edge of chaos), so allows the applied voltage to perturb the system into the full range of different chaotic states. A sine wave carrier at fixed frequency of 500 Hz is used for the input voltage. An improved PCB design can be seen in Figure 39(b). This introduces the missing voltage port, fixing the design.

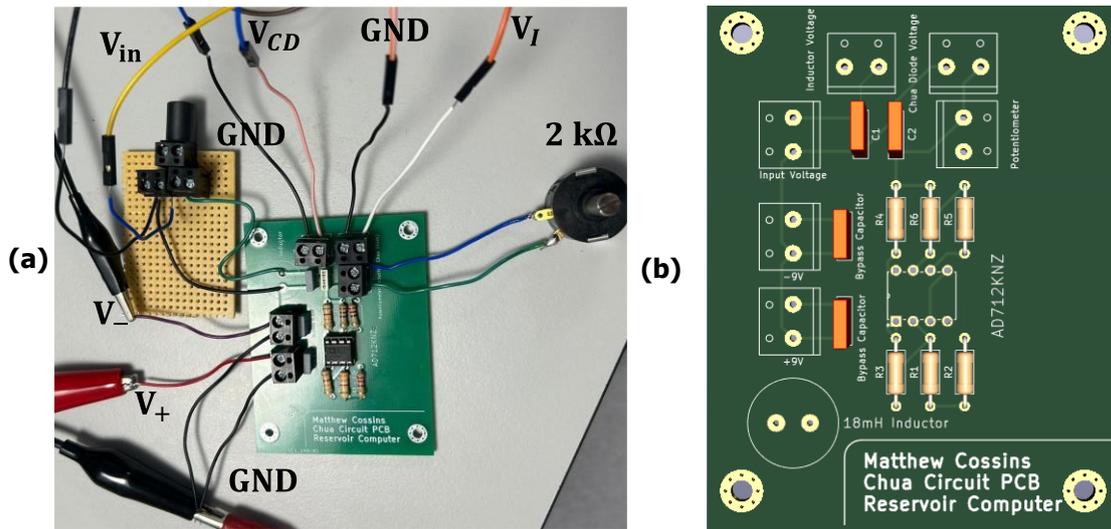

**Figure 39- (a)** PCB with daughter board for input voltage in serial with inductor. Resistance achieved through potentiometer, **(b)** Improved PCB design with input voltage port in series with inductor.

Figure 40 shows the bifurcation diagrams for varying input voltage amplitude. These have much higher resolution than the prior resistance and capacitance bifurcation diagrams (Figure 36 and Figure 38). This is because the AD2 can be controlled programmatically via MATLAB. This enabled writing a script to automate the process of generating different amplitudes, acquiring the output, and storing it. Manual intervention on each test was no longer required. This is also the reason why adjusting the voltage amplitude is the ideal method of the three for performing Reservoir Computing (however it is important to understand the other parameters so that ideal fixed conditions can be obtained). Figure 40(a) and Figure 40(b) clearly agree with the simulated bifurcation diagrams in Figure 13(c) and Figure 13(f). The key difference is that the experimental diagrams are much cleaner – particularly the Inductor Voltage in Figure 40(b). This could be due to the low-pass filter and sampling frequency of the AD2 (100 kHz was used) filtering out higher frequency voltages. It could also be due to the AD2 accuracy being less than that of MATLAB, so smaller differences in voltage aren't recorded.

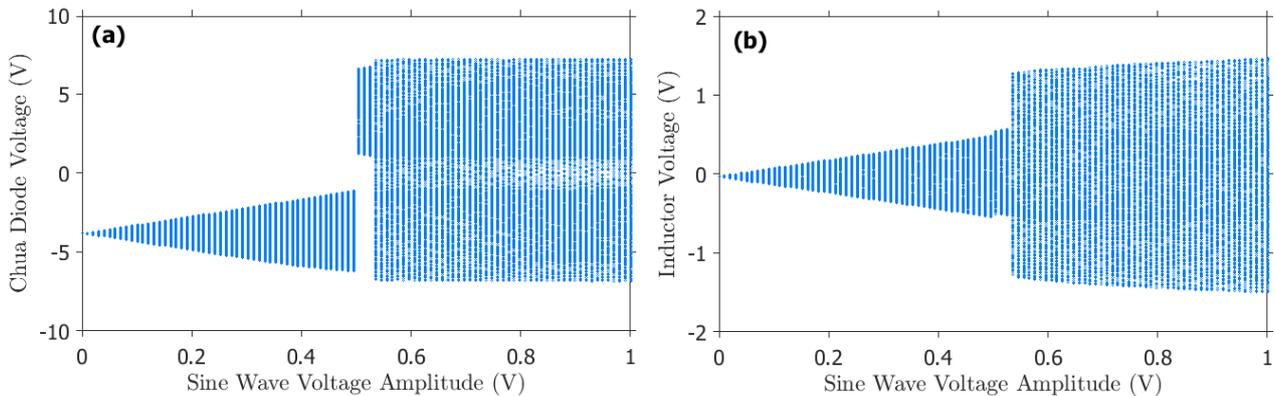

**Figure 40-** Bifurcation diagram of physical circuit for varying input voltage amplitude, **(a)** Chua Diode Voltage bifurcation – fixed parameters of 1.85 kΩ and 10 nF, **(b)** Inductor Voltage bifurcation – fixed parameters of 1.85 kΩ and 10 nF.

### iv. Assessing Noise and Bandwidth

Figure 41(b) shows the FFT of the time-domain voltage in Figure 41(a). Figure 41(a) is the Chua Diode Voltage at the simplest chaotic state of the physical Chua circuit. This conforms with the previously simulated FFT in Figure 9. There are slight differences, which is to be expected between simulation and actual circuit, such as slightly reduced amplitude across most frequencies. It shares a similar spike just above 1 kHz, and, aside from a small spike between 10 and 100 Hz, the highest amplitudes are in the range of 1 to 10 kHz.

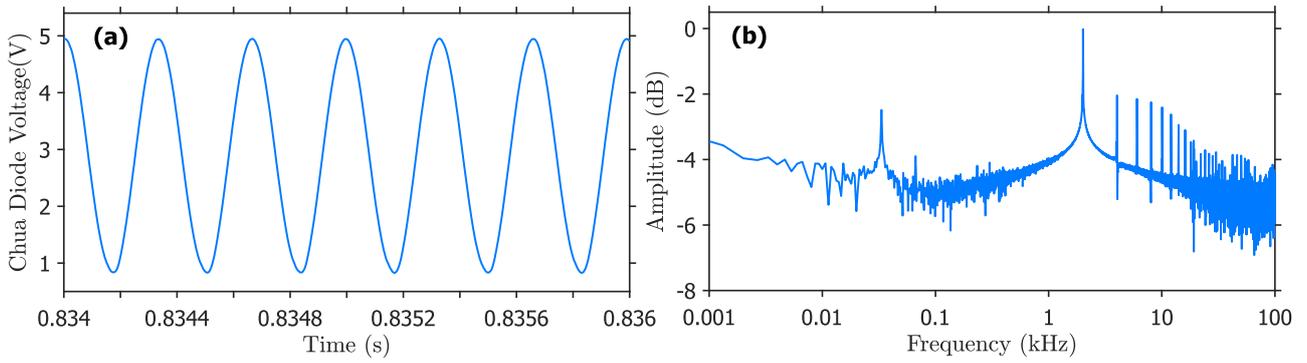

**Figure 41- (a)** Time domain of Chua Diode Voltage (physical Chua circuit) at simplest unstable period, **(b)** FFT of Chua Diode Voltage.

The in-built SNR function of the AD2 was utilised to obtain an estimate for the SNR at the output of the system. This varied depending upon frequency. The bandwidth was determined to be between approximately 1 to 10 kHz, so SNR was assessed at both frequencies. At each frequency, both the Chua Diode Voltage and the Inductor Voltage achieved approximately 30 dB. The results can be seen in Figure 42(a-b).

| #Digilent WaveForms Spectrum Measurements | | | |
|---|---|---|---|
| #Device Name: Discovery2 | | | |
| #Serial Number: SN:210321B2021D | | | |
| #Date Time: 2024-03-19 15:48:55.429 | | | |
| | | | |
| Channel | Name | Value | Frequency |
| T1 | SNR | 29.66 dBc | 1 kHz |
| T2 | SNR | 31.86 dBc | 1 kHz |

| #Digilent WaveForms Spectrum Measurements | | | |
|---|---|---|---|
| #Device Name: Discovery2 | | | |
| #Serial Number: SN:210321B2021D | | | |
| #Date Time: 2024-03-19 15:46:26.744 | | | |
| | | | |
| Channel | Name | Value | Frequency |
| T1 | SNR | 29.11 dBc | 10 kHz |
| T2 | SNR | 31.26 dBc | 10 kHz |

**Figure 42- (a)** SNR reading from AD2 at 1 kHz, **(b)** SNR reading from AD2 at 10 kHz.

This once again indicates that there are no issues with noise in the circuit. A final test to characterise noise, is assessing the V-I characteristic of the Chua diode in multiple readings across different days. This demonstrates the variance in the V-I characteristic due to thermal Johnson noise. This is provided as part of the Final System Testing in Section 3.8.

### 2.2.2 Reservoir Computer Development
### a) Learning with Errors Implementation
### i. MATLAB Code Development and Translation to Chua-RC Input

As outlined in Section 2.1.3b), the values required for the encryption of a message bit $\phi$, in addition to the message bit itself, are $A_{\text{samples}}$ and $B_{\text{samples}}$ [30]. The possible LWE values are all reduced modulo $q$, restricting them to the range 0 to $q-1$. Since the RC kernel is analogue, the LWE range can be normalised to a desired analogue input voltage range. This thesis will focus on single-bit encryption, attempting to establish a proof-of-concept. Figure 43 shows how the parameters required to encrypt a single bit can be constructed into an input for the Chua-RC. The cryptosystem in Figure 43 is defined with modulus $q = 7$, restricting the range of both input and output values to 0 to 6. The vector size is set $n = 1$, making the private key, in this case $s = 2$, a scalar, and public keys $A$ and $B$ vectors rather than matrices. This is done, as discussed in Section 2.1.3b), to restrict the scope of the project to objectives achievable within the time constraints. 5 samples are taken from each of $A$ and $B$, creating $A_{\text{samples}}$ and $B_{\text{samples}}$. Along with the message bit $\phi$, and dummy message bit $D$, as discussed in Section 2.1.2b)ii, the input buffer can be formed, see Equation (11).

$$inputBuffer = [A_{\text{samples}}, B_{\text{samples}}, \phi, D] \qquad (11)$$

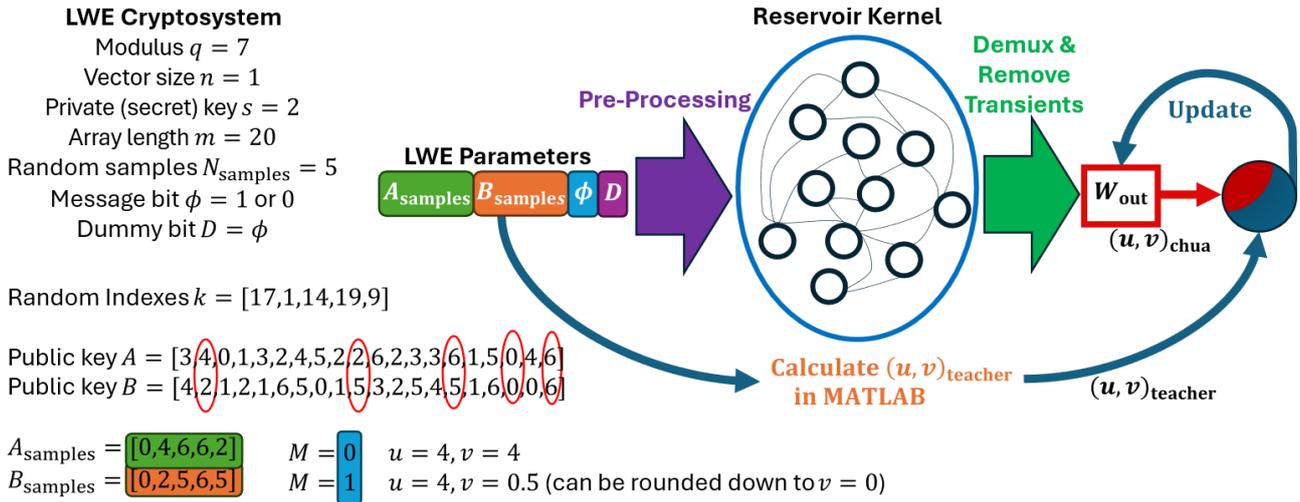

**Figure 43-** RC process for LWE encryption. The teacher is adapted to be a two-dimensional output $[u, v]$ and the input is a time-varying signal containing $A_{\text{samples}}$, $B_{\text{samples}}$, the message bit $\phi$, and a dummy message bit $D$. A brief example input is provided for an example LWE cryptosystem.

As discussed in Section 2.1.2a), the AD2's resolution presents a constraint on the input to the Chua-RC. Another constraint is posed by the knowledge that the optimal voltage amplitude range will likely be around the magnitude of 0 to 1 V, as per the bifurcation diagrams in Figure 40(a-b). It is known from Section 2.1.2a)v, that the AD2 is accurate only to $\pm 10$ mV $\pm 0.5\%$ [13]. This means that the number of different possible values that can be encoded without overlap is limited. By fixing the LWE modulus $q$, the potential values can be restricted. This is because all matrices involved in LWE are reduced modulo $q$, limiting their values to the range 0 to $q - 1$. In the example in Figure 43, the input message will contain values in the range 0 to 6. This range is encodable within the limited voltage range, without requiring value intervals beyond the accuracy and resolution of the AD2. For example, if 0.1 V is taken as the minimum, and 0.5 V taken as the maximum, the voltage values 0 to 6 will be normalised as shown in Table 8. For this reason, $q = 7$ is used in Section 3.4, as it ensures each value is distinct when normalised into a specific voltage range around the 1 V magnitude.

| Message Value | 0 | 1 | 2 | 3 | 4 | 5 | 6 |
|---|---|---|---|---|---|---|---|
| Normalised Voltage (V) | 0.1 | 0.167 | 0.233 | 0.3 | 0.367 | 0.433 | 0.5 |

**Table 8-** Example normalisation of input message values into voltage within 0.1 to 0.5 V range.

Since $q = 7$ is a relatively low modulus, there is a small probability that a test case is not correct (the decrypted value is not equal to the plaintext value) [30]. When test cases are generated, decryption is performed to identify any incorrect results. For example, of 500 randomly generated test cases, 493 may correctly encrypt both a 1 and a 0, and the remaining 7 that performed at least one of the bit encryptions incorrectly, are discarded. Of $N_{\text{testCases}}$, it is appropriate to randomly partition into a proportion for training $N_{\text{train}}$ and a smaller proportion for validation $N_{\text{val}}$. This reduces overfitting to the training sample as performance is assessed on unseen cases; the goal is for the Reservoir Computer to "learn" how to perform the LWE encryption operation in general, not for it to only be "taught" specific cases. Good options range from 10-25% validation, 75-90% training, such as in [9, 11], and these are used throughout Section 3. The developed code allows for an automated generation of as many test cases as required.

LWE had to be implemented in software to provide training teachers and validation targets for the RC. Since the RC input and output code was to be developed within MATLAB, it was a logical choice for this task to also use MATLAB. This was achieved utilising Regev's proposal for an LWE cryptosystem [30], along with a python implementation developed by Buchanan [33]. The final developed code is demonstrated in Figure 44, to make for easy reproduction across similar programming languages or systems.

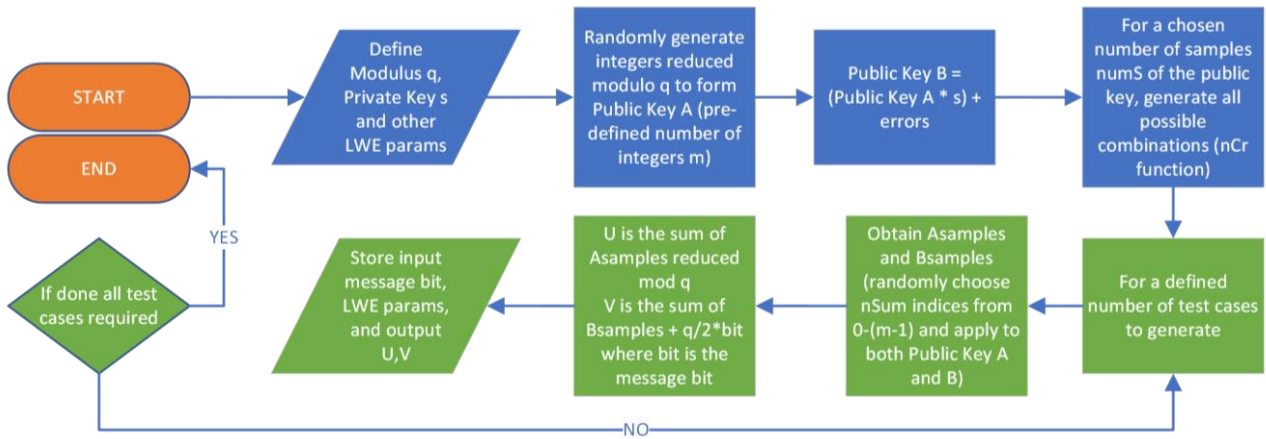

**Figure 44-** Flowchart to demonstrate basic structure of process to generate LWE test cases to act as teachers when training the Reservoir Computer. Green section highlights the for loop that generates as many test cases as required. Blue section is LWE parameters.

As stated in Section 2.2.1c), it was initially believed that buffer size would present an additional limitation. The input buffer shown in Equation (11) has the following size. The message bit $\phi$ will be a fixed size of 1 (as will the dummy bit $D$), $A_{samples}$ and $B_{samples}$ will each have size $N_{samples}$, where $N_{samples} \leq m_{numberOfEquations}$. Samples in this context refers to sampling each public key. A reasonable $m$ was chosen as 20, with $N_{samples}$ as 5. This ensures $20C5 = 15504$ possible test cases for training and validating the Chua-RC. This leads to a message size of $5 + 5 + 1 + 1 = 12$. Further pre-processing increases this size. For example, multiplexing with $N_{masks} = 25$, and applying a sample hold of $\theta = 10$ increases the size to $12 * 25 * 10 = 3000$. Prior to the use of the new MATLAB toolbox, see Section 2.2.1c), there was no way in MATLAB to use a buffer size greater than 4096. Whilst this was acceptable for the case above, a different choice for $\theta$ would easily exceed the limit. Moreover, it prevented the option of future work extending the proof-of-concept single-bit encryption of this project to a multi-bit solution. Although this issue has been fixed, it is still important to not fill the input buffer with unnecessary data. The larger the buffer, the longer the operation time at a fixed frequency. In a real-world solution, operation time may be a key reason to choose Chua-RC, so it should be minimised where possible.

The input buffer in Equation (11) consists of the minimum data that must be provided to the Chua-RC to enable encryption. An alternative approach could have been to send the entire public key, along with the message bit, and a random clocking signal to enable the random sampling of the public key. However, this unnecessarily increases the complexity of the RC's task and requires a much larger message size. This could become desirable for multi-bit systems as it would negate the need to send $A_{samples}$ and $B_{samples}$ for every message bit. After a certain number of message bits, the message size would be smaller with this second approach. However, the focus of this thesis is a proof-of-concept, so this is an area that is reserved for future scaling-up work.

The equations the Chua-RC attempts to learn are seen in Section 2.1.3b). Other than $A_{samples}$, $B_{samples}$, and $\phi$, the modulus $q$ is also required. However, the modulus $q$, along with parameters such as $N_{samples}$ and the private key $s$, should remain fixed across all messages using a particular LWE cryptosystem definition. As such, these parameters are not required for each input buffer. For each differently defined cryptosystem (e.g. different modulus $q$), the Chua-RC can be retrained, generating a weight unique to that cryptosystem.

## ii.  Learning with Errors Sub-System Testing

This sub-system was first tested to generate a single example of encrypting both a 1 and a 0. The error array was fixed to random integers ranging from 0 to 3. Table 9, demonstrates successful, correct, encryption. Values of $v$ are rounded down to integers (e.g., 23.5 to 23, which still results in correct encryption/decryption). The modulus used is $q = 29$. This demonstrates that whilst a modulus $q = 7$ will primarily be used in this project, for the reasons discussed in Section 2.2.2a)i, the MATLAB implementation of LWE allows the user to define the cryptosystem in code. This will support future work to scale-up the project to more complex cryptosystems.

| Modulus ($q$) | Private key ($s$) | $A_{samples}$, $B_{samples}$ | Plaintext bit $\phi$ | ($u, v$) | Decryption |
|---|---|---|---|---|---|
| 29 | 11 | [0,4,20,21,11], [1,15,17,0,5] | 1 | 27, 23 | 1 |
| 29 | 11 | [26, 22, 12, 17, 21], [26, 10, 17, 14, 0] | 0 | 11, 9 | 0 |

**Table 9-** Sub-system testing of LWE Encryption.

Subsequent testing involved the generation of multiple test cases. The modulus used in this example is $q = 7$. Table 10 shows the first two lines of a 1000 test case set of data generated by the MATLAB code. This stores all the parameters used to generate a ($u, v$) pair, enabling the training and validation of the Chua-RC.

| Bit $\phi$ | Decrypt Equation Value | $u$ | $v$ | $A_{samples}$ | $B_{samples}$ | $q$ | $s$ | $M$ | Divisor of M | Public key A | Public key B |
|---|---|---|---|---|---|---|---|---|---|---|---|
| 0 | 3 | 4 | 4 | [4,2,6,0,6] | [2,5,5,0,6] | 7 | 2 | 20 | 4 | 1x20 double | 1x20 double |
| 1 | 6 | 4 | 0 | [4,2,6,0,6] | [2,5,5,0,6] | 7 | 2 | 20 | 4 | 1x20 double | 1x20 double |

**Table 10-** Output from LWE test cases generation.

## b) Input Signal Generation
## i. MATLAB Code Development

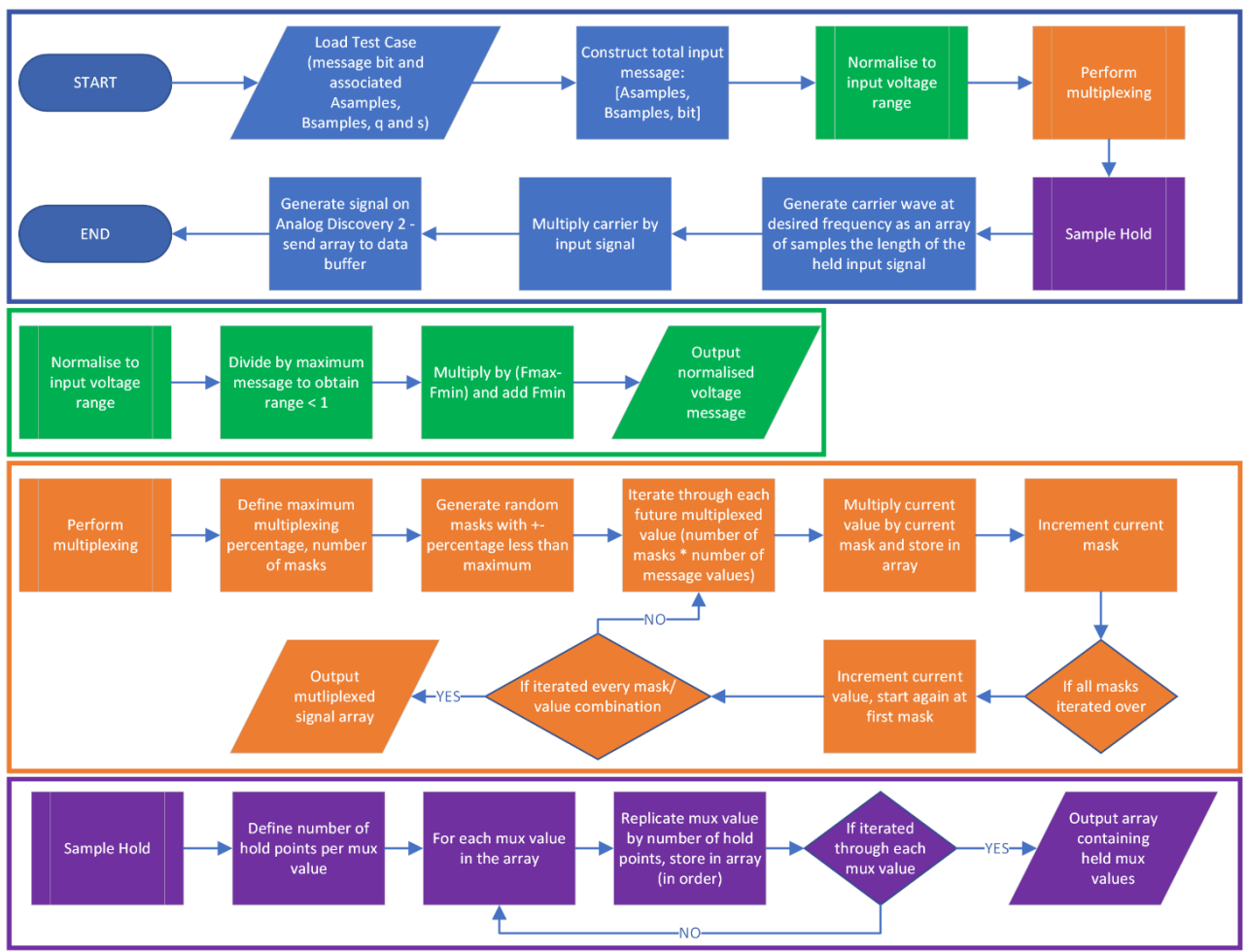

**Figure 45-** Flowchart to demonstrate pre-processing of input and generation using AD2. The chart is split into four sections. The first, in blue, is the main code. This loads a test case, constructs the input, performs pre-processing, and generates the result. The pre-processing consists of the three functions in green, orange, and purple (performing normalisation, multiplexing, and sample hold), followed by AM using a carrier wave of fixed frequency.

To implement the input to the Chua-RC, MATLAB and the AD2 were utilised as discussed in Section 2.1.2a). The construction of the input from LWE parameters is as discussed in Section 2.2.2a)i. The key processes in the MATLAB code are illustrated in Figure 45. This flowchart enables easy reproduction across other programming languages. It should be noted that, in addition to generating the input waveform experimentally (using the AD2 and the physical Chua circuit), it was also practical to adapt the previously developed Simulink model to support an arbitrary waveform input. This enables Chua-RC to be performed in simulation – a simulation model that has been demonstrated in Section 2.2.1c) to be consistent with the actual performance of the physical circuit. As is discussed in Section 3), this allows for more practical automated testing that can be left unmonitored. In contrast, whilst the AD2 can be automated, it would be unwise to leave long tests running unattended on a live circuit, due to health and safety and fire risks.

### ii.   Signal Generation Sub-System Testing

Successful input signal generation is demonstrated in Figure 46. A sine wave carrier is shown, but this can be changed to any carrier, including square waves. The frequency in this demonstration is 100 Hz, but as with the carrier wave, this is customisable, along with other parameters such as $N_{\text{mask}}$ (which is 4 in Figure 46). Figure 46(a) shows the input constructed from LWE parameters. This is processed to create the desired waveform in Figure 46(b). The upper envelope is shown. This waveform is generated on the AD2 in Figure 46(c).

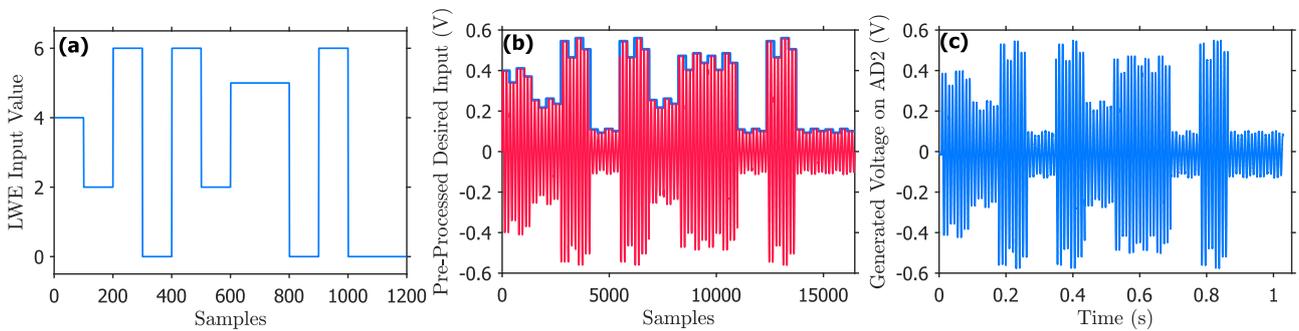

**Figure 46- (a)** Example input $[4,2,6,0,6,2,5,5,0,6,0,0]$ consisting of $\left[A_{\text{samples}}, B_{\text{samples}}, \phi, D\right]$, **(b)** Pre-processed input (normalisation to 0.1 to 0.5 V, multiplexing with four masks, sample hold, 100 Hz sine wave carrier), with upper envelope shown, **(c)** Generated voltage using AD2 (acquired with AD2 oscilloscope).

As discussed in Section 2.2.1c), there was initially an issue in the DAC operation that resulted in the generated signal not achieving the desired waveform. This was solved with the new MATLAB toolbox, and no further issues were encountered. There is a slight discrepancy between the desired amplitude, and the generated amplitude, but this is negligible. Figure 46(c) clearly shows that each multiplexed value is distinctly reproduced.

## c)   Output Signal Processing and Training
### i.   MATLAB Code Development

The key processes in the read-out from the Chua-RC are illustrated in Figure 47, enabling reproduction of the code in alternative programming languages to MATLAB. The flowchart specifies AD2 acquisition, but as with the input code, this is flexible and can be exchanged for interfacing with the Simulink model. Since Simulink does not produce a fixed number of output samples for each simulation, the code is designed to support a variable number of output samples. This means that when using the AD2, it can be operated at different sampling rates, without edits required to the code. The demultiplexing follows the design specified in Section 2.1.2b)ii, and synchronisation to remove initial delay is performed. A threshold of 0.1 V was chosen, as the noise signal received during initial delay was consistently below 0.1 V. The code searches for the first instance of Inductor Voltage greater than 0.1 V and assumes this is the start of the signal. The code iterates through only the desired multiplexed values, removing any data corresponding to the dummy bit. This ensures that data corresponding to the whole message, excluding the dummy bit, is acquired, despite the initial delay occupying buffer space. The training algorithm follows the pseudo-code discussed in Section 2.1.2b)iii, adapted from [10].

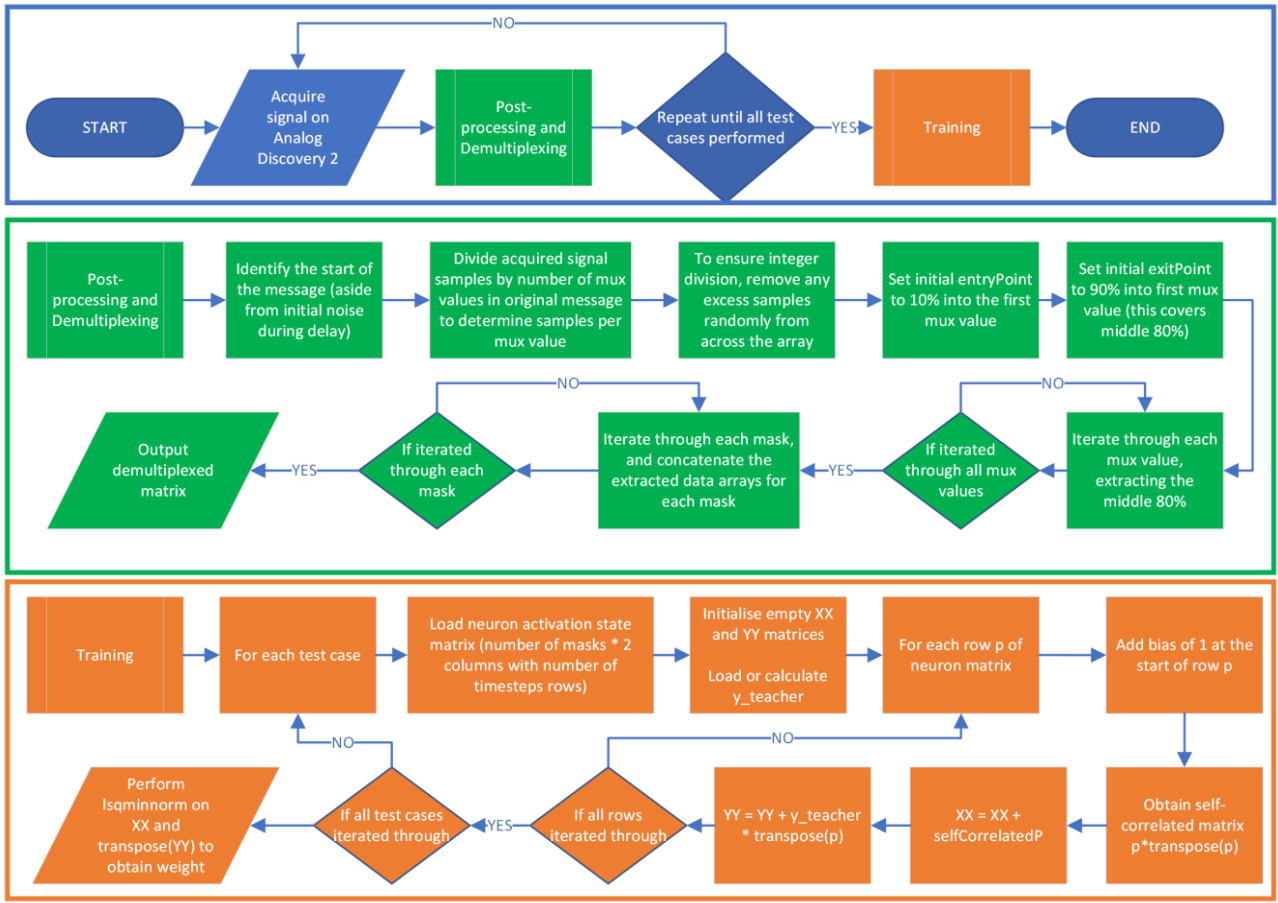

**Figure 47-** Flowchart to demonstrate signal acquisition using AD2, post-processing and demultiplexing, followed by training on the chosen application. The chart is split into three sections. The first, in blue, is the main code, with two sub-systems in green and orange (post-processing/demultiplexing, and training respectively).

## ii. Signal Acquisition and Demultiplexing Sub-System Testing

Successful signal acquisition and demultiplexing are demonstrated in Figure 48. In Figure 48(a-b), both the Inductor Voltage and Chua Diode Voltage are obtained, and any initial delay has been removed. The data corresponding to the dummy bit has also been removed, leaving 11 distinct sections corresponding to the 11 distinct pieces of the input $[A_{samples}, B_{samples}, \phi]$. Figure 48(c-d) shows how each are demultiplexed (according to four masks), into four similar but distinct output signals per physical tap. Each are clearly derived from the output but are a slightly different variant of it. This has multiplied the number of output states by four, which provides higher output dimensions and, in theory, greater Chua-RC performance [1]. This is assessed in Section 3.2.2.

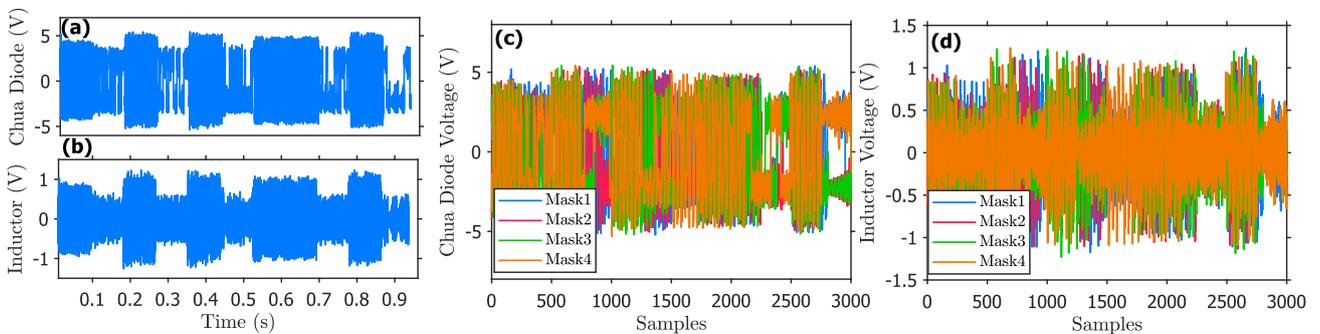

**Figure 48- (a)** Acquired Chua Diode Voltage (with initial delay removed), **(b)** Acquired Inductor Voltage (with initial delay removed), **(c)** Demultiplexed Chua Diode Voltages (four masks), **(d)** Demultiplexed Inductor Voltages (four masks).

### d) Further Improvements, Adaptations, and Stretch Deliverables
#### i. Real-time System

With the individual sub-systems developed, and the AD2 configured to work completely programmatically, it was possible to develop code that would utilise the system as a real-time solution. Assuming a training weight had been pre-determined and could be supplied to the code, the input signal generation, reservoir output acquisition, and subsequent transformation into an estimate, using the weight, was practical. This could utilise either the Simulink simulation as the reservoir, or the experimental circuit combined with the AD2.

#### ii. Multi-bit Messages

This thesis has focussed on single-bit messages. Should single-bit encryption be achieved, it would be possible, and desirable, to extend this to multi-bit messages. A typical AES private key is 128, 192, or 256-bits [29]. The private key of a symmetric cryptosystem (e.g., AES) is a typical message encrypted using PKE techniques. It is important that each bit, whilst being encrypted with the same LWE public and private keys, utilises different random samples of the public key, as discussed in Section 2.1.3b). Therefore, multi-bit messages will require a full input buffer per each bit, consisting of $[A_{\text{samples}}, B_{\text{samples}}, \phi, D]$. This is theoretically achievable with the current system, if single-bit encryption is successful, however encryption speed would be a concern. To scale this in future development, a high frequency Chua circuit could be used, discussed in Section 4.1.

#### iii. Decryption

The stretch deliverable **D6** is to utilise the system to perform decryption. Assuming that there would be a differently trained weight for each modulus $q$ and private key $s$, the current system can be adapted to generate an input buffer as follows: $[(u, v), D]$. The system only requires $u$ and $v$ to decrypt, and a dummy bit is sent to avoid data loss. To make this work for any system regardless of $q$ or $s$, $q$ and $s$ would also need to be included in the input signal. The target teachers for training would be 0 or 1, the correctly decrypted bit. Whilst this is feasible, it is contingent on the Chua-RC's performance. This is tested in Section 3.5.

#### iv. Reservoir Computer Benchmarks

It is also possible to adapt the system to support a range of benchmarking applications, including regression and classification. For example, polynomial regression can be performed on input values within a specific range. These values can be normalised to the desired voltage range and pre-processed as normal. At the output, the teachers can be calculated using the correct polynomial within MATLAB. On a single dataset of inputs transformed by the Chua circuit, the polynomial teacher function can be exchanged for any number of different equations, allowing for benchmarking on a series of regression tasks of different complexity. In the case of classification, classes can be assigned according to $x$ and $y$ coordinates within a specific range. The coordinates, after normalisation and pre-processing, can be independently transformed by the Chua circuit, with outputs combined to create a final dataset of Chua-RC outputs related to the target classes. These tasks are performed in Section 3.2.

# 3) Final System Testing and Specification Validation

Sections 2.2.1 and 2.2.2 respectively demonstrate successful completion of both the reservoir kernel (the Chua circuit), and the required I/O MATLAB code. Section 2.2.2 demonstrates a successful implementation of an LWE cryptosystem in MATLAB and applies it to generating test cases for both training and validation. Final testing consists of several steps to validate the overall system, assess performance, and apply it to LWE. Initially the Chua-RC is tested using the Simulink model. Firstly, it is benchmarked on classification and regression tasks, as discussed in Section 2.2.2d)iv. This demonstrates that the system can successfully learn to solve tasks in line with the current standard achieved by Jensen [9]. Parameter tuning is performed to sweep different operating regions for optimal performance, assessing $R_{\text{variable}}$ and input voltage range (a further work call in Jensen's paper). The amount of multiplexing used is also assessed. The capacitance, however, is fixed, as it is a difficult value to vary, and produces a very similar bifurcation diagram to changing resistance.

The testing continues by extending Jensen's work through the original application of the Chua-RC to more challenging problems, including modulo and time-varying problems. These form the "sub-problems" required for LWE, so indicate the areas where the Chua-RC performance is limited. This is then followed by original application to both LWE encryption and decryption, including parameter tuning for the encryption problem. The results demonstrate that at all tested regions, LWE is too high-dimensional and complex for the Chua-RC.

Further testing is conducted to investigate additional parameters such as the carrier wave. With a full picture of the capabilities of the Chua-RC determined, physical testing is performed on the implemented circuit Chua-RC, with the AD2 for signal generation/acquisition, using the classification and regression benchmarks. The effect of varying thermal noise on the circuit is characterised by plotting the Chua diode V-I characteristic at multiple readings across different days.

Each test is performed using large datasets consisting of training and validation cases. This ensures repeatability. The NMSE is obtained across many different validation cases and is averaged to give an accurate estimate of performance. Quantitative statistical analysis is performed through a variety of different figure types, including histograms to assess the distribution of the NMSE across test cases, and a confusion matrix to easily demonstrate classification performance. The different figure types are used as appropriate, with justification.

## 3.1  Testing Set-up

Initial testing is performed using the Simulink model rather than the physical implemented circuit. This is for two reasons. The first is good practice. It is important to first get the system working in simulation. That way, if it does not work in physical testing, it will be known that it is not caused by a flaw in the system design. Secondly, it is more practical to run simulations as a computer can be left running, whereas it would not be safe to leave the physical circuit running unattended; this would be a clear health and safety risk as it is only a prototype device.

**Table 11** - Testing set-up Chua-RC-1.

| Resistance $R_{variable}$ | Capacitance $C_1$ | Frequency $f$ | $N_{masks}$ | Carrier Wave | Voltage Range |
|---|---|---|---|---|---|
| 1.92 kΩ | 10 nF | 100 MHz | 50 | Square | 0.4 V ≤ $V_{in}$ ≤ 1 V |

The circuit used is a Kennedy-based Chua circuit, as in Sections 2.1.1 and 2.2.1. The testing set-up is summarised in Table 11. It is set with a resistance of 1.92 kΩ, and a capacitance of 10 nF. These were chosen due to the bifurcation performances previously demonstrated in simulation and on the physical circuit – they provide a point just on the edge of chaos. Whilst not true in general, it is common in RC for optimal performance to arise at the edge of chaos [25]. Using voltage frequency $f = \omega/2\pi RC$ and $\omega = 0.7$ [9], input voltage frequency can be calculated as $f \approx 5.8$ kHz. The carrier wave used is square, and the Chua output is recorded for 5 periods – a total execution time per test of 0.86 ms. Multiplexing of 50 masks with a maximum 1% deviation is applied. The assumption was made, that the more masks, the better the performance [1]. This is validated in Section 3.2.2. With the use of two physical taps, 100 output channels are obtained. This is simulated in Simulink with a maximum timestep of 10 ns – a sampling frequency of 100 MHz. The voltage range in the tests below is 0.4 to 1 V. This testing set-up can be referred to as Chua-RC-1. Further work is undertaken in Section 3.2.2 to sweep different voltage ranges, and resistances to confirm the best performance region. NMSE is used as the main performance metric – in-line with other RC papers [2, 10, 11]. However, it should be noted that Jensen utilised NRMSE so 1:1 comparisons of exact performance figures between this paper and Jensen cannot be made. Additionally, the datasets used in this paper are not identical to those used by Jensen. The alterations made for the physical circuit testing are discussed in Section 3.7.

## 3.2  Reservoir Computer Benchmark

### 3.2.1  Classification and Regression Tasks

Regression and classification tasks can be used to validate the Chua-RC design. Jensen utilised a non-linear polynomial for regression and a concentric circles dataset for classification [9]. This paper uses similar datasets to demonstrate the performance. Figure 49 shows the successful classification of two concentric rings. Figure 49(a) shows the Chua-RC trained for approximately 5750 datapoints. These datapoints either exist within class 0 or class 1. A further approximately 1150 datapoints were tested as validation data, classified by the Chua-RC, and are displayed on the graph. An 80% training, 20% validation model was utilised. This can be called the Chua-RC-1 classification dataset. This resulted in 100% classification accuracy, and a 0.0075 mean average NMSE for the validation data. For the purposes of avoiding infinite NMSE when the goal was class 0, every datapoint was trained for $classValue + 1$, i.e. class 0 datapoints were trained for 1, and class 1 datapoints were trained for 2. Figure 49(b) shows that the NMSE never exceeds 0.15. No misclassifications occur, and no datapoints are generated with greater than 15% error from the desired class. The use of the histogram allows for further statistical analysis of the NMSE. Almost 100% of validation cases have an NMSE less than 0.05, and approximately 60% have an NMSE less than 0.005 – <u>this is very good performance.</u>

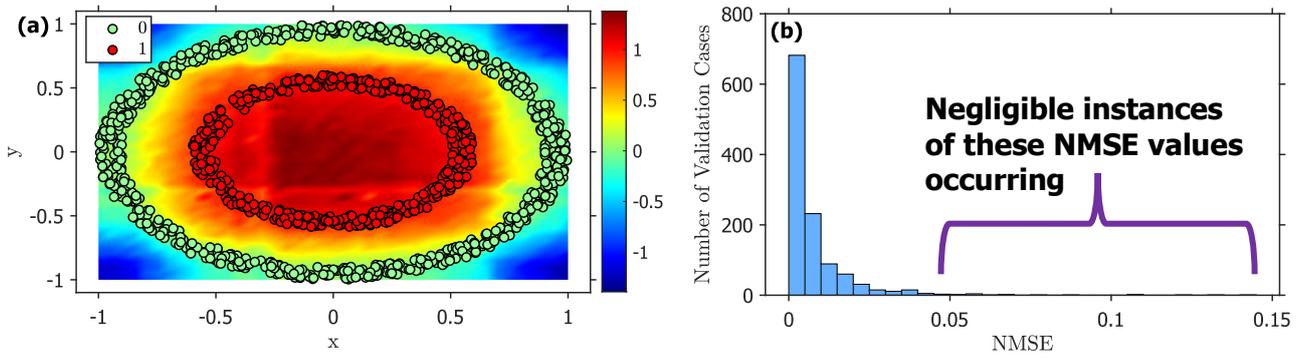

**Figure 49- (a)** Concentric circles classification and decision surface, **(b)** Histogram showing number of validation cases resulting in NMSE within defined ranges.

Further validation was completed by sweeping $x$ and $y$ with an additional 1600 datapoints to generate the decision surface. This is demonstrated by the heatmap in Figure 49(a). There is clear distinction between the central red zone for class 1, and the outer layer of green/cyan for class 0. It is also able to differentiate further, identifying the corners as distinct from class 0. In the program, with just two classes, the Chua-RC read-out would designate a $-1$ as a 0 as it is closer to 0 than to 1. This shows the capacity, however, for more classes.

This demonstrates that a Kennedy-based Chua circuit, with differing Chua diode implementation to the Jensen circuit, different $R_{\text{variable}}$, and differing physical outputs, can achieve high performance at classification tasks. The Chua-RC is a single hidden-layer neural network (with the Chua reservoir being the hidden-layer). Normally, a multi-layer neural network would be required to correctly classify concentric circles since they are non-linearly separable, however the higher dimensions of the non-linear Chua kernel negate this requirement [16]. The next benchmark assessed is non-linear polynomial regression. The chosen polynomial can be seen in Equation (12). It is similar to the polynomial used by Jensen but offers a greater range of $y$ outputs.

$$y = x(x-4)(x-3)(x-2)(x-1)(x+1)(x+2)(x+3)(x+10) \tag{12}$$

Figure 50(a) shows both the target polynomial, and the estimate for the validation dataset. Approximately 2900 test cases were used, with an 80% training, 20% validation split. This is the Chua-RC-1 regression dataset. Figure 50(b) shows the NMSE for each $x$ value rather than as a histogram. In this case, this is a superior method for showing where the large sources of error occur.

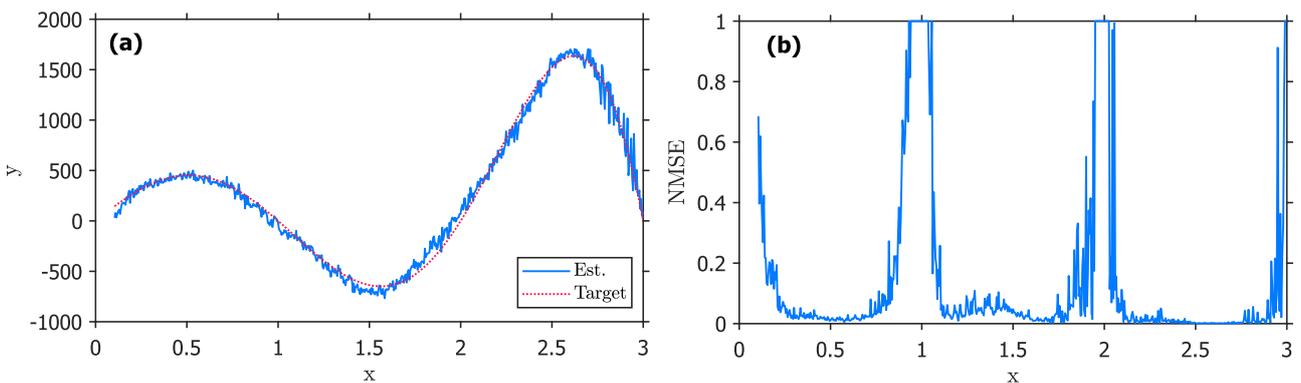

**Figure 50- (a)** Non-linear polynomial regression Equation (12), **(b)** NMSE of validation dataset.

Maximum NMSE is again restricted to 1. This prevents obtaining infinite NMSE at zero-crossing points due to divide-by-zero error. The mean average NMSE of the validation dataset is 0.1346. From Figure 50(a), it is clear that <u>high performance is achieved.</u> Figure 50(b) shows more clearly that the four areas of performance degradation are at points where $x$ results in $y = 0$. This is a result of even a small deviation from a small target leading to a very large NMSE. For example, achieving 0.02 instead of 0.01 provides the following NMSE:

$$\text{NMSE} = \frac{(0.02 - 0.01)^2}{0.01^2} = 1 \tag{13}$$

This is the maximum possible NMSE and shows that the difference between the obtained value and the target value is at least equal to the target value. However, for a polynomial that has $y$ values ranging from approximately $-500$ to $+1700$ for $0.1 \leq x \leq 3$, obtaining 0.02 for a target of 0.01 is an extremely good result. To exclude these outliers, the median NMSE can be taken, rather than the mean. Doing so provides a median average NMSE of 0.0266 across the validation dataset. Visually, the result has greater error than the mostly exact curve achieved in simulation on a similar regression task performed in [9]. However, the simulation performed by Jensen is purely normalised equations, shown in Equation (2), whereas the simulation performed here uses Simulink to accurately model circuit components, so this difference is expected.

The two benchmarks demonstrate that a Kennedy-based Chua-RC can achieve high performance on typical machine learning tasks. These benchmarks have met expectations and as such have achieved a PASS. They demonstrate that a successful RC has been developed, achieving **D3**.

### 3.2.2  Parameter Tuning for Reservoir Computer Optimisation

Jensen did not assess voltage range and called for investigation of this parameter [9]. It was a logical choice to do so with the new Kennedy-based Chua-RC. Additionally, Jensen did not assess resistance's effect on bifurcation, whereas this paper has characterised the resistance bifurcation diagram for $R_{\text{variable}}$. It is logical to include resistance as a second tuneable parameter. The testing set-up was as follows. With a total range of 0.6 V fixed, the central voltage was varied from 0.4 V to 1.2 V. The resistance was varied, in intervals of 80 $\Omega$, from 1.6 k$\Omega$ to 2 k$\Omega$, which includes the previously assessed 1.92 k$\Omega$. This was applied to the same polynomial regression task. The NMSE results are provided in the heatmap in Figure 51(a).

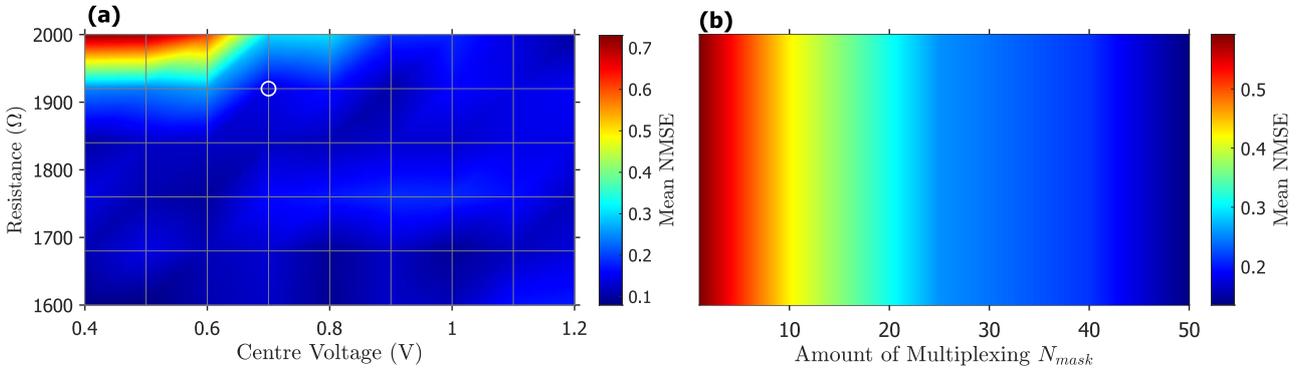

**Figure 51**- **(a)** NMSE for varying resistance and the central voltage of the input amplitude range, **(b)** NMSE for increasing $N_{\text{mask}}$ (the amount of multiplexing).

Figure 51(a) shows there is a large region of high performance. Performance degradation occurs only when the resistance is set too high for the corresponding voltage amplitude. The regions of worse performance are where the input voltage is too low to perturb the system into the chaotic region. This is logical – whilst it is common that the optimal performance in RC occurs on the edge of chaos [25], the perturbation across the range of input voltage must be sufficient to enter the chaotic region and not remain above it. Interestingly, this demonstrates that a position on the edge of chaos is not necessary for superior performance in this application. This agrees with Carroll that whilst for some RC the concept is true, in general RC does not require edge of chaos for optimal performance [25]. The primary region chosen for testing, Chua-RC-1, is marked with a white circle (1.92 k$\Omega$ and $0.4 \text{ V} \leq V_{in} \leq 1 \text{ V}$).

Figure 51(b) validates the assumption that greater multiplexing leads to higher performance. The system was tested at $N_{\text{mask}} = 1, 10, 25, 40, 50$, and the mean average NMSE results were interpolated using MATLAB to produce the heatmap. Of the tested values, 50 achieved the lowest NMSE and is therefore used throughout the testing. However, to reduce computational load incurred by greater multiplexing, $N_{\text{mask}}$ could be reduced to 25 without a significant reduction in performance.

## 3.3  More Challenging Tasks

### 3.3.1  Modulo and Polynomials

LWE cryptography is highly dependent on the use of modulo, a highly non-linear function. For the Chua-RC to be able to compute LWE encryption or decryption, it must be capable of computing modulo. The Chua-RC-1 testing set-up is used. The target function used by the MATLAB training code is customisable. Where previously this function was defined as Equation (12), it can be changed to any other equation and the training re-ran. It was chosen to assess $mod(x, 1.3)$. Since $0.1 \leq x \leq 3$, performing $mod(x, 1.3)$ will result in two "wrap-arounds" of the modular arithmetic.

Figure 52(a) shows the target overlaid with validation data estimate. Since the Chua-RC-1 regression dataset is used, this is again an 80% training, 20% validation split – with approximately 2900 datapoints. The mean average NMSE is 0.168, and the median NMSE is 0.0206. Partly this low NMSE is due to the problem involving large sections of linear behaviour $y = x$, where the Chua-RC naturally performs extremely well. However, the good performance extends beyond this. The estimate clearly tracks the target function and performs both non-linear modulo "wraparounds". This is a clear indication that the Chua-RC can perform regression on highly non-linear functions. This achieves Jensen's call for assessing more challenging problems [9].

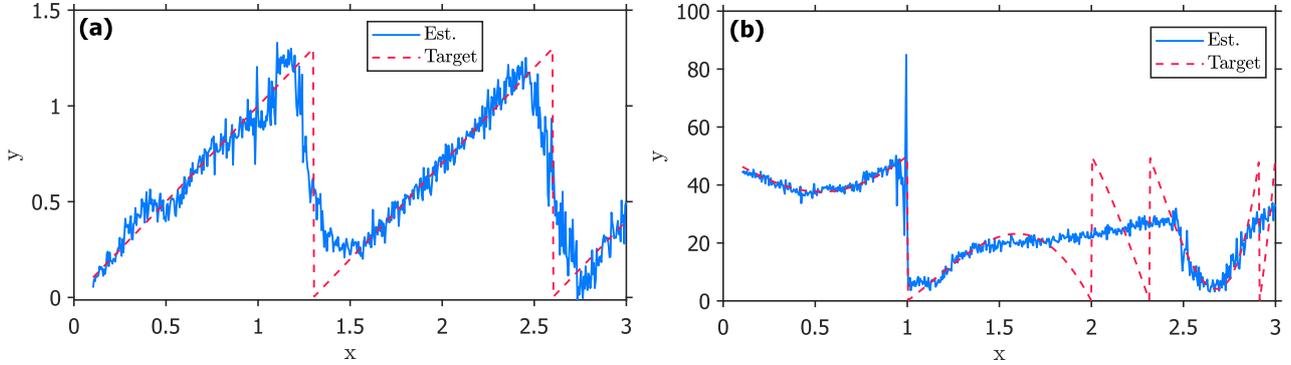

**Figure 52- (a)** Modulo regression $mod(x, 1.3)$, **(b)** Modulo and polynomial regression $mod(f(x), 50)$.

The next task to assess was combining modulo regression with polynomial regression. LWE involves performing multiplication and arithmetic on input values, and then reducing the result $mod\ q$. Some forms of LWE, such as Ring-LWE, utilise polynomials as well as modulo [30]. Therefore, performance of $mod(f(x), 50)$ was assessed, where $f(x)$ is Equation (12). Figure 52(b) shows the validation data estimate overlaid onto the target function – it achieves a mean average NMSE of 0.17. The target function is a mixture of "smooth" and "spiky" areas. "Smooth" areas are those where neighbouring $x$ values produce more similar $y$ values and "spiky" areas are the opposite. In other words, "spiky" areas demonstrate higher non-linearity. The Chua-RC performs very well on "smooth" areas such as $0 < x < 1$. It also performs well at the first "spike" at $x = 1$, and at the areas $1 < x < 1.5$ and $2.5 \leq x < 2.8$. However, the performance is poor at spikes such as $x = 2$. In these areas, the Chua-RC is unable to correctly discriminate between datapoints; the function is too complex and non-linear for it to learn. Instead, it tends to an average result. This shows that whilst a Chua-RC has good performance too on both modulo functions and on non-linear polynomials, combining the two can result in problems too challenging for the RC to learn. It is likely that the non-linear kernel of the Chua-RC does not sufficiently increase the number of dimensions to account for the difficulty of this problem, and that consequently, there is significant overlap even after inputs are transformed by the kernel, preventing discrimination from occurring. The Chua-RC achieves a PASS for performing modulo and a PASS for performing modulo with a non-linear polynomial – but only up to a limit in the degree of non-linearity. This indicates that LWE cryptography may be too challenging for Chua-RC.

### 3.3.2 Time-varying Signals
Jensen called for assessing time-varying inputs to a Chua-RC [9]. So far, all benchmarks have been achieved with single input data (albeit with each piece being time-multiplexed). With classification, where there are $x$ and $y$ inputs for each datapoint, the $x$ and $y$ values were computed separately and then combined into a larger output matrix in post-processing. For LWE, it would not be practical to apply each input required for encryption of a single bit separately to the Chua kernel. As discussed in Section 2.2.2a)i, this paper will assess LWE with 5 samples of each public key. Along with the message and dummy bits, this results in an input message of $[A_{\text{samples}}, B_{\text{samples}}, \phi, D]$, with a size of 12. Even for this small size, it would be impractical to perform 12 different simulations per message bit. With increased $N_{samples}$ it would only become more impractical. Instead, the message should be sent as a single time-varying input to the RC. Several benchmarks are performed in Figure 53 to test time-varying inputs. A dataset of approximately 1600 test cases was used, with 80% for training and 20% for validation. The inputs $x_1$ and $x_2$ are varied from 0 to 40 and combined in each possible way: $41 \cdot 41 = 1681$ different combinations (this results in some repeats depending on whether the order of $x_1$ and $x_2$ matters – which depends on the target function). The Chua-RC-1 testing set-up is used.

Figure 53(a) shows a linear target function $y = x_1 + x_2$ and the estimate from the validation dataset. The otherwise simple process is complicated by the fact that $x_1$ and $x_2$ are sent as one time-varying signal. This achieves a mean average NMSE of $0.017$ and Figure 53(a) shows that the estimate tracks the target function.

A more complex task is seen in Figure 53(b): $y = x_1 * x_2$. The mean average NMSE for this task is much greater at $0.3219$, and this can be seen visually as the validation estimate is very noisy. However, it still achieves the general trend of the curve, and does so more than fitting just straight line. It is clear, however, that this is a more challenging task for the Chua-RC.

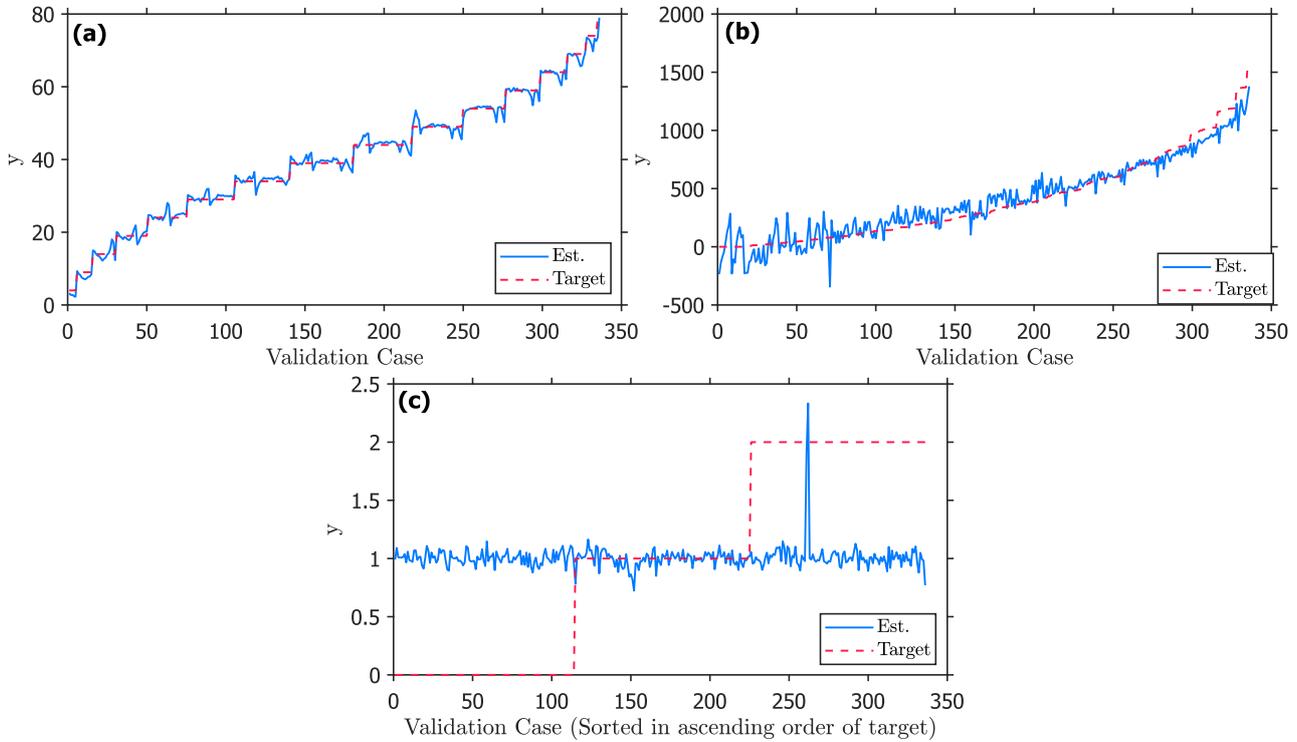

**Figure 53- (a)** Input of time-varying signal $[x_1, x_2]$ regression to $x_1 + x_2$, **(b)** Regression to $x_1 \cdot x_2$, **(c)** Regression to $mod(2x_2 - x_1, 3)$. Validation cases are sorted in each plot by ascending order of target.

The final benchmark can be seen in Figure 53(c). The target function utilised is $y = mod(2x_2 - x_1, 3)$. This task is very challenging as it involves a time-varying input where the order of the values in the input matters, and the result is reduced $mod\ 3$. It is acceptable to benchmark with a fixed modulo parameter as for a defined cryptosystem, the modulo value $q$ will also be fixed. Whilst it purports to have a mean average NMSE of $0.429$, visual inspection of Figure 53(c) clearly shows that the problem is too challenging for the Chua-RC to compute. In Figure 53, the plots are ordered in terms of increasing target value. In the case that Figure 53(c) was ordered in the original order of the dataset, it would appear very non-linear, similar to Figure 54 in Section 3.4. Figure 53(c) shows, similarly to Figure 52(b), that when the Chua-RC is incapable of discriminating between datapoints, it instead averages them. The use of $mod(x, 3)$ and integer data means there are 3 possible results: $0, 1, 2$. Unable to ascertain a pattern that results in a particular datapoint being assigned to a value, the Chua-RC averages to the middle and every datapoint, aside for a single anomalous spike, is assigned to approximately 1. Due to the result averaging, the NMSE implies a better performance than is achieved.

The Chua-RC achieves a PASS for performing operations on time-varying signals, however its performance is limited by the non-linearity of the task. Performing a task involving modulo, another function, and time-varying signals is beyond the Chua-RC capability, further indicating that LWE too will be beyond the ability of the Chua-RC. It should be noted that this is not a FAIL, as the work aimed to investigate the capability of a Chua-RC system – which it has achieved. Developing the system further to increase complexity and allow for solving these more challenging problems is beyond the project scope, discussed in Section 4.1.3.

## 3.4 Learning with Errors Encryption

### 3.4.1 System Testing

Since LWE encryption is expected to be too challenging for the Chua-RC, it is first performed with a single case – used for both training and validation. This is used to demonstrate that, in theory, the RC can encrypt an input message to a $(u, v)$ pair. To obtain two output values means increasing the teacher from one value to two, and thereby generating a weight with two rows. The Chua-RC-1 testing set-up is used.

Training for the case [6,3,1,3,0,5,6,2,6,1,1], which produces a target $(u, v) = (6,2)$, and then validating provides the estimate [6.00000000100844, 1.999999999998365], which is an exact match to any reasonable number of significant figures. With this demonstrated, a larger dataset can be used. Figure 54 shows LWE encryption performed with a dataset of 2000 test cases. A 90% training, 10% validation split is used. This is higher than normal to maximise the opportunity for the Chua-RC to learn the LWE method.

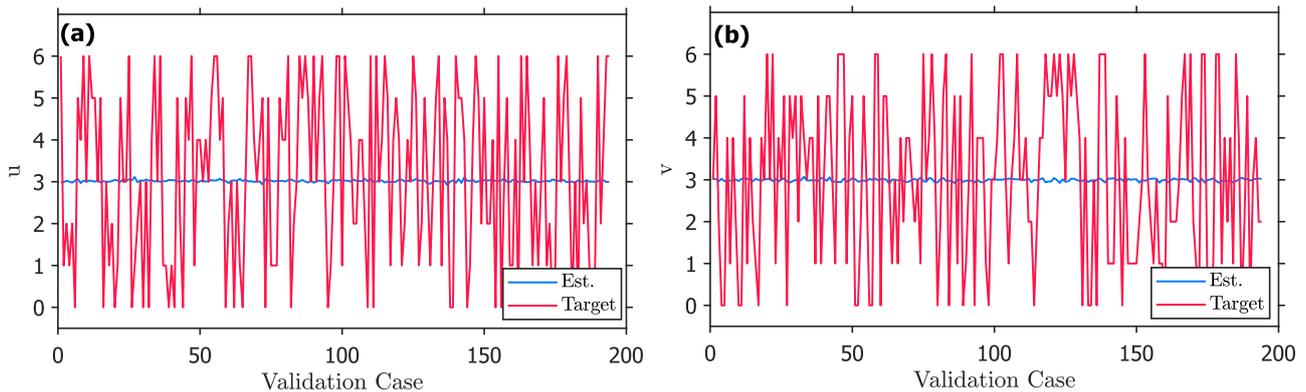

**Figure 54-** Estimate vs target for LWE encryption **(a)** $u$, **(b)** $v$.

Figure 54(a) shows the target and estimate for $u$ and Figure 54(b) shows the target and estimate for $v$. The red target functions are clearly highly non-linear. Whilst a mean average NMSE of 0.2651 is obtained, the visual inspection shows that the problem, as suspected from the prior investigations and benchmarks, is too difficult for the Chua-RC. As with Figure 53(c), the integer data and use of modulo restricts the number of outputs. In this case, the outputs are restricted to 0,1,2,3,4,5,6. The Chua-RC performs the same averaging seen before. Unable to discriminate, every test case is forced to become approximately $(3,3)$, leading to a deceptive NMSE.

### 3.4.2 Parameter Tuning for Reservoir Computer Optimisation

To verify that the problem is too challenging for Chua-RC, and that it is not due to using a sub-optimal region, parameter tuning is performed for again for LWE. parameters were chosen for investigation – normalised voltage range, and resistance $R_{\text{variable}}$. As previously, capacitance was fixed at 10 nF. For a modulus $q = 7$, the potential input values are 0,1,2, … ,6. The AD2 accuracy is 10 mV ± 0.5% at ≤ 1 V [13], so 0.3 V is chosen as a an acceptable voltage range as all seven values can be encoded without overlap.

14 different potential centre voltages were considered (ranging from 0.2 V to 0.85 V) and 10 different resistances ranging from 1700 Ω to 2100 Ω. This meant 140 regions were tested. 500 test cases were used, and validation and training were performed on the same test cases. This was done to give the best chance of successful discrimination. Additionally, two different types of training were performed – one with a bias and offset applied, and another with only an offset. The second was performed as an attempt to prevent the $(3,3)$ averaging by increasing the variance of the estimate.

Figure 55(a) shows the NMSE at each region for the attempt without a bias applied to the training. Figure 55(b) shows the conventional training method with a bias applied. Both attempts, at every region of the tuning parameters, resulted in the Chua-RC failing to discriminate and instead averaging. In the case of Figure 55(a), this averaging is less precise and does result in greater output variance – leading to the greater range in NMSE results – but the system is still unsuccessful.

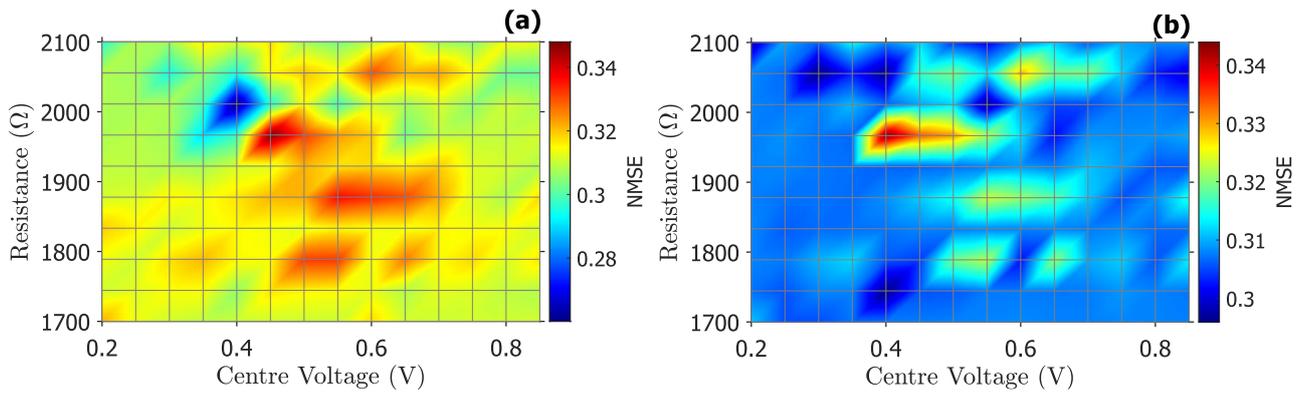

**Figure 55-** NMSE for tuning voltage range and resistance - LWE Encryption **(a)** No bias but offset applied, **(b)** Conventional training system (bias and offset).

Ultimately, Figure 55 shows that the LWE encryption problem is too challenging for the Chua-RC, even after different tuning regions are swept to optimise the system. This is likely due to two main reasons:

- The number of different potential test cases for LWE encryption for $q = 7$ and $N_{samples} = 5$ is extremely high. This is because each test case consists of 11 values $[A_{samples}, B_{samples}, \phi]$ – the dummy bit is irrelevant. The contents of $A_{samples}$ and $B_{samples}$ are from 1 to 6 and $\phi = 0$ or 1. This leads to $7^{10} \cdot 2$ potential test cases. Even accounting for repeats the number of test cases is far higher than a reasonable number that could be trained. This means the Chua-RC performs significant interpolation.

- The LWE problem is highly non-linear and challenging (utilising both modulo and time-varying inputs where order matters). As such it is difficult for the Chua-RC to interpolate and learn the result for an unseen input. The Chua-RC, even when multiplexed, does not sufficiently increase the number of dimensions to account for this non-linearity.

As in Section 3.3.2, this is considered a PASS. This is because it was unknown if it was possible to perform LWE encryption using a Chua-RC. This project demonstrates that the current state-of-the-art of Chua-RC does not have the capability, and therefore further development is required. This achieves **D4** – applying the Chua-RC to LWE encryption.

## 3.5  Learning with Errors Decryption

### 3.5.1  System Testing

This thesis also attempted LWE decryption. As with LWE encryption, this can be demonstrated first for a single case, and then for multiple cases. Where training and validation occurs on the same single case, the decryption is performed correctly. For example, for the case $(u, v) = (0,3)$, the correct decrypted bit of 0 is obtained. For the case $(0,4)$, the correct decrypted bit of 1 is obtained.

Figure 56 demonstrates the performance when trained with approximately 1900 test cases and validated with approximately 500 – an 80% training, 20% validation split. With a modulus $q = 7$, there are only 7P2 = 42 distinct cases. Therefore, there are many repeats to artificially increase the training and validation dataset. This means that the Chua-RC is given the opportunity to perform training on a full dataset of possible test cases. LWE decryption is a time-varying input problem involving arithmetic, multiplication, and modulo. This is very similar to the benchmark in Figure 53(c). The decryption equation is as follows:

$$decrypted = mod(v - u \cdot s, q) \qquad (14)$$

In this case, where $q = 7$, and $s = 2$, this is $mod(v - 2u, 7)$. It is also complicated further by the $if/else$ check: if the result is greater than $q/2$, the decrypted bit is a 1, otherwise it is a 0.

Figure 56 shows that the problem is too challenging for the Chua-RC, despite the full dataset being available. Whilst the accuracy is around 72%, this cannot be considered correct performance. Figure 56(a) shows the confusion matrix for the Chua-RC attempting to perform the whole decryption process, and Figure 56(b) shows the Chua-RC attempting to perform just Equation (14). Figure 56(b) shows that, unlike with Chua-RC attempting to perform LWE encryption, the RC is capable of discriminating. In other words, it does not exclusively average to (3,3), though it does do so for most of the validation test cases. This is interesting as it demonstrates that the Chua-RC is better at performing decryption than encryption. This is expected as decryption is less complex, and the full dataset is available. For many test cases it cannot discriminate and instead averages, but for some test cases it discriminates correctly. For others, it attempts to discriminate but misclassifies. Overall, the problem is easier for Chua-RC than encryption, but still too challenging. This achieves stretch **D6** – applying the Chua-RC to LWE decryption. It is logical that the Chua-RC struggles to perform these operations. They are highly non-linear by nature in order to make good cryptosystems, so it is inherently difficult to interpolate the correct output from a provided input.

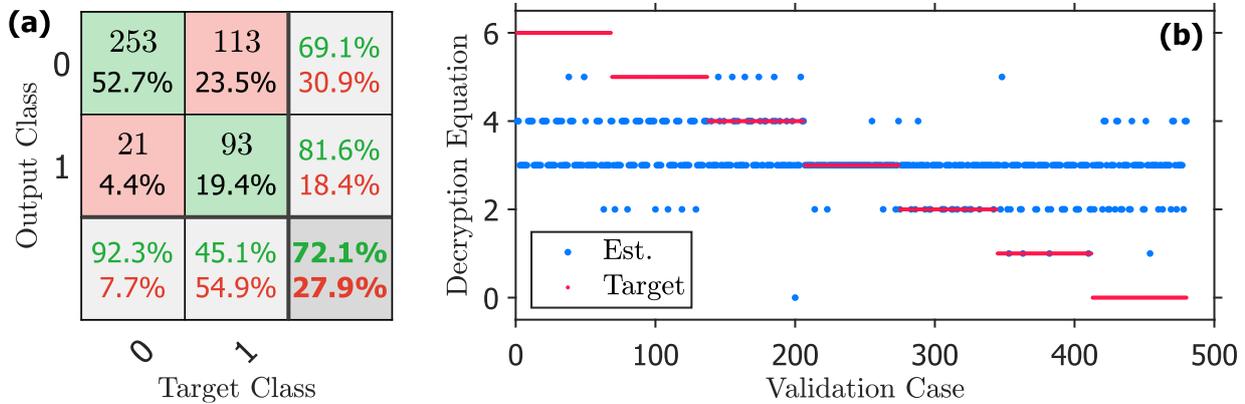

**Figure 56-** Reservoir computer estimate vs target for LWE Decryption **(a)** Confusion matrix of decryption, **(b)** Result of decryption equation.

## 3.6  Additional Testing

Several additional areas were briefly investigated. Square waves were used as the carrier throughout the prior testing, but this project also performed initial testing using sine waves and using a carrier of $V_{DC} = 1$ V, equivalent to no carrier. To assess these, the polynomial regression task from Section 3.2 was used.

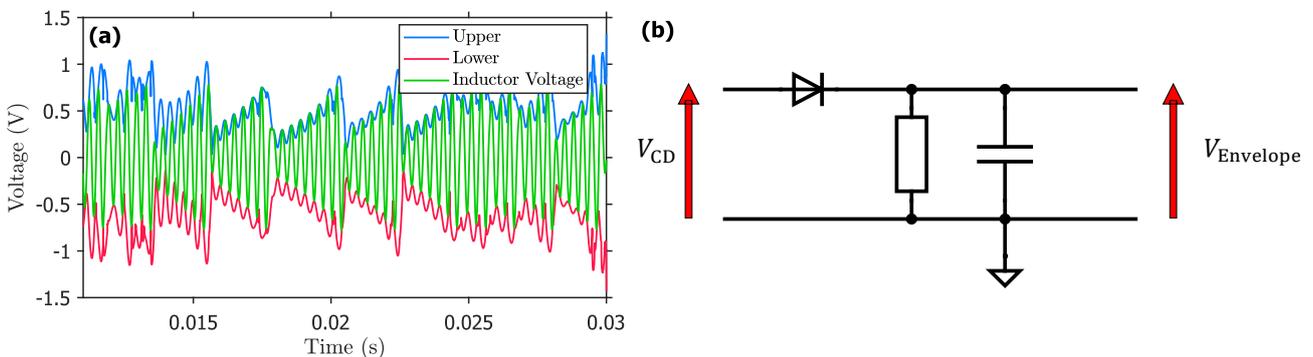

**Figure 57- (a)** Envelope function performed on Inductor Voltage (upper and lower envelopes shown), **(b)** Envelope detection (demodulation) circuit that can be implemented on the output of the Chua circuit, shown for $V_{CD}$ (Chua Diode Voltage).

Another area investigated was using the envelope of the Chua-RC output. In theory, taking the envelope output may reduce overlap between different outputs associated with distinct inputs. It would also reduce the amount of data involved in the training stage. This is beneficial for reducing computational load. Figure 57(a) shows how the upper and lower envelopes of the output Inductor Voltage can be plotted. This has been achieved using MATLAB, but Figure 57(b) shows how this can be implemented in hardware through a demodulation circuit, reducing the computational load on the post-processing. This was tested with all three carrier types and the results are summarised in Table 12.

**Table 12**- Mean average NMSE for each carrier type using normal output and envelope output. Tested using polynomial regression task.

|  | **Square Wave** | **Sine Wave** | **No Carrier** |
| --- | --- | --- | --- |
| **Normal Output** | 0.1346 | 0.1875 | 0.1297 |
| **Envelope** | 0.1307 | 0.1849 | 0.1268 |

There is a small but notable reduction in performance when changing from square to sine waves. This is likely because when square waves are used, the modulated wave values are equal to the magnitude of the modulation amplitude. With sine waves, most of the voltage in the modulated wave is at values less than the desired amplitude, and there will be more similarity between different input waveforms. With a carrier of $V_{DC} = 1$ V, it was found that performance improved slightly. This is likely due to it being the most exact representation of the desired amplitude. This suggests, aside from practical signal generation reasons, that there is no advantage to using AM and a signal carrier. Using an envelope at the output demonstrated only marginally better performance. However, as will be discussed in Section 3.7, it was effectively utilised in the physical circuit.

Another avenue of investigation was adding memory to the Chua circuit. This was only very briefly considered, as it does not directly relate to the PQC application or the previously tested benchmarks. The Kennedy Chua circuit does not have memory and common RC applications require non-linear memory, such as speech-recognition [1, 9]. It is thought that memory could be introduced to the circuit through feedback, requiring an attenuating gain and a time-delay, shown in Figure 58(a). This can be adapted into hardware components through an RC (resistor and capacitor – not reservoir computer) delay circuit (a low-pass filter) and an amplifier, shown in Figure 58(b). This could be benchmarked using Nonlinear Autoregressive Moving-average (**NARMA**) as in [11]. However, this is beyond the current project's scope, and its left for further work.

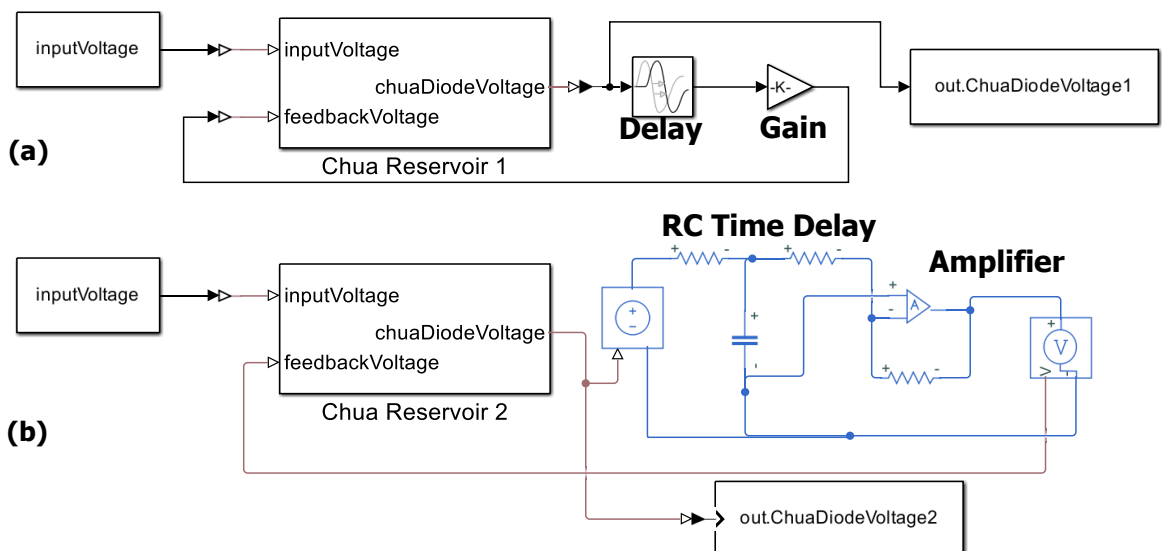

**Figure 58- (a)** Chua circuit with feedback, using Simulink, **(b)** Chua circuit with feedback implemented with electronics.

## 3.7  Physical Circuit Testing

The results achieved above in simulation can be reproduced with the physical circuit. Section 2.2.1c) has already demonstrated that the experimental circuit corresponds to the Simulink simulation, with some minor deviations. For example, where $1.92$ kΩ is at the edge of chaos for the Simulink model, the experimental circuit only requires approximately $1.85$ kΩ.

Despite the AD2 being rated for $100$ MSamples/s [13], the same frequency as used in the Simulink simulations, in practice, the AD2 was incapable of achieving this sampling rate. At any higher than approximately $5$ MSamples/s, the MATLAB functions for the AD2 returned that all samples were lost or corrupted, and the data array returned was empty. To work-around this issue, the AD2 was operated at a lower sampling rate of $10$ kSamples/s and run for longer – approximately $1$ s rather than $1$ ms. The input signal was adapted to include the desired input for only $0.01$ s, and otherwise to send $0$ V. This means that the output can be trimmed to just the first $0.01$ s, which is approximately $50$ periods of the resonant frequency, still $10x$ greater than previously used but as close as possible. Using this output directly was unsuccessful, suggesting that outputs for distinct inputs are, when obtained over a longer period, less distinct. This would make discrimination more challenging. In this case, applying the envelope function in MATLAB sufficiently improved the performance, likely because it removed overlapping content in the Chua-RC outputs. If this project did not have budgetary restrictions, higher specification devices could be used, enabling the experiment to be conducted the same at the desired sampling frequency, and removing the requirement for using the envelope function.

The circuit used had a 10 nF capacitance and 1.85 kΩ resistance. The voltage range was set to 0.4 V ≤ $v_{in}$ ≤ 1 V. Ideally, the time-averaging features of the AD2 would be used, however these are not supported on the MATLAB toolbox. Using the envelope function however removes most of the need for time-averaging. In simulation, no reset was required between test cases as the circuit simulation always starts from the same conditions. To emulate this in physical testing, the reset procedure in [9] was used to force the circuit to a stable state. A 1 V fixed signal is applied for 1 s after the test case concludes, followed by 1 s with 0 V.

Figure 59 shows the physical circuit being used for the same polynomial regression task as in Figure 50. 2900 test cases are used, with an 80% training, 20% validation split. The mean NMSE is 0.1511. Figure 59(a) shows that the performance is very good with clear tracking of the target polynomial. The performance is slightly poorer than in simulation – particularly around $x = 0.5$. However, this is to be expected from a real system and the data generation/acquisition is being operated t non-ideal sampling rates and runtimes. Figure 59(b) shows the NMSE for each validation case and there is clear similarity to Figure 50(b). Poor performance areas are zero-crossing points, where the NMSE score implies performance is far worse than reality, due to the small target value. A slight performance reduction also occurs at lower values of $x$, as previously identified. This could be a result of the input voltage amplitudes being slightly lower than optimal. Shifting the range higher would result in greater perturbation at the lower $x$ values, potentially reducing the performance loss.

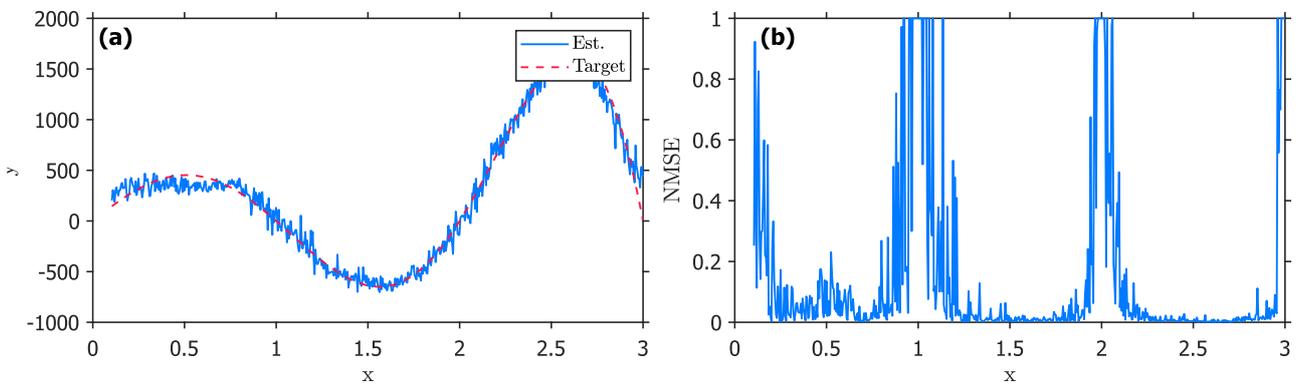

**Figure 59**- Physical circuit polynomial regression **(a)** Estimate vs target, **(b)** Validation case NMSE.

Figure 60 shows the physical circuit being used for the same classification task as in Figure 49. The system is trained with approximately 450 validation cases and 1800 training cases. As with the simulation, 100% classification accuracy is achieved, despite the number of train and validation cases being smaller. The decision surface is very similar as well to that seen in Figure 49(a). The mean average NMSE is slightly higher than that achieved by the simulation – at 0.0098 rather than 0.0075. Both are negligible. Figure 60(b) shows that again, close to 100% of the validation cases have an NMSE less than 0.05, and again 60% of the validation cases have an NMSE less than 0.005.

The physical circuit closely resembles the simulation in performance. There is a slight reduction in performance, likely due to real-world noise and to the compromise in data generation/acquisition parameters. However, this performance reduction is negligible in the classification task and extremely minimal in the regression task. This achieves a PASS and is a significant result – <u>a substantial achievement for the project.</u>

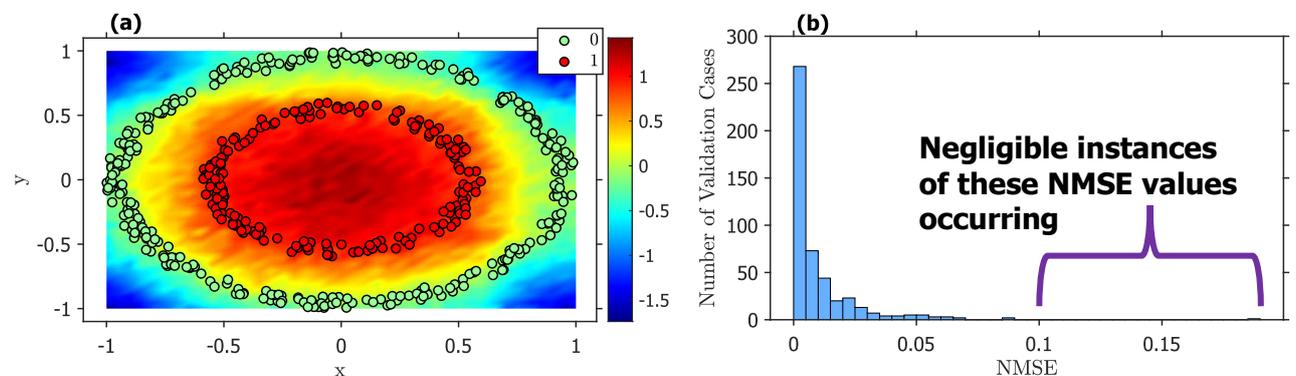

**Figure 60**- Physical Chua circuit classification of concentric circles, **(a)** Validation data classification and decision surface, **(b)** Histogram showing number of validation cases per NMSE range.

One source of concern is the consistency of the circuit. A use-case could be obtaining training data once and generating a weight that would be used going forwards. The circuit must therefore be consistent from one day to the next so that the same input voltage will produce the same output state. As the circuit is highly sensitive, small variances due to thermal noise could cause a previously generated weight to be poorly fitted to the circuit. The following section characterises the consistency of the circuit by assessing how noise variance from day-to-day impacts the Chua diode V-I characteristic.

## 3.8 Noise Performance

Figure 61(a) shows the Chua diode V-I characteristic across five different days, sampled using the AD2 at 10 kSamples/s. It demonstrates that variance due to changing system noise (such as gaussian thermal noise) has little impact on the resultant V-I characteristic. Figure 61(b) shows the same readings at 100 kSamples/s. Whilst at a higher sampling rate the readings are noisier, the V-I characteristic retains its slopes. The error bars show the extent of the change, approximately $\pm 17\%$ in the worst case. Chua diode slope gradients can impact performance, however the vast majority of gradients assessed in [9] are within the optimal region. This suggests that small changes in gradient due to noise are unlikely to shift the circuit into a non-optimal region.

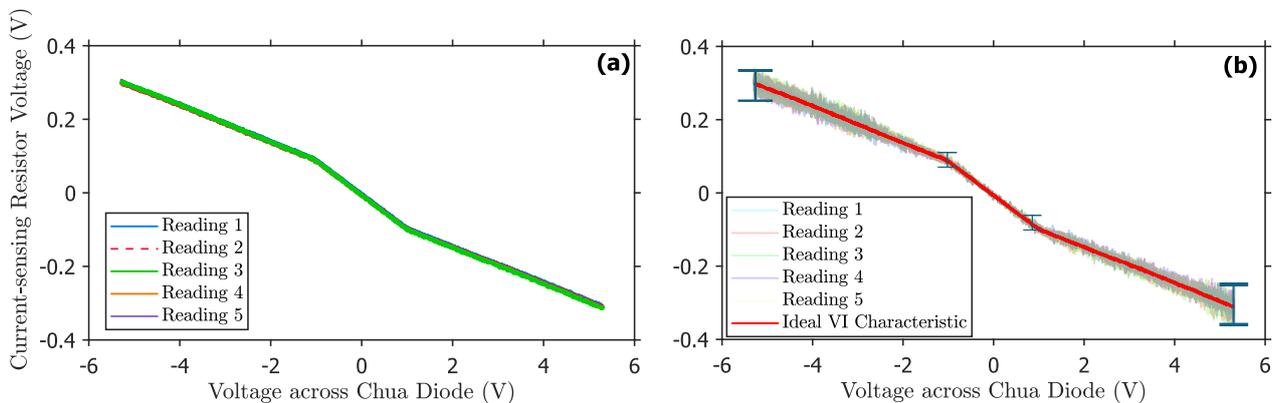

**Figure 61- (a)** V-I characteristic of Chua diode plotted across different days to see variance due to noise (10 kSamples/s), **(b)** Same V-I Chua diode characteristic at higher sampling rate (100 kSamples/s).

However, the small change could be sufficient to cause a weight previously calculated from data collected over several hours (where the circuit components will be hot from use) to be inaccurate when using with the same circuit at room temperature. This warrants further investigation. Should it prove to be a significant issue, FPGA implementations of the Chua circuit, such as [22], would likely rectify the problem by removing components that can be significantly impacted by thermal noise. This section demonstrates that whilst noise is not an immediate concern for the Chua-RC performance, it could prove a small barrier to applications of the Chua-RC where the objective is to train once, obtain a weight, and then re-use the circuit with the fixed weight. The suggested solution of implementing the Chua circuit via FPGA is an ideal source of further development.

# 4) Conclusion

## 4.1 Consideration of System Within a Wider Context

### 4.1.1 Project State Review

Table 13 lists all the core and stretch deliverables, alongside an explanation of what work has been completed and, consequently, a PASS/FAIL decision (or "not attempted – NA" in the case of some stretch objectives).

**Table 13-** State of the project.

| Core Deliverable | Work Completed | P/F |
|---|---|---|
| **D1.** Characterise Chua circuit (**Sim**) | This deliverable was completed with no issues. The work is presented in Section 2.1.1c) – a detailed characterisation of the Kennedy-based Chua circuit covering time-domain and parametric outputs, bifurcation responses to changing initial conditions, and noise performance. | PASS |
| **D2.** Chua experiment with simulation comparison (**Exp**) | This deliverable was also completed without issue. The work is presented in Section 2.2.1. The work experimentally realised the **D1** results. Where minor discrepancies were present, these are discussed and explained. | PASS |
| **D3.** Chua Reservoir Computer with MATLAB (**App**) | This deliverable was achieved – evidenced by the successful benchmarks for regression and classification tasks in Sections 3.2 and 3.7. This was achieved in both simulation and on the physical circuit. They confirmed that a new Chua-RC had been successfully developed. The work completed has extended Jensen's work by utilising a different variant of the Chua circuit, an additional output, and assessing different parameters [9]. | PASS |
| **D4.** Apply RC to performing PQ encryption. Investigate whether this is within the capability of Chua-RC (**App**) | It was empirically demonstrated (Section 3.4) that the LWE encryption problem is too challenging for the current Chua-RC system. Single-case training and validation showed that the system is correct, but it was demonstrated that the overall operation is too non-linear and high-dimensional. This deliverable has been achieved since the RC has been applied to PQ encryption, and it has been determined that a more complex RC beyond the scope of this project would be required. Applying to PQ Encryption has further extended Jensen's work by investigating time-varying inputs and more complex non-linear tasks [9]. | PASS |
| **D5.** Ensure noise is kept minimal through methods as required (**Sim/Exp**) | This deliverable has been achieved through characterising noise in simulation and experiment and taking steps to mitigate noise such as the development of a PCB. It has been demonstrated that noise is not a significant problem in the simulated and realised circuit. | PASS |
| **Stretch** | **Work Completed** | **P/F** |
| **D6.** Apply RC to performing PQ decryption (**App**) | In addition to PQ encryption, it has been empirically demonstrated (Section 3.5) that PQ decryption, whilst easier, is still too non-linear and high-dimensional for the Chua-RC to learn. Further development, beyond the scope of this project, is required. This is a PASS because the RC has been applied to performing PQ decryption and has demonstrated why a Chua-RC cannot compute this problem. | PASS |
| **D7.** Apply final system to additional PQC encryption algorithms (**App**) | As predicted, this deliverable was beyond the scope of the project and therefore was not attempted. It is likely, however, that since the high non-linearity of LWE prevents a single Chua-RC from being able to learn the operation, that other PQC algorithms would also be too challenging. This is because encryption methods tend to rely on high non-linearity, and many utilise functions such as modulo. Therefore, the work in this project on LWE can be used to infer similar results with other encryption algorithms. | N/A |
| **D8.** Publish results to an academic journal | Publishing to an academic journal was an ambitious goal. Whilst currently not attempted, this can be considered after the project's end. | N/A |

All core specification deliverables have been met, along with a stretch deliverable. An original RC with a Kennedy-based Chua kernel has been developed and benchmarked on a series of regression and classification tasks – with results taken from both simulation and a physical experiment. This is believed to be only the second Chua-RC in literature, the first to use the Kennedy design, and the first to assess more complex non-linear problems involving time-varying inputs. The system has been applied to both LWE encryption and decryption and has demonstrated in each case the limitations of the Chua-RC when applied to such challenging problems. This paper calls for further investigation into the use of multiple Chua-RCs in a single system. This could be in parallel to create additional neurons or in series to create a multi-layer neural network. This project's aim was as a research exercise to extend current work in the fields of chaotic computing and reservoir computing – with a focus towards a PQC application. In this, it has succeeded, and developed a physical prototype system, with clear benchmarks that demonstrate up to what levels of complexity the system can compute.

The challenge posed throughout the project was significant. It required obtaining a high level of knowledge about three distinct subject areas: Chua, RC, and PQC. All these subject areas were unfamiliar prior to the commencement of the project. Moreover, the project consisted of both simulation and practical work, and in addition to extensive research aims, obtaining a working system was a priority. This meant that the project was a significant piece of work and a substantial challenge. Whilst **D1** and **D2** were achievable after understanding of Chua had been obtained, there was no guarantee of success with **D3** and **D4**, particularly as Chua-RC is under-researched and has only been demonstrated once previously, and for limited applications. It is very rewarding to have achieved strong performance and to have obtained useful results about the capability of Chua-RC.

### 4.1.2  Wider Societal Context

This project impacts three areas of study. Its primary impact is in the field of Reservoir Computing, a relatively new concept in AI [11]. This is believed to be only the second example of using a Chua circuit as a basis kernel for RC and the first to assess time-varying inputs. As stated above, it is also original in other features such as the use of the Kennedy implementation and the use of multiple physical outputs taps. One of the main areas of originality is the application to PQC, specifically LWE. This means that the project also impacts on the field of Cryptography, a highly socially relevant research area. Since the basis kernel is a Chua circuit, the work is also relevant to wider research into Chua circuits and chaos theory.

The commercial potential of these fields is significant. The global AI market was estimated at \$200B in 2023 [41], and the PQC market is expected to be worth \$160M in 2024 [6]. AI is a significantly larger market and is expected to undergo 36.6% Compound Annual Growth Rate (**CAGR**) up to 2030. However, PQC is predicted to undergo an astounding 95.2% CAGR up to 2030 [6]. This is due to the threat of Quantum Computing and the public work by national and international standardisation organisations to improve cybersecurity in response to this threat. With cybersecurity underpinning $21^{st}$ century life (financial transactions, secure communications, data privacy) [5], there is significant commercial potential. This translates into more funding available for research. This is believed to be the first RC applied to PQC. Whilst it has shown that PQC is too complex for the current system, it has also shown that in theory it is possible with a higher-dimensional reservoir or more complex network. This project paves the way for future development spanning two critical, contemporary, and highly funded fields. Dedicated Graphics Processing Units (**GPU**) for AI are already the norm [41]. Although an ambitious extrapolation, it is not out of the realm of possibility to consider a future with dedicated hardware analogue RCs used in conjunction with traditional digital computing. These could handle tasks better suited to analogue processing than digital. These could include performing encryption and decryption for communications.

In the case of AI, the market growth is largely related to the advent of Large Language Models (**LLM**) such as GPT-4, and the global prominence of companies such as Google's DeepMind and OpenAI [41]. Businesses and organisations want to develop superior architectures for neural networks. Through its high-dimensional mapping, the Chua-RC can compute difficult non-linearly separable tasks that would otherwise require multi-layer neural networks [16]. Given that the Chua circuit developed in this project is inexpensive to manufacture and easy to mass produce on PCB, there is a clear financial incentive to use Chua-RCs. GPUs are designed to enable parallelism of digital operations [16]. This is simply not required by a Chua-RC as it relies on the intrinsic computing of the chaotic circuit [9].

Moreover, since the Chua circuit has low power consumption (estimated at < 0.02 W with typical circuit current < 2 mA and Op-Amp supply voltages of $\pm 9$ V) is an ideal choice to alleviate environmental concerns. One of the largest issues in the AI industry is the large power consumption for processing large machine learning models on GPUs. In 2022, data centres for AI and cryptocurrencies accounted for 2% of global energy consumption: 460 TWh [42]. If the number of layers can be significantly reduced and replaced with a simple low-power circuit, there is potential for large energy savings. Reservoir Computing, with any basis kernel, offers the opportunity to exploit a non-linear system to reduce the components and processing involved in a neural network, offering speed and power advantages.

Finally, the project impacts research into the Chua circuit and into chaotic systems. This provides results for other academics to reproduce and may promote further development into Chua-RCs. Since Reservoir Computing is a relatively new concept, and Chua-RCs are extremely under-researched, there are a vast number of potential future avenues of investigation. Some, related directly to Chua-RCs are discussed in Section 4.1.3. Others include investigating other chaotic kernels such as memristors, as suggested by Jensen [9].

In addition to the fields impacted, it is important to consider health and safety. This project is a research task that has culminated in a prototype demonstration system and is not an endeavour to develop a commercial product. However, there is commercial potential if the system can be upgraded to perform PQC. In this eventuality, health and safety standards would be critical. These standards should be achievable since the project deals with low voltages and currents. The Op-Amp power supply used is $\pm 9$ V, the typical input voltage is around 1 V and the typical circuit current is around 1 to 2 mA. As part of this project, proper PCB designs have been developed, in addition to an appropriate 3D-printed case. Whilst a commercial case would be upgraded to IP65 protection (dust/liquid), the current case provides a good baseline by reducing exposed circuitry.

### 4.1.3  Further Developments

There are two main areas that require further development and investigation. These concern the complexity of the system, and the scalability. It has been demonstrated that the current Chua-RC is incapable of transforming the input to high-enough dimensions to learn the LWE problem. One option is to attempt to add kernel methods in software, to improve the performance. These are additional software calculations performed on the Chua circuit output that can increase the overall output dimensions [1]. This, however, would conflict with the aim of the project – namely to handle computation in hardware, using the chaotic properties of the Chua circuit. A superior approach is to consider using multiple Chua-RCs in parallel, increasing the number of neurons in the hidden layer. Another method to increase the number of hidden layer neurons could be to increase the number of physical outputs from the Chua circuit [1], perhaps considering the voltages across resistors in the Chua diode. Alternatively, using multiple Chua-RCs in series could develop the system into a multi-layer neural network, as opposed to the current single hidden-layer. A final option for increasing complexity is to investigate incorporating feedback in a single Chua-RC; something briefly discussed in Section 3.6. Adding delayed feedback could add non-linear memory to the circuit. This could allow the application of Chua-RCs to other benchmarks in Reservoir Computing such as NARMA as in [11], which would unlock many other potential applications, including speech recognition [1].

The other key consideration, as mentioned, is the scalability of the system. The AD2 is a bottleneck – both in terms of sampling rate and, more crucially for scalability, voltage resolution and accuracy. In this paper, the LWE cryptosystem was fixed to have a modulus $q = 7$. This fixed the possible inputs between 0 and 6. This meant that across small voltage ranges of $< 1$ V, each distinct input number could be represented by a distinct voltage. The AD2 has limited resolution and accuracy, but at $q = 7$, it could be guaranteed that no two distinct numbers would produce overlapping voltages. If attempting this at a cryptosystem with a higher modulus $q$, a much greater accuracy and resolution would be needed to keep the input voltages distinct.

Another scalability concern is the speed of the system when a large amount of data is required for encryption. In a real-world usage of the cryptosystem, it is likely that the size of private key $s$ would be $n > 1$, leading to the private key $s$ being a vector of size $n$ and public key A being a matrix $n \times m$. In this example, the private key was fixed as a scalar, and public key A was fixed as a vector of size $m$. This change would lead to $u$ being a vector of size $n$. This would require an adaptation to the read-out training to support a vector output for $u$. More importantly, it would require additional data at the input for each new row $n$ of the $A_{\text{samples}}$. For larger values of $N_{\text{samples}}$, the amount of data at the input also increases. For each bit to be encrypted, the size of the input (prior to multiplexing, modulation, or sample hold) is shown in Equation (15).

$$size\ of\ input = (N_{\text{samples}} * n) + N_{\text{samples}} + 1 \qquad (15)$$

On the currently working benchmarks with inputs of size 1, the ideal simulation system uses $\frac{1}{5800\ \text{Hz}} \cdot 5$ periods $\approx 0.87$ ms. It can be assumed that in a more complex system where LWE is achieved, the timing would be scaled by at least the size of the input: $0.87 \cdot input\ size$ (ms). This would repeat for each encrypted bit. With $N_{\text{samples}} = 10$, $n = 5$, and a 128-bit message, the timing estimate would be $61 \cdot 128 \cdot 0.87$ ms $\approx 7$ s. There would also be additional processing time. The use of high-frequency Chua circuits, such as [12], could enable faster encryption. This would, however, be restricted by the sampling rate of the available peripheral equipment. Therefore, for both resolution and sampling rate, higher specification equipment would be required.

Whilst noise has been demonstrated in this project to be a minor concern, it is possible that in a high-frequency Chua circuit, or in a multi-layer architecture or parallel network of Chua circuits, noise becomes more significant and is problematic. It has also been identified that noise could be sufficient to prevent a previously calculated weight from consistently achieving the same performance. In either case, noise could be further alleviated using envelope detector circuits or low-pass filters built in hardware on the output of each circuit. Further mitigation could be obtained through implementing the Chua circuits directly in FPGAs as in [22].

It is arguable that LWE and PQC are a misapplication of a Chua-RC, even if a future system is sufficiently complex to compute LWE with high performance. The use of a Chua-RC removes the need for a digital computer to perform matrix multiplication and modulo. However, the training of the Chua-RC still requires large in-software matrix calculations. This training would need to be repeated for each different configuration of the cryptosystem modulus $q$, and size $n$. In this case, that may result in over-engineering a solution.

Alternatively, it could be argued that since the Chua reservoir is a black box, only the read-out training need be altered when the task is altered. Compared to other neural network approaches, this is highly beneficial. Moreover, after training it is possible to configure the fixed weight in hardware by using tuned amplifiers on the output. This further minimises software processing. Ultimately, LWE was a good test case for investigating the application of a new Chua-RC to complex non-linear time-varying input problems. The main recommendation of this project for further work would be to continue to restrict the modulus $q$ to low values and the size $n$ to 1, to not focus on system speed, and instead to focus on developing a Chua-RC-based neural network capable of solving the LWE problem. A side goal would be to investigate adding memory feedback to the Chua circuit, as this broadens the number of possible applications considerably. After these are considered, scaling up can be tackled. Figure 62(a) shows four Further Work (**FW**) objectives.

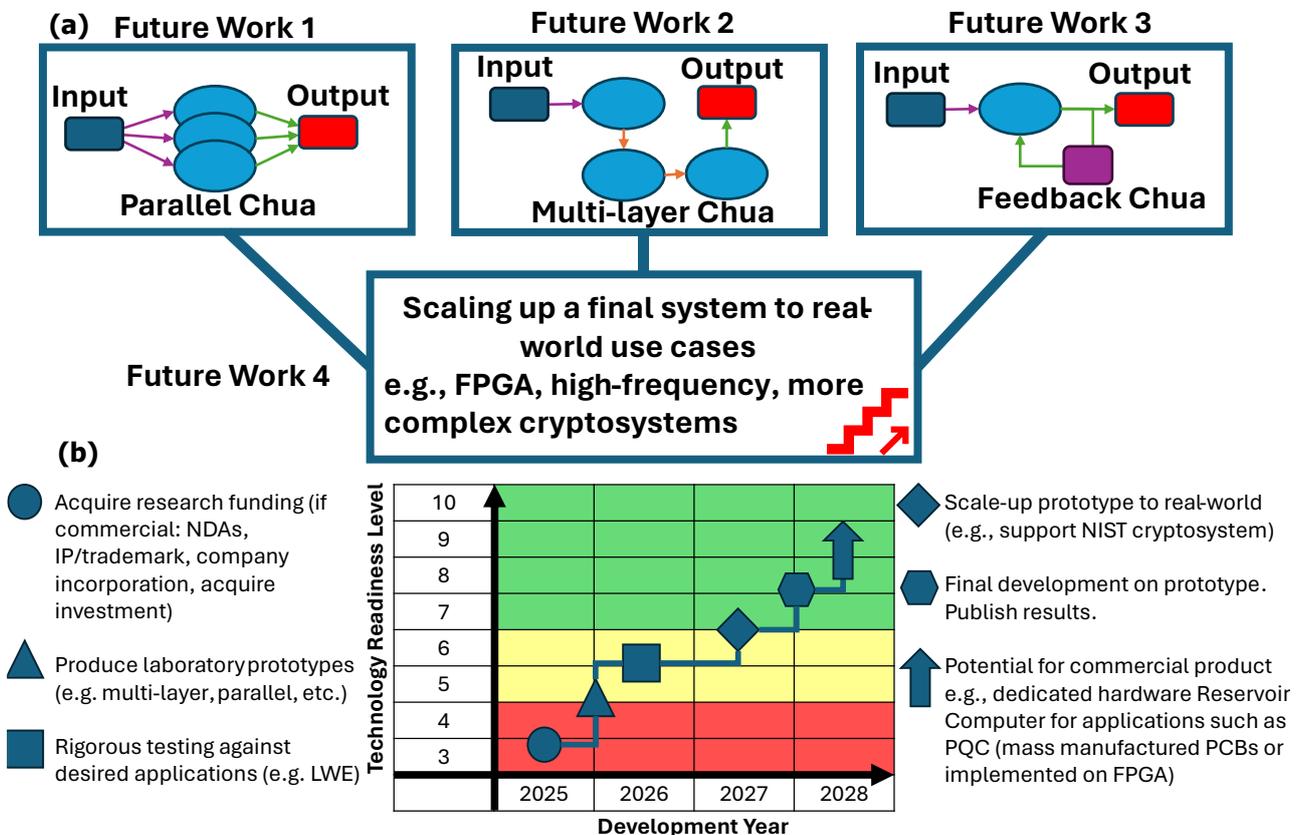

**Figure 62**- **(a)** The four main areas of future work, **(b)** Technology Readiness Level (**TRL**) roadmap.

**FW 1.** Developing a parallel Chua-RC system, with a single-hidden layer comprised of multiple Chua circuits. Applying to LWE or similar time-varying input highly non-linear problems.

**FW 2.** Developing a multi-layer neural network, first comprised of layers with a single Chua circuit, and then of layers of multiple parallel Chua circuits. Applying to LWE or similar time-varying input highly non-linear problems.

**FW 3.** Developing a Chua circuit with internal memory through applying delayed feedback – assessing its application to typical benchmarks such as NARMA prediction as in [11].

**FW 4.** Further investigation with high sampling rate/high resolution equipment into scaling the Chua-RC for real-world use-cases. FPGA implementation or use of high-frequency Chua circuit. Other investigation areas include the addition of new outputs to the Chua-RC.

For those unexperienced with Chua or RC, these could form 3 or 4 separate 1-year projects. For an experienced person, this is feasible in a single 1 or 2-year project. It is not advised to consider other cryptosystems or more complex PQC algorithms until a baseline is achieved with LWE. There is currently extremely limited research into Chua-RCs, but photonic RCs have been demonstrated to be $10^6$-fold faster and $10^3$-fold more energy efficient than traditional processors [11] and are widely used within RC [1]. Similar assessments should occur for a finalised Chua-RC system. This will enable comparisons against alternative RC implementations and can identify the areas in which Chua-RC is the optimal choice.

FW 1 and 2 would both require additional Chua circuits. Since a minimum of five PCBs can be purchased, it is reasonable to suggest these objectives could use up to five Chua circuits each. It costs approximately £30 per set of five PCBs. Only a single AD2 should still be required, but additional circuitry will be needed for connecting the different Chua circuits, likely with filtering, amplification, and possibly envelope detection as well. This could be designed in another PCB, costing another approximately £30. FW 3 would require similar additional circuitry to achieve the feedback. FW 4 would require the most expense. To achieve the desired sampling rates and resolution, very high specification equipment would be required. For example, if an application required a 0.5 V to be split into 1000 distinct, non-overlapping inputs, accurate resolution of 500 µV is required. This is achievable with highly accurate AFGs such as the Applicos 16-bit 400 MHz arbitrary waveform generator [43]. A quote has been requested and it is expected to cost significantly more than an AD2.

## 4.2  Reflection on Management

The project was well-managed, from start to completion. However, there are areas that, upon reflection, I could have approached differently. This section will cover what aspects of the project management and planning were successful and will also highlight some key changes that could have aided with the development.

The first success in the project management was the weekly use of presentation slides for each supervisor meeting, coupled with task management using Jira. At the start of the project, the objectives were not clearly defined. The brief was to develop something related to Chua circuits, possibly in the field of Reservoir Computing. This meant that in the early stages of the project, each week involved many disparate tasks as well as separate research into different applications and subject areas. Jira, a tool I previously used whilst on industrial placement, made it straightforward to group tasks, assign priority, consider dependence, and provide a sense of completion. Following Christmas, the simulation stage had been completed and the objectives and specification had been finalised. The tasks each week became fewer (though each required more time). At this point, I stopped using Jira. With such a small number of tasks, it would have become an unnecessary and cumbersome tool, superfluous in addition to the weekly presentation slides. The slides maintained a detailed log of what was achieved during the project, including any roadblocks faced and overcome, and the key decisions taken. They focussed the discussion for each week's meeting, leading to in-depth examination of ideas for next steps. These slides were provided in advance of each meeting to the supervisor and meant that each meeting was extremely useful. A high-level Gantt chart, developed for the project proposal, was effective for identifying which tasks should be considered each week. However, as the project progressed, it was followed less strictly.

One of the main difficulties with the project was the uncertainty on specific objectives. Chaotic systems, RC, and PQC were new and unfamiliar topics to me. It was initially unclear what was achievable. In large part, this project has formed organically across the year through research into the three fields. This has both been a success, and something that could be improved. The flexible way in which the project has been approached has been beneficial for both personal learning and development, and for achieving a good result. However, it would have been better to have spent more time planning and to have obtained a clearer picture of what was required to achieve objectives sooner. This would have made the Gantt chart more useful and would have caused it to be followed more strictly. In future projects I would attempt to establish a longer planning phase.

At the start of the project, the potential scope was vast. It was important to develop vague aims into a solid proposal. Progress Review 1 shows the state of the project after submission of the project proposal. The goal at this point was to have a good understanding, at a high-level, of the overall project aims. This was to be a Chua-RC applied to PQC. Progress Review 1 shows that the supervisor believed progress to be "excellent", but that important next steps included assessing the OSMZ, and determining a plan for the simulation stage. The chosen I/O device needed to meet several required specifications and was a potential blocker to future work if not determined and ordered in advance.

By Progress Review 2, this feedback had been addressed. The OSMZ was assessed and determined to be non-viable. The AD2 had been chosen instead. This was completed in advance of the plan according to the project

Gantt chart because of the supervisor feedback. With the device being critical to the project, it was important to sort early, or risk future delays. If it was determined that a device of the required specification could not be obtained, the project objectives would need to be altered. This was a risk mitigation success. Progress Review 2 also details how the simulation stage was able to exceed the original aims, by characterising capacitance bifurcation. Initial work on the RC read-out was also completed and the experiment stage was ready to begin. However, at this point the work on the application to PQC was limited. One of the difficulties with the project was that due to each of the three areas being unfamiliar, it was not possible to complete the project in a traditional linear fashion. In this case, the Chua-RC was being developed for application to PQC. Simultaneously, work was taking place to identify exactly what the application would consist of.

In hindsight, it would have been better to develop the application for use with the simulated reservoir immediately after performing circuit characterisation with the simulation. Once the reservoir application was completed in simulation, the physical circuit could be developed, characterised, and then used as a reservoir. However, at the time this was not feasible. Once the simulation and characterisation of the Chua circuit had been completed, it was necessary to develop the RC and its PQC application alongside the experiment stage. Because of Chinese New Year, it was necessary to order components and PCBs prior to Christmas to avoid potential delays. It was also challenging to properly understand the three complex new topics. At the time of starting the experiment stage, the understanding of Reservoir Computing was not sufficient to focus solely upon it. More time was required to understand the literature. Moreover, it was known that the AI module in spring semester would be very useful for better understanding the topic. As such more of the work developing the RC was pushed to after Christmas.

Progress Review 3 demonstrated a significant amount of progress. The goal was to have begun testing a physical breadboard Chua circuit. In actuality, the testing was near completion and on PCBs rather than breadboard. The choice to use PCBs was a good risk management decision. They reduce both noise and the amount of soldering. This removes opportunities for human error. The Chua circuit is very sensitive to perturbation, and initial performance on breadboard showed poor connections could prevent chaotic behaviour from occurring. Moreover, static being introduced to the circuit could damage and destroy the ICs. Therefore, the decision to use PCBs and, later, 3D-printed PCB cases, ensured superior connections and reduced the chance of static. Additional risk management was completed by ordering multiple spares of each component. This was critical when multiple ICs were destroyed by static. An oversight when designing the PCB caused the manufactured boards to be missing the vital input voltage port. This had the potential to be a major costly mistake in the project. However, the back-up was to use Veroboard instead (with a Veroboard design already created). This was not actually required, as it was possible to instead create a small "daughterboard" providing an additional port in series with the inductor. Whilst the experiment stage was significantly ahead of schedule, the work on the application was slightly behind. This was acceptable but needed to be rectified by the next Progress Review.

Progress Review 4 shows that this was addressed. The full RC I/O was developed and applied to LWE. The initial results suggested that LWE was too difficult for the Chua-RC to compute, so the next step was to perform a sweep of parameters to check if instead an optimal region was required. There was also consideration of the decryption stretch objective. The risk at this stage was that if the Chua-RC could not compute LWE, there would be no evidence as to why, and no proof that the Chua-RC worked properly at all. Benchmarks were needed to assess the system. This was also required to obtain an estimate of the system performance. This was achieved by Progress Review 5. The simulated Chua-RC was successfully benchmarked on both regression and classification tasks, in-line with the results achieved by Jensen [9]. Building-up from these tasks was able to empirically demonstrate the limitations of the system and why LWE was too challenging.

The thesis work at this point was on schedule, enabling a comprehensive draft exceeding 50 pages to be submitted for formative review. This example of excellent time management meant that the feedback on the draft would be highly relevant and useful, as opposed to submitting only a skeleton structure. However, it is also arguable that starting the thesis work so early led to wasted time as sections were rewritten and adjusted multiple times as developments occurred in the project. In future projects, I would be wary of over-committing to writing report sections too early, as substantial amounts of work may become redundant.

Progress Review 4 also includes the decision to not consider any more complex PQC algorithms. This included both scaling up the LWE cryptosystem, and other PQC algorithms such as those standardised by NIST. It was clear at this point that this was a completely unachievable task and was beyond the scope of the project. Taking this decision allowed me to focus exclusively on LWE, and to limit specific parameters so that the goal became achievable.

Progress Review 5 also shows that due to the limitations of the AD2, it was believed it would not be possible to recreate the simulated Chua-RC results experimentally. However, the step was taken to solder a 2-pole switch $\pm 9V$ battery power supply that could be utilised at home over the Easter break. This provided the flexibility to trial different workarounds whilst away from the project workshop and resulted in a compromise on AD2 settings and Chua-RC parameters that enabled successful regression and classification benchmarking on the physical Chua-RC.

As the project progressed it was important to be prepared to restrict avenues of exploration. For example, it was chosen to fix the PQC exploration to an LWE cryptosystem, and then to further fix parameters such as modulus $q$ to provide a usable test case. Being prepared to recognise what compromises must be made in the scope of a project is a very valuable attribute for project management. It is often not possible to investigate every parameter or develop something applicable to all use cases. It requires a firm decision to choose to focus on achievable goals, rather than be spread too thin. This will be brought forward to future projects.

One area for improvement would be in gaining a greater understanding of the new topics earlier. The approach taken, due to the time constraints, was to balance learning and understanding with pressing ahead with development. On the one hand, this is very beneficial as actively pushing the development of the project aided with understanding. Moreover, it was, to an extent, necessary, to identify problems early and to avoid delays when ordering components. On the other hand, it caused tasks to take longer because not fully understanding the subject area led to wasting time exploring "blind alleys". There were several instances of returning to previously read literature and realising that if the contents had been better understood the first time, far less time would have been wasted. However, this is somewhat unavoidable, particularly in the case of this project where better understanding was achieved after participating in the AI module in Spring Semester. In future projects, it would be sensible to better record the research undertaken – i.e. writing up in one's own words what in each paper is relevant to the project. Keeping a log of this would enable a better record of what was understood about each of the three topics and may have led to earlier identification of stumbling down the wrong path.

A small improvement would have been to invest a small amount of time in making the simulations more efficient. Parallel techniques in MATLAB such as parfor and parsim could have significantly sped-up running many test cases on the Chua-RC Simulink model. It is likely that the initial time investment to parallelise this process would have resulted in a net time gain. Considering trade-offs such as these and taking the decision to act is an important trait that should be applied to future projects.

As stated, the management of the project was overall completed well. The main areas of improvement relate to better fixing of objectives and specification earlier, and a more protracted research stage enabling project areas such as RC development to start earlier. Put simply – greater consolidation of information before beginning system modelling and development. This would be very useful to apply to future projects as it is likely that many industry projects will similarly involve unfamiliar fields. However, the flexibility with regards to project scope, and the structure and time-plan have been very successful features of this project's management – the ability to adjust the work as new information becomes available and to make decisions on scope and focus is a very valuable skill. The risk mitigation conducted was also very effective – with both noise and experimental difficulties alleviated through the additional work to design a PCB. Overall, the additional time spent designing the PCB was easily recouped through how much more practical using the PCB rather than breadboard and Veroboard was. Additionally, the AD2 was ordered in advance of schedule after appropriate time was spent assessing the OSMZ, as were the required components, including spares. The main mistake, failing to add an input voltage port to the PCB design, had two contingencies: the back-up Veroboard design and adding a daughterboard. Finally, good document management procedures were used, ensuring regular back-ups of this thesis. This meant that no data loss occurred in the project.

# 6) Appendix

## 6.1 Progress Reviews

### a) Progress Review 1

| Summarised Planned State of Project: | Actual Progress Since Last Review |
|---|---|
| *[This section should summarise where you expected to be with the project at this review stage according to your Time Plan.]* | *[Summarise the actual progress you have made on the project since previous Progress Review (or start of project). You should consider relevant deliverables, milestones, and Gannt chart tasks.]* |
| "Proposal complete and defence prepped for. Clear idea of next steps (simulation stage), and overall aims. Discussion of further research into PQC algorithm implementation and RC I/O." | Proposal is complete. First draft of defence slides are complete. I have outlined in my weekly slides what the next steps are for the following week: the beginning of the simulation stage and starting to working on MATLAB implementation of RC I/O and PQC algorithms. |
| About to commence work for Deliverables 1, 3, 4, and 5 | Milestone 1 is therefore successfully complete. |

**Next Steps and Supervisor Feedback**

*[In this section you should consider the differences between where you planned to be and where you are. How will this affect the project? What are you going to do to get back on track? Are there on-going problems that will have a knock-on effect to future tasks? Is there additional support or resource that is required? You should include feedback from your supervisor as discussed in the meeting and their approval of changes.]*

Milestone 1 has been reached on time and the project is on track. No additional support/resources are required currently. The simulation stage will now be begun in earnest, along with initial MATLAB work for the application stage.

Defence is prepped for.

Supervisor Feedback:
- Excellent progress so far. Next critical step is to make sure OpenScope can be used for the experiment.
- There are currently several simulation options (circuit/mathematical and MATLAB/LTSpice); these options must be finalised.

### b)  Progress Review 2

| Summarised Planned State of Project: | Actual Progress Since Last Review |
|---|---|
| *[This section should summarise where you expected to be with the project at this review stage according to your Time Plan.]* | *[Summarise the actual progress you have made on the project since previous Progress Review (or start of project). You should consider relevant deliverables, milestones, and Gannt chart tasks.]* |
| Simulation stage (**D1**) finishing up. Transitioning into Experimentation stage (**D2**), with appropriate risk mitigation (**D5**) selected. RC I/O explored and initial development on both MATLAB control system and implementing PQC algorithms begun. A next step is deciding if the OpenScopeMZ is sufficient or if an Analog Discovery 2 (AD2) is required. | **D1** is finishing up as expected and consideration of the experiment stage has begun. The simulation stage went beyond what was planned, covering capacitance and voltage bifurcation, not just resistance bifurcation. |
| | Extensive work has been completed ahead of time by determining the OpenScopeMZ was unviable early, acquiring an AD2, and developing the MATLAB-AD2 input system. |
| | Methods for RC read-out have been thoroughly investigated and implemented in MATLAB. |

| Next Steps and Supervisor Feedback |
|---|
| *[In this section you should consider the differences between where you planned to be and where you are. How will this affect the project? What are you going to do to get back on track? Are there on-going problems that will have a knock-on effect to future tasks? Is there additional support or resource that is required? You should include feedback from your supervisor as discussed in the meeting and their approval of changes.]* |
| More work on PQC within MATLAB is needed and will ramp up as expected in parallel with experimentation stage. |
| A little more work is needed on exploring noise to determine risk mitigation before experimentation stage starts |

### c)  Progress Review 3

| Summarised Planned State of Project: | Actual Progress Since Last Review |
|---|---|
| *[This section should summarise where you expected to be with the project at this review stage according to your Time Plan.]* | *[Summarise the actual progress you have made on the project since previous Progress Review (or start of project). You should consider relevant deliverables, milestones, and Gannt chart tasks.]* |
| Expected to have either just started breadboard/circuit construction or be about to start. This would allow for circuit testing, bifurcation diagram generation, etc. over next fortnight. Expected to be just starting **D2.** | **D1** (simulation) completed. **D5** risk completed through PCB design. |
| | Almost completed experiment stage – initial testing with breadboard as well as fully constructed PCB circuits. Initial testing complete, and initial bifurcation diagrams generated. **D2** almost complete. |
| Expected to be under-development of MATLAB PQC and to have completed MATLAB read-out program. | MATLAB PQC development underway/almost complete. |
| | Read-out program still under development. |
| Expected to have completed initial documentation of results and procedures | Results documentation and initial work on thesis begun |

**Next Steps and Supervisor Feedback**

*[In this section you should consider the differences between where you planned to be and where you are. How will this affect the project? What are you going to do to get back on track? Are there on-going problems that will have a knock-on effect to future tasks? Is there additional support or resource that is required? You should include feedback from your supervisor as discussed in the meeting and their approval of changes.]*

Ahead when it comes to experiment/laboratory work – **D2** almost finished – just requires additional data readings for bifurcation diagrams, and noise/bandwidth measurements.

MATLAB PQC development in-line with schedule.

Read-out program behind schedule, however this is reasonable – should be a focus going forwards, however. Previous work has been done on training methods. Potential problem in confusion of how to implement this and concern about successfulness of complete system.

Results documentation has occurred, and initial work on thesis has begun – much more is required going forwards.

## d)  Progress Review 4

| Summarised Planned State of Project: | Actual Progress Since Last Review |
|---|---|
| *[This section should summarise where you expected to be with the project at this review stage according to your Time Plan.]* | *[Summarise the actual progress you have made on the project since previous Progress Review (or start of project). You should consider relevant deliverables, milestones, and Gannt chart tasks.]* |
| **D2** achieved – Chua circuit experiment constructed and tested. <br> **D3** achieved – reservoir computer input/output developed and applied to Chua circuit to create Reservoir Computer | Finished off **D2**. Completed **D3**. |
| **D4** finishing up – apply Reservoir Computer to PQC task and optimise tuneable parameters to obtain best result. | Finishing up **D4** – however major roadblock in getting optimal results from Reservoir Computer that would make it a viable implementation of LWE. The objective of using more complex algorithms was moved to stretch objective as software implementations, particularly MATLAB implementations, of PQC are limited so would extend scope of project (**D7**). |
| Consideration of stretch objectives <br><br> Thesis writing begun | Stretch objectives such as real-time encryption, multi-bit messages, decryption (**D6**) considered with viable methods for these developed. Contingent, however, on **D4** completed with working training. |
| | Thesis writing begun and on track |

| Next Steps and Supervisor Feedback |
|---|
| *[In this section you should consider the differences between where you planned to be and where you are. How will this affect the project? What are you going to do to get back on track? Are there on-going problems that will have a knock-on effect to future tasks? Is there additional support or resource that is required? You should include feedback from your supervisor as discussed in the meeting and their approval of changes.]* <br><br> Mostly on-track as **D2** and **D3** achieved. This brings total objectives achieved so far to **D1, D2, D3, D5**. <br><br> Consideration of stretch objective **D6** and additional stretch objectives has been made, so on track. However, this is contingent on **D4** finishing well. <br><br> Overall – am on track with the plan, however the issue with **D4** is concerning and requires work and potentially an overhaul of training method to resolve. Once this is achieved, the rest of the project should be relatively "plain sailing". |

## e)  Progress Review 5

| Summarised Planned State of Project: | Actual Progress Since Last Review |
|---|---|
| *[This section should summarise where you expected to be with the project at this review stage according to your Time Plan.]* | *[Summarise the actual progress you have made on the project since previous Progress Review (or start of project). You should consider relevant deliverables, milestones, and Gannt chart tasks.]* |
| **D4** completed – Reservoir Computer applied to PQC. | **D4** completed – applied RC to PQC task and by process of benchmarking its capability on sub-tasks (non-linear polynomial, modulo, multi-value, etc.) demonstrate that PQC is too "hard" for Chua-RC. |
| Draft Thesis completed. | Very good results for classification and non-linear regression tasks |
| Concluded work on stretch goals. | Issues with achieving these experimentally, however this is due to sampling rate of equipment in practice not achieving the theoretical potential (which is required). This is not a problem as circuit characterisation plus simulation proves it would work given more expensive equipment. |
| Discussed presentation with supervisor | The same work is performed for PQC Decryption, achieving **D6.** |
|  | Thesis draft completed with 50/60 pages of final thesis done – very good progress. |
|  | Discussion of presentation with supervisor occurred with plan for how to present complicated topics in accessible way without over-simplifying. |

**Next Steps and Supervisor Feedback**
*[In this section you should consider the differences between where you planned to be and where you are. How will this affect the project? What are you going to do to get back on track? Are there on-going problems that will have a knock-on effect to future tasks? Is there additional support or resource that is required? You should include feedback from your supervisor as discussed in the meeting and their approval of changes.]*

Project has been successful – it is a shame that the PQC is too difficult for the Chua-RC, however the project has demonstrated this effectively and has pushed the boundary with Chua-RCs by extending to time-varying inputs, more complex tasks, a different circuit structure including an additional physical tap, and assessing different optimisation parameters.

Stretch goal of decryption also successfully completed, with the same evidence as with PQC Encryption.

Thesis work ahead of schedule so everything on track.